%% file: HIG-17-025_temp.tex
\pdfoutput=1

\documentclass[11pt,twoside,a4paper,cmspaper,final,collab]{cms-tdr}
\def\svnVersion{488197P}\def\cmsCernNoTag{CERN-EP-2018-166}\def\cmsCernDate{\today}\def\cmsMessage{Published in the Journal of High Energy Physics as \href{http://dx.doi.org/10.1007/JHEP01(2019)183}{\doi{10.1007/JHEP01(2019)183.}}}

\begin{document}\cmsNoteHeader{HIG-17-025}

\hyphenation{had-ron-i-za-tion}
\hyphenation{cal-or-i-me-ter}
\hyphenation{de-vices}
\RCS$HeadURL: svn+ssh://svn.cern.ch/reps/tdr2/papers/HIG-17-025/trunk/HIG-17-025.tex $
\RCS$Id: HIG-17-025.tex 481702 2018-11-19 09:29:38Z vtavolar $
\newlength\cmsFigWidth
\ifthenelse{\boolean{cms@external}}{\setlength\cmsFigWidth{0.85\columnwidth}}{\setlength\cmsFigWidth{0.4\textwidth}}
\ifthenelse{\boolean{cms@external}}{\providecommand{\cmsLeft}{top\xspace}}{\providecommand{\cmsLeft}{left\xspace}}
\ifthenelse{\boolean{cms@external}}{\providecommand{\cmsRight}{bottom\xspace}}{\providecommand{\cmsRight}{right\xspace}}
\newlength\cmsTabSkip\setlength{\cmsTabSkip}{1ex}
\providecommand{\cmsTable}[1]{\resizebox{\textwidth}{!}{#1}}

\newcommand{\integratedlumi}{\ensuremath{35.9\fbinv}\xspace}
\newcommand{\mgg}{\ensuremath{m_{\gamma\gamma}}\xspace}
\newcommand{\mH}{\ensuremath{m_{\PH}}\xspace}
\newcommand{\ptgg}{\ensuremath{\pt^{\gamma\gamma}}\xspace}
\newcommand{\Hgg}{\ensuremath{\PH\to\Pgg\Pgg}\xspace}
\newcommand{\Zee}{\ensuremath{\cPZ\to\Pep\Pem}\xspace}
\newcommand{\Zmm}{\ensuremath{\cPZ\to\Pgmp\Pgmm}\xspace}
\newcommand{\Zmmg}{\ensuremath{\cPZ\to\Pgmp\Pgmm\Pgg}\xspace}
\newcommand{\Wenu}{\ensuremath{\PW\to\Pe\Pgn}\xspace}
\newcommand{\ttH}{\ensuremath{\ttbar\PH}\xspace}
\newcommand{\mylumi}{\integratedlumi}
\newcommand{\sqrts}{\ensuremath{\sqrt{s}}\xspace}
\newcommand{\jone}{\ensuremath{j_{\text{1}}}}
\newcommand{\jtwo}{\ensuremath{j_{\text{2}}}}
\newcommand{\fid}{\ensuremath{\text{fid}}}
\newcommand{\njet}{\ensuremath{N_{\text{jet}}}\xspace}
\newcommand{\nbjet}{\ensuremath{N_{\text{jet}}^{\cPqb}}\xspace}
\newcommand{\nlep}{\ensuremath{N_{\text{lepton}}}\xspace}
\newcommand{\absrapgg}{\ensuremath{\abs{y^{\gamma\gamma}}}\xspace}
\newcommand{\abscosthetast}{\ensuremath{\abs{\cos(\theta^*)}}\xspace}
\newcommand{\ptjone}{\ensuremath{\pt^{\jone}}\xspace}
\newcommand{\absrapjone}{\ensuremath{\abs{y^{\jone}}}\xspace}
\newcommand{\absDphiggjone}{\ensuremath{\abs{\Delta\phi^{\gamma\gamma,\jone}}}\xspace}
\newcommand{\absDrapggjone}{\ensuremath{\abs{\Delta y^{\gamma\gamma,\jone}}}\xspace}
\newcommand{\ptjtwo}{\ensuremath{\pt^{\jtwo}}\xspace}
\newcommand{\absrapjtwo}{\ensuremath{\abs{y^{\jtwo}}}\xspace}
\newcommand{\absDphijonejtwo}{\ensuremath{\abs{\Delta\phi^{\jone,\jtwo}}}\xspace}
\newcommand{\absDphiggjonejtwo}{\ensuremath{\abs{\Delta\phi^{\gamma\gamma,\jone\jtwo}}}\xspace}
\newcommand{\zeppenf}{\ensuremath{\abs{\overline{\eta}_{\jone\jtwo} - \eta_{\gamma\gamma}}}\xspace}
\newcommand{\mjonejtwo}{\ensuremath{m^{\jone\jtwo}}\xspace}
\newcommand{\absDetajonejtwo}{\ensuremath{\abs{\Delta\eta^{\jone,\jtwo}}}\xspace}
\newcommand{\sigmaMoM}{\ensuremath{\sigma_{m}}\xspace}
\newcommand{\sigmaMoMdecorr}{\ensuremath{\sigma_{m}^{D}}\xspace}
\newcommand{\DeltaR}{\ensuremath{\Delta R}\xspace}
\newcommand{\Peta}{\ensuremath{\eta^0}\xspace}
\newcommand{\Lk}{\ensuremath{\mathcal{L}}\xspace}
\newcommand{\Lt}{\ensuremath{\widetilde{\mathcal{L}}}\xspace}
\cmsNoteHeader{HIG-17-025}

\title{Measurement of inclusive and differential Higgs boson production cross sections in the diphoton decay channel in proton-proton collisions at $\sqrt{s}=13$\TeV}

\date{\today}

\abstract{
Measurements of the inclusive and differential production cross sections for the Higgs boson in the diphoton decay channel are performed using the data set of  proton-proton collisions at $\sqrt{s}=13\TeV$ collected by the CMS experiment at the LHC in 2016 and corresponding to an integrated luminosity of $35.9\fbinv$.
The cross sections are measured in a fiducial phase space defined by a set of requirements on the isolation and kinematic variables of the photons.
Differential cross sections are measured as functions of the kinematic properties of the diphoton system and the event.
A subset of the measurements is performed in regions of the fiducial phase space, where relative contributions of specific Higgs boson production mechanisms are enhanced. The total cross section in the chosen fiducial phase space is measured to be $84\pm11\stat \pm7\syst\unit{fb} = 84\pm13\unit{fb}$, to be compared with a theoretical prediction of $73\pm4\unit{fb}$. All measurements are found to be in agreement with the theoretical predictions for the standard model Higgs boson with a mass of $125.09$\GeV  within the experimental and theoretical uncertainties.}

\hypersetup{%
pdfauthor={CMS Collaboration},%
pdftitle={Measurement of inclusive and differential Higgs boson production cross sections in the diphoton decay channel in proton-proton collisions at sqrt(s) = 13 TeV},%
pdfsubject={CMS},%
pdfkeywords={CMS, physics, Higgs, diphoton, differential, cross section}}

\maketitle

\section{Introduction}
\label{sec:Introduction}

The discovery of a Higgs boson (\PH) was announced in 2012 by the ATLAS and CMS
Collaborations\,\cite{Aad:2012tfa,Chatrchyan:2012ufa,Chatrchyan:2013lba} based on proton-proton (\Pp\Pp) collision data
collected at the CERN LHC at center-of-mass energies of $7$ and $8$\TeV. Since its discovery, an extensive
campaign of measurements~\cite{Aad2016} has been underway  to characterize the new particle and
test its properties against those predicted by the standard model (SM) of particle
physics.
By comparing measured cross sections with predictions, as functions of the kinematic properties of the diphoton system and of the particles produced in association with the Higgs boson,  it is possible to investigate the dynamics of Higgs boson production, decay, and accompanying jet activity.

 These investigations are expected to give insights into the nature of the Higgs boson
 and enable testing of  the perturbative quantum chromodynamics (QCD) predictions for Higgs boson production. Both
the ATLAS and CMS Collaborations have presented results on the measurement of inclusive and differential cross
sections for production of the Higgs boson in {\Pp\Pp} collisions at $\sqrt{s}=8$\TeV
in the diphoton \cite{Aad2014,Khachatryan2016}, four-lepton \cite{2014234,Khachatryan:2118088}, and $\PW\PW$ \cite{Aad:2145362,Khachatryan:2158105} decay channels. Both Collaborations have also presented measurements of inclusive and differential production cross sections in the four-lepton final state at $\sqrt{s}=13$\TeV \cite{Aaboud:2277731,Sirunyan:2272260}.

Production of the Higgs boson in {\Pp\Pp} collisions at the LHC occurs via four main mechanisms: gluon-gluon fusion
($\Pg\Pg\PH$), vector boson fusion (VBF), associated production with a {\PW}/{\cPZ} boson (V\PH), and
associated production with a top quark-antiquark pair (\ttH). At the center-of-mass energy of $13$\TeV, $\Pg\Pg\PH$ production is about one order of magnitude larger than the sum of the other production mechanisms.
The SM prediction of the branching fraction for the \Hgg decay is only about 0.2\%~\cite{LHCHXSWG:YR4} but this channel has a clean signature and it is possible to reconstruct the
diphoton invariant mass  with high precision.  The most precise measurements of differential cross sections of Higgs boson production can be made in this decay channel.
The dominant sources of background are irreducible prompt diphoton production, and the reducible processes $\Pp\Pp\to \gamma + \text{jets}$  and $\Pp\Pp\to \text{multijets}$, where the jets are misidentified as
 photons.

In this paper we report the measurement of the inclusive and differential cross
sections for Higgs boson production in the diphoton decay channel using
data corresponding to an integrated luminosity of \mylumi of {\Pp\Pp} collisions at $\sqrts=13$\TeV recorded by the CMS experiment
in 2016. The aim of the analysis is to perform measurements of the Higgs boson production cross section in a fiducial phase space, to be compared with theoretical predictions. The methods used closely follow those developed for the \Hgg
differential cross section measurements at $\sqrts=8$\TeV~\cite{Khachatryan2016} and are designed to measure the Higgs boson production as a function of the  final state kinematic observables with a minimal dependence on theoretical assumptions, allowing a direct comparison between the experimental results and the theoretical predictions. In contrast, the complementary approach adopted in~\cite{CMS-PAS-HIG-16-040} aims at maximizing the observation sensitivity for the SM Higgs boson by explicitly relying on theoretical predictions and their uncertainties.

For each bin of the
differential observables, the signal is extracted by fitting to a narrow signal peak on top
of the steeply-falling background spectrum of the diphoton invariant mass distribution. To improve the precision of the measurements,
 the events are categorized using a diphoton mass resolution
estimator.
Both inclusive and differential cross sections are
measured and unfolded within a fiducial phase space defined by the requirements on the photon kinematic variables and
 isolation.
Differential cross sections are measured as
functions of several observables, describing the properties of the diphoton system and of ({\cPqb} quark) jets, leptons,
 and  missing transverse  momentum accompanying the diphoton system.
A double-differential cross section measurement is also performed as a function of
the transverse momentum (\pt) of the diphoton system and the number of
additional jets in the event.
Cross section measurements are also performed in regions of the fiducial phase space. The regions are chosen to enhance the contribution of specific production mechanisms to the signal composition, based on the additional particles produced in association with the diphoton system and on the topology of the event.

\section{The CMS detector}
\label{sec:CMS}
The central feature of the CMS apparatus is a superconducting
solenoid of 6\unit{m} internal diameter, providing a magnetic
field of 3.8\unit{T}. Within the solenoid volume are a silicon
pixel and strip tracker, a lead tungstate crystal
electromagnetic calorimeter (ECAL), and a brass and scintillator
hadronic calorimeter (HCAL), each composed of a barrel and two
endcap sections. Forward calorimeters extend the pseudorapidity ($\eta$)
coverage provided by the barrel and endcap detectors.
Muons are detected in gas-ionization chambers embedded in the
steel flux-return yoke outside the solenoid.

The electromagnetic calorimeter consists of 75\,848 lead tungstate
crystals, which provide coverage in pseudorapidity
$\abs{\eta} < 1.48 $ in a barrel region (EB) and
$1.48 < \abs{\eta} < 3.0$ in two endcap regions (EE).
Preshower detectors consisting of two planes of silicon
sensors interleaved with a total of $3 X_0$ of lead are
located in front of each EE detector.

In the region $\abs{\eta} < 1.74$, the HCAL cells have
widths of 0.087 in pseudorapidity and 0.087 in azimuth
($\phi$). In the $\eta$-$\phi$ plane, and for
$\abs{\eta} < 1.48$, the HCAL cells map on to $5\times5$
arrays of ECAL crystals to form calorimeter towers projecting
radially outwards from close to the nominal interaction point.
For $\abs{ \eta} > 1.74$, the coverage of the towers increases
progressively to a maximum of 0.174 in $\Delta \eta$ and
$\Delta \phi$.

The forward hadron (HF) calorimeter uses steel as an absorber
 and quartz fibers as the sensitive material. The two halves of the
 HF are located 11.2\unit{m} from the interaction region, one on each
 end, and together they provide coverage in the range $3.0 < \abs{\eta} < 5.2$.
 They also serve as luminosity monitors.

Events of interest are selected using a two-tiered trigger system~\cite{Khachatryan:2016bia}. The first level (L1), composed of custom hardware processors, uses information from the calorimeters and muon detectors to select events at a rate of around 100\unit{kHz} within a time interval of less than 4\mus. The second level, known as the high-level trigger (HLT), consists of a farm of processors running a version of the full event reconstruction software optimized for fast processing, and reduces the event rate to around 1\unit{kHz} before data storage.

A more detailed description of the CMS detector, together
 with a definition of the coordinate system used and the
 relevant kinematic variables, can be found in
Ref.~\cite{Chatrchyan:2008zzk}.

\section{Data samples and simulated events}
\label{sec:samples}

The events used in the analysis were selected by a diphoton trigger with asymmetric \pt thresholds of 30 (18)\GeV on the leading (sub-leading) photon, a minimum invariant diphoton mass \mgg  of 90\GeV, and loose requirements on the calorimetric isolation and electromagnetic shower shape of the photon candidates. The trigger selection is $>$99\% efficient at retaining events passing the selection requirements described in Section~\ref{sec:selection}.

A detailed simulation of  the CMS detector response is based on a model implemented using the \GEANTfour~\cite{Agostinelli:2002hh} package.
Simulated events include the effects of pileup (additional {\Pp\Pp} interactions from the same or nearby bunch crossings)  and
are weighted to reproduce the distribution of the number of interactions in data.

The signal samples are simulated with $\MGvATNLO$ v2.2.2~\cite{Alwall:2014hca} at next-to-leading order (NLO) in perturbative QCD  with FxFx merging~\cite{Frederix:2012ps} for the $\Pg\Pg\PH$, VBF, V\PH, and \ttH production processes. These samples include production of up to two additional jets in association with the Higgs boson. The parton-level samples are interfaced to $\PYTHIA 8.205$~\cite{Sjostrand:2014zea} with the CUETP8M1~\cite{CUETP8M1} underlying event tune, for parton showering, underlying event modeling, and hadronization. In order to match the prediction for $\Pg\Pg\PH$ production mechanism from the {\textsc{nnlops}} program~\cite{Hamilton:2013fea,PowhegMinlo,Kardos:2014dua}, the generated events are weighted according to the Higgs boson \pt and the number of jets in the event.
The {\textsc{nnlops}} program has the advantage of predicting at next-to-next-to-leading-order (NNLO) accuracy, both the differential cross section with respect to the QCD radiative effects and the normalization of the inclusive cross section.
The $\Pg\Pg\PH$ samples are also generated with the \POWHEG v2 program~\cite{powheg1,powheg2,powheg3,powheg-ggH,Bagnaschi2012}, which includes production of one additional jet, in order to provide an alternative theoretical prediction  for inclusive measurements and measurements involving the highest-\pt jet in the event.
The NNPDF3.0 set~\cite{Ball:2014uwa} is used for parton distribution functions (PDFs).
The SM Higgs boson cross sections and branching fractions are taken from the LHC Higgs Cross Section Working Group report~\cite{LHCHXSWG:YR4}.

Simulated background samples are used for training multivariate discriminants, and to define selection and classification criteria.
The irreducible prompt diphoton background events are generated using the \SHERPA v2.2.1 program~\cite{Gleisberg:2008ta}. This program includes the tree-level matrix elements with up to three additional jets and the box diagram at leading order accuracy. The reducible background arising from $\gamma+\mathrm{jet}$ and multijet events is modeled with $\PYTHIA$.

Samples of \Zee, \Zmm, and \Zmmg simulated events are generated with \MGvATNLO
and used for comparison with data and for the derivation of energy scale and resolution corrections.

\section{Event reconstruction}
\label{sec:reco}

Photon candidates are reconstructed from clusters of energy deposited in the ECAL and merged into superclusters~\cite{CMS:EGM-14-001}. The reconstruction algorithm for photon clusters allows almost complete recovery of the energy from photons that convert to an electron-positron pair in the material upstream of the ECAL. A detailed description of the algorithm can be found in Ref.~\cite{1748-0221-10-06-P06005}. Changes in the transparency of the ECAL crystals due to irradiation
during the LHC running periods and their subsequent recovery are monitored continuously and corrected for, using light injected from the laser and LED systems \cite{Chatrchyan:2013dga}.

A multivariate regression technique is used to correct for the partial containment of the shower in a supercluster, the shower losses for photons
that convert in the material upstream of the calorimeter, and the effects of pileup. Training is performed on simulated events
using  shower shape  and position variables of the photon as inputs. The photon energy response distribution is parametrized by an extended form of the Crystal Ball
function~\cite{Gaiser:1982yw} built out of a
Gaussian core and two power law tails. The regression provides a per-photon estimate of the function
parameters, and therefore a prediction of the distribution of the ratio of
true energy to the uncorrected supercluster energy. The most probable value of this
distribution is taken as the photon energy correction. The width of the Gaussian core is
used as a per-photon estimator of the relative energy resolution $\sigma_E/E$.

In order to obtain the best energy resolution, the calorimeter signals are calibrated and corrected for several detector effects~\cite{Chatrchyan:2013dga}. Calibration of the ECAL uses photons from $\PGpz\to\gamma\gamma$ and $\Peta\to\gamma\gamma$ decays, and electrons from \Wenu and \Zee decays. The energy scale in data is aligned to that in simulated events, while an additional smearing is applied to the reconstructed photon energy in simulation in order to reproduce the resolution observed in data, through a multistep procedure exploiting electrons from \Zee decays.

In  the ECAL barrel section, an energy resolution of
about 1\% is achieved for unconverted or late-converting
photons, \ie, photons converting near the inner face of the ECAL, that have energies in the range of tens of GeV.
The remaining photons reconstructed in the barrel have a resolution of about
1.3\% up to a pseudorapidity of $\abs{\eta} = 1$, rising
to about 2.5\% at $\abs{\eta} = 1.4$. In the endcaps, the
resolution of unconverted or late-converting photons is
about 2.5\%, while the remaining  endcap photons have a
resolution between 3 and 4\%~\cite{CMS:EGM-14-001}.

The global event reconstruction (also called particle-flow event reconstruction~\cite{CMS-PRF-14-001}) aims to reconstruct and identify each individual particle in an event, with an optimized combination of all subdetector information. In this process, the identification of the particle type (photon, electron, muon, charged hadron, neutral hadron) plays an important role in the determination of the particle direction and energy. Photons (\eg, coming from \Pgpz\ decays or from electron bremsstrahlung) are identified as ECAL energy clusters not linked to the extrapolation of any charged particle trajectory to the ECAL. Electrons (\eg, coming from photon conversions in the tracker material or from \PQb quark semileptonic decays) are identified as a primary charged particle track and potentially many ECAL energy clusters, corresponding to this track extrapolation to the ECAL and to possible bremsstrahlung photons emitted along the way through the tracker material. Muons (\eg, from \PQb quark semileptonic decays) are identified as a track in the central tracker consistent with either a track or several hits in the muon system, associated with an energy deficit in the calorimeters. Charged hadrons are identified as charged particle tracks neither identified as electrons, nor as muons. Finally, neutral hadrons are identified as HCAL energy clusters not linked to any charged hadron trajectory, or as ECAL and HCAL energy excesses with respect to the expected charged hadron energy deposit.

The energy of photons is obtained from the ECAL measurement. The energy of electrons is determined from a combination of the track momentum at the main interaction vertex, the corresponding ECAL cluster energy, and the energy sum of all bremsstrahlung photons attached to the track. The energy of muons is obtained from the corresponding track momentum. The energy of charged hadrons is determined from a combination of the track momentum and the corresponding ECAL and HCAL energy, corrected for zero-suppression effects and for the response function of the calorimeters to hadronic showers. Finally, the energy of neutral hadrons is obtained from the corresponding corrected ECAL and HCAL energy.

For each event, hadronic jets are clustered from either particle-flow candidates (for data and simulation)
or stable particles excluding neutrinos (for generated events) using
the infrared and collinear-safe anti-\kt algorithm~\cite{Cacciari:2008gp, Cacciari:2011ma} with a distance parameter of 0.4.
The jet momentum is determined as the vectorial sum of momenta of all objects
clustered into the jet. Extra proton-proton interactions within the same or nearby bunch crossings
can contaminate the jet reconstruction. To mitigate this effect,
particle-flow candidates built using tracks originating from pileup vertices are discarded and an
offset correction is applied to account for remaining contributions~\cite{CMS-PAS-JME-16-003}.
Additional selection criteria are applied to each jet to remove jets potentially
dominated by anomalous contributions from various subdetector components
or reconstruction failures. The momenta of jets reconstructed using particle-flow candidates in simulation
are found to be within 5 to 10\% of particle-level jet momenta
over the whole jet \pt spectrum and detector acceptance,
and corrected on average accordingly. In situ measurements of the momentum balance in dijet,
$\text{photon} + \text{jet}$, $\cPZ + \text{jet}$, and multijet events are used to account for any residual differences
in jet energy scale in data and simulation~\cite{Khachatryan:2016kdb}. The jet energy resolution amounts typically
to 15\% at 10\GeV, 8\% at 100\GeV, and 4\% at 1\TeV.

Jets originating from the hadronization of {\cPqb} quarks are identified using the combined secondary vertex (CSV) \cPqb-tagging algorithm \cite{Sirunyan:2017ezt}. The algorithm converts information on the displaced secondary vertex into a numerical discriminant, assigning high values to jets whose properties are  more likely to be originating from {\cPqb} quarks. A tight working point on this discriminant is used in this analysis, which provides a misidentification rate for jets from light quarks and gluons of $0.1$\% and an efficiency for identifying {\cPqb} quark jets of about $55$\%.

The missing transverse momentum \ptvecmiss, whose magnitude is referred to as \ptmiss,
is defined as the negative vectorial sum of the transverse momenta of all reconstructed
particle flow candidates in the global event reconstruction.

Because no tracks are associated to photons, the assignment of the diphoton candidate to a vertex can only be done indirectly by exploiting the properties of each reconstructed vertex. Three discriminating variables are calculated for each reconstructed vertex: the sum of the squared transverse momenta of the charged-particle tracks associated with the vertex, and two variables that quantify the vector and scalar balance of \pt between the diphoton system and the charged-particle tracks associated with the vertex. In addition, if either photon has an associated charged-particle track that has been identified as originating from a photon conversion to an electron-positron pair, the conversion information is used. The variables are used as the inputs to a multivariate classifier based on a boosted decision tree (BDT) to choose the reconstructed vertex to be associated with the diphoton system. The average vertex finding efficiency of this algorithm is about 81\%~\cite{CMS-PAS-HIG-16-040}. The vertex is considered to be correctly identified if it is within 1\unit{cm} of the true vertex in the longitudinal direction. The contribution to the diphoton mass resolution from vertex displacements smaller than 1\unit{cm} is found to be negligible compared to the contribution from the photon energy  resolution of the calorimeters.

A photon identification algorithm separates prompt photons from photon candidates resulting from the misidentification of jet fragments~\cite{Khachatryan:2014ira}. These are mostly collimated photons from neutral-hadron decays (\PGpz, \Peta). The algorithm is implemented with a BDT trained on simulated events. The input variables of the BDT are: the pseudorapidity and energy of the supercluster corresponding to the reconstructed photon, several variables characterizing the shape of the electromagnetic shower, and the isolation energy sums computed with the particle-flow algorithm~\cite{CMS-PRF-14-001}. Further information on the photon identification BDT can be found in~\cite{CMS-PAS-HIG-16-040}.

\section{Event selection}
\label{sec:selection}

Each photon of the candidate pair entering the analysis is required to have a supercluster within
$\abs{\eta}<2.5$, excluding the region $1.4442<\abs{\eta}<1.566$, which corresponds to the ECAL barrel-endcap transition region, and to satisfy selection criteria, described in Ref.~\cite{CMS-PAS-HIG-16-040},
slightly more stringent than the trigger requirements, based on
transverse momentum, isolation, and shower shape variables. The
transverse momentum scaled by the invariant mass of the diphoton candidate
($\pt/\mgg$) has to be greater than 1/3 (1/4) for the \pt-leading (\pt-subleading) photon. The use of thresholds in $\pt/\mgg$, rather than fixed thresholds in \pt,  prevents
the distortion of the low end of the \mgg spectrum.
Furthermore, each photon must
 fulfill a requirement based on the output of the photon identification classifier, chosen as explained in Section~\ref{sec:categorization}.

Jets are selected if they fulfill the pileup rejection criteria \cite{CMS-PAS-JME-13-005} and have $\pt>30$\GeV. To avoid double counting of photon candidates as jets, the minimum distance between each photon and a jet is required to satisfy $\DeltaR(\gamma,\mathrm{jet}) = \sqrt{\smash[b]{\abs{\Delta\eta(\gamma, \mathrm{jet})}^2 + \abs{\Delta\phi(\gamma, \mathrm{jet})}^2}} > 0.4$, where $\Delta\eta(\gamma, \mathrm{jet})$ and $\Delta\phi(\gamma, \mathrm{jet})$ are the pseudorapidity and azimuthal angle differences between the photon and the jet. Two collections of jets are selected in different pseudorapidity regions: $\abs{\eta}<2.5$ and $\abs{\eta}<4.7$. The two collections are used to study differential observables requiring  at least one or two selected hadronic jets in the event, respectively.
The jets in the $\abs{\eta}<2.5$ collection benefit from tracker information and this results in better reconstruction quality and energy resolution;  when requiring two jets in the same event, the  $\abs{\eta}$ range is extended to $4.7$ to increase the selection acceptance.
 The same kinematic selection is applied to generator-level jets.
Jets with $\abs{\eta}<2.4$ are identified as {\cPqb} jets at the reconstruction level if they satisfy the requirements
described in Section~\ref{sec:reco}. At the generator level, at least one {\PB} hadron has to be clustered in a jet to be called a {\cPqb} jet.

Leptons (electrons and muons) are selected if they have $\pt>20 $\GeV and  $\abs{\eta}<2.4$. The angular separation between the photon and the lepton $\DeltaR(\gamma,\mathrm{lepton})$ is required to be greater than 0.35. Electrons must satisfy a set of loose requirements as described in Ref.~\cite{CMS-DP-2015-067} and they are not selected in the pseudorapidity region $1.4442<\abs{\eta}<1.566$. Furthermore, the invariant mass of the candidate electron and either of the two photons is required to be at least $5$\GeV  from the nominal {\cPZ} boson mass, in order to reject $\cPZ+\Pgg\to\Pep\Pem\Pgg$ events with a misidentified electron. Muons are required to pass a tight selection based on the
quality of the track fit, the number of associated hits in the tracking detectors, and the longitudinal and transverse impact parameters of the track with respect to the event vertex and to satisfy a requirement on the relative isolation, corrected for pileup effects,  calculated as the sum of the transverse energy of charged hadrons, neutral hadrons and photons in a cone of radius $0.4$ around the muon.
Generator-level leptons, required to satisfy the same kinematic selection, are ``dressed'', \ie, the four-momenta of all photons in a cone of radius $\DeltaR=0.1$ around the lepton
are added to the four-momentum of the lepton.

The identification and trigger efficiencies are measured using data events containing a {\cPZ} boson decaying to a pair of electrons, or to a pair of electrons or muons in association with a photon~\cite{CMS:EGM-14-001}.
After applying corrections, based on control samples in data, to the input of the photon identification classifier, the efficiencies measured in data are found to be 3 (5)\% lower than in simulation for photons in the barrel (endcap) regions with $\RNINE < 0.85$ ($0.9$), where $\RNINE$ is defined as the sum of the energy measured in a $3\times3$ crystal matrix, centered on the crystal with the highest energy in the ECAL cluster of the candidate, divided by the energy of the candidate. Photon candidates undergoing a conversion before reaching the ECAL have wider shower profiles and lower values of $\RNINE$. A correction factor is applied to simulated events to take into account the discrepancy in the efficiency between data and simulation. For the remaining photons, the predicted efficiencies are compatible with the ones measured in data.

\section{Mass resolution estimator}
\label{sec:sigmaM}

The selected photon pairs are categorized according to their estimated relative mass resolution. For the typical energy range of the photons used in this analysis, corresponding to tens of GeV, the energy resolution estimator depends on the energy itself because of the stochastic and noise terms in the energy resolution of the ECAL~\cite{Chatrchyan:2013dga,CMS:EGM-14-001}. The nature of these two terms is such that the energy resolution improves at higher energy. This dependence is propagated to the relative mass resolution estimator \sigmaMoM, which is thus dependent on the mass of the diphoton pair, with events characterized by a larger diphoton mass more likely to have better mass resolution. An event categorization simply based on such a variable would distort the shape of the mass distribution  in the different categories and it would make the background distribution more complex to parametrize. In particular, a deficit of low-mass events would be observed in categories corresponding to low values of \sigmaMoM, invalidating the assumption of a smoothly falling mass distribution on which the background model, described in Section~\ref{sec:bgmodel}, is based.
To avoid such an effect, the correlation between \sigmaMoM and the diphoton mass is removed, following the methods in Ref.~\cite{Khachatryan2016}, and a new relative mass resolution estimator is built, \sigmaMoMdecorr.

\begin{figure}
 \begin{center}
   \includegraphics[width=0.45\textwidth]{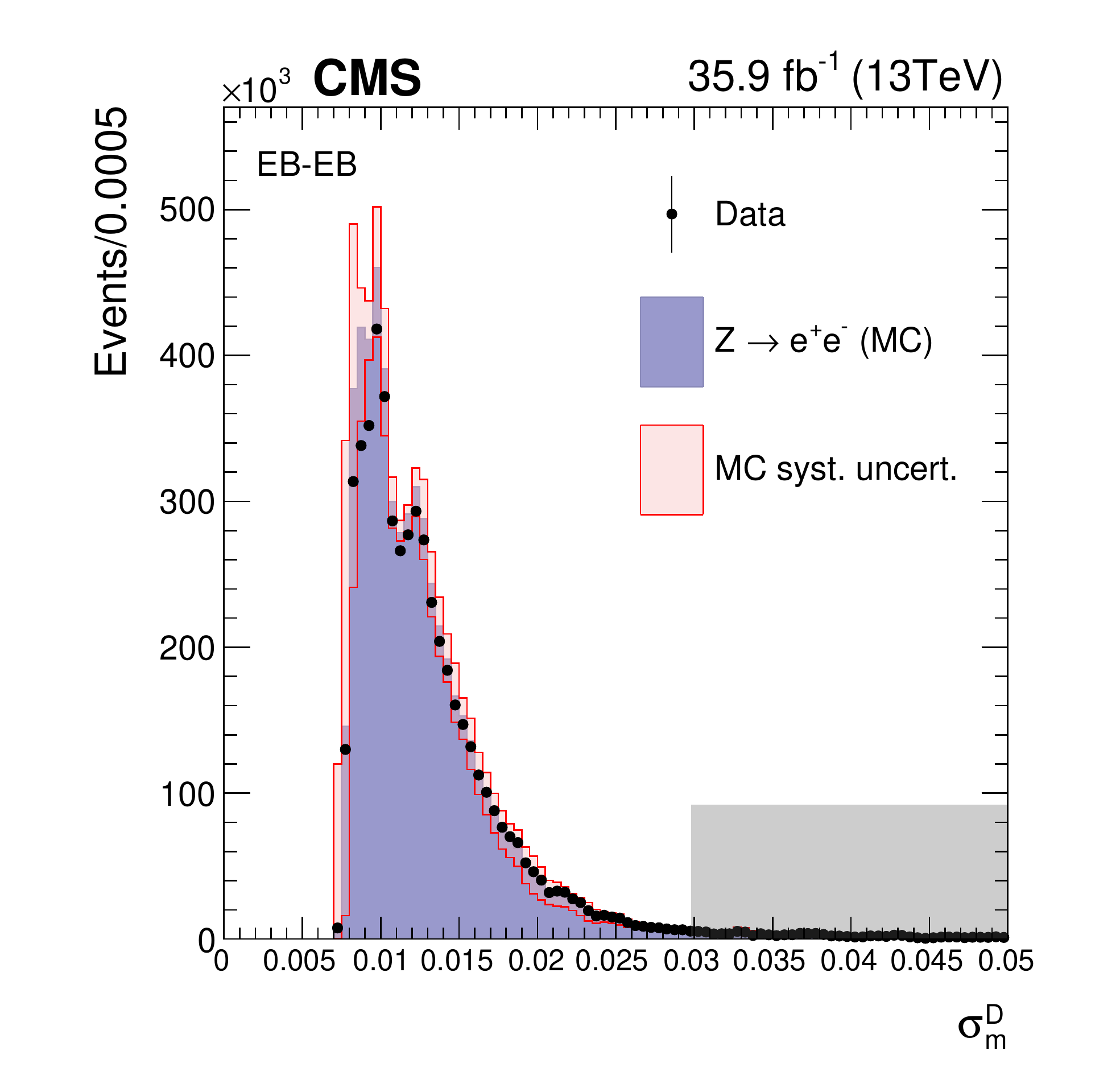}
   \includegraphics[width=0.45\textwidth]{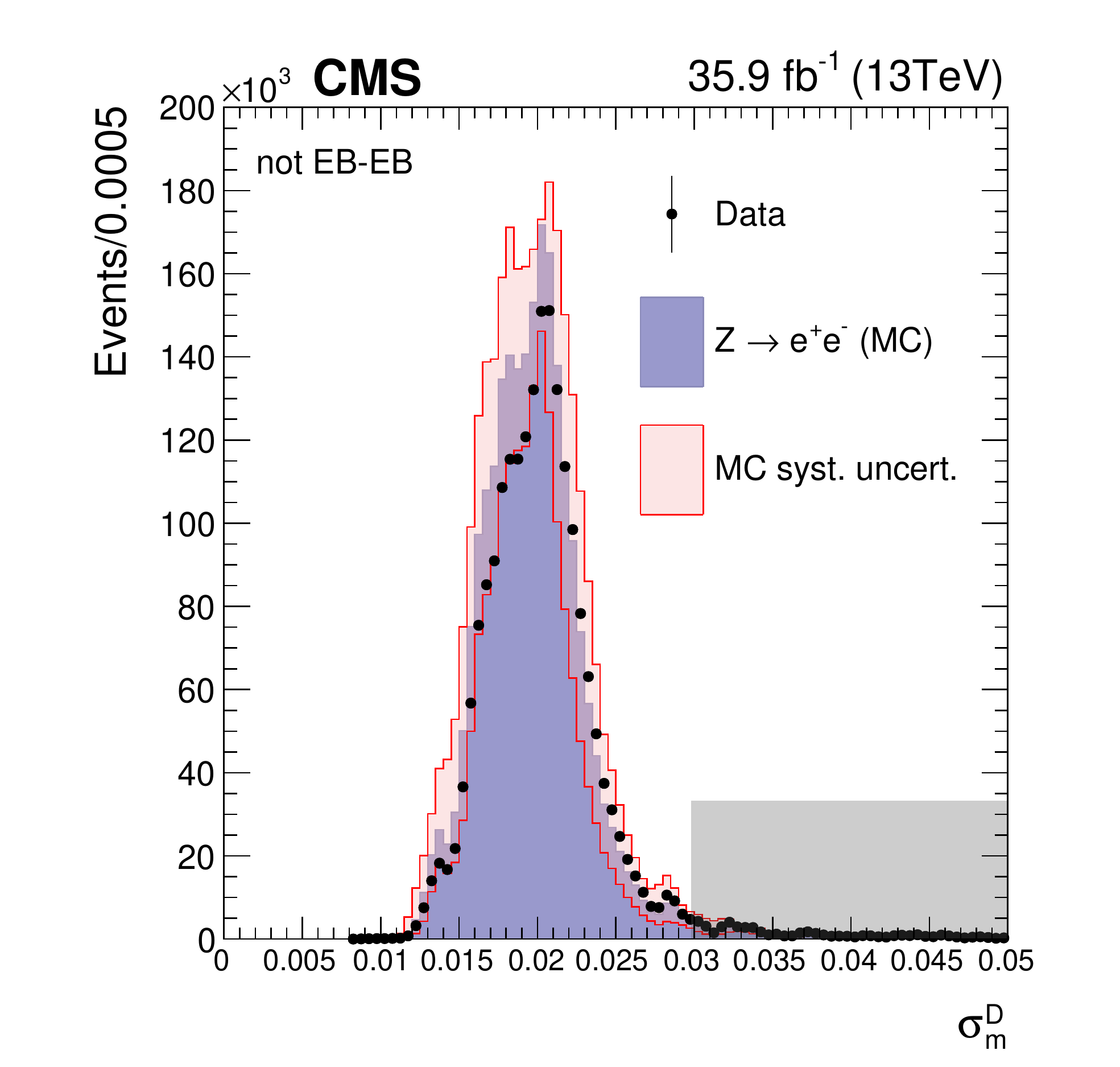}
   \caption{Comparison of the decorrelated mass resolution estimator \sigmaMoMdecorr distributions in data and simulation for \Zee events where both electrons are reconstructed as photons, passing the selection defined in Section~\ref{sec:selection}. The impact of the systematic uncertainty in the $\sigma_E/E$ is indicated by the red band. The distributions are shown separately for events with both electrons  in the EB (left) and the remainder of the events, \ie, events with at least one photon in the EE (right). Events in the shaded gray region are discarded from the final analysis.}
  \label{fig:sigmaMoverM}
 \end{center}
\end{figure}

The modeling of the decorrelated mass resolution estimator is studied with simulated \Zee events, where
electrons are reconstructed as photons. The per-photon resolution estimate $\sigma_{E}/E$ is affected by the imperfect modeling of the electromagnetic shower shape variables in simulation, which are among the inputs of the regression used to estimate  $\sigma_{E}/E$, as described in Section~\ref{sec:reco}.
To minimize the disagreement~\cite{CMS-DP-2017-004}, the per-photon resolution estimate is recomputed using  as input
simulated shower shapes corrected to match those observed in data. A systematic
uncertainty of 5\% is assigned to the value of
$\sigma_{E}/E$ for each photon candidate, to cover the residual
discrepancy. Figure \ref{fig:sigmaMoverM} shows the comparison between data (dots) and simulation (histogram) for the
decorrelated mass resolution estimator \sigmaMoMdecorr, with the impact of the
systematic uncertainty in the $\sigma_E/E$ (red band). Events with a value of \sigmaMoMdecorr in the region depicted in gray are discarded from the final analysis.

\section{Event categorization}
\label{sec:categorization}
Events with both photons passing a minimum requirement on the output of the photon identification classifier, are sorted into categories of \sigmaMoMdecorr to maximize the analysis sensitivity
to the SM Higgs boson. The number of categories and the  positions of their \sigmaMoMdecorr boundaries are optimized
simultaneously with the lower bound on the output of the photon identification
classifier.
Three categories, labeled 0, 1, and 2 in ascending order of \sigmaMoMdecorr values,  are found adequate to saturate the maximum sensitivity achievable
with this method for the present data set.
The boundaries of the \sigmaMoMdecorr categorization are found to be $0$, $0.0084$, $0.012$, $0.030$, with a minimum requirement on the output of  photon identification classifier.
Events with $\sigmaMoMdecorr>0.030$ are discarded (shaded gray region in Fig.~\ref{fig:sigmaMoverM}).
The efficiency of the photon identification selection is roughly 80\% for
signal events in the fiducial phase space, discussed in Section~\ref{sec:observables}. The categories obtained from the
optimization process correspond approximately to the configurations where both photons are
reconstructed in the central barrel ($\abs{\eta}<1$) for the first category, both photons
are reconstructed in the barrel with at least one falling outside the central barrel for
the second category, and at least one photon reconstructed in the endcap regions of the ECAL for the
last category.

\section{Observables and fiducial phase space}
\label{sec:observables}

The  analysis  provides  measurements of  the production cross section of the
Higgs boson in a fiducial phase space. This is defined by a set of selection criteria at generator level based on kinematic, geometrical and isolation variables, as well as on the topology of the event.
By defining a fiducial phase space, the measurements are compared to the theoretical predictions while avoiding the extrapolation to the full phase space and the consequent uncertainty.
In order to extend such a comparison to future and alternative theoretical calculations, it is important to have a simple definition of the fiducial phase space  so that it can be easily reproduced.
Furthermore, the selection criteria in data, described in Section~\ref{sec:selection}, are necessarily defined at the reconstruction level, while the
fiducial phase space, for which theoretical predictions are computed,  is defined without
considering the effect of the detector response on the generator-level quantities. Because of the finite detector resolution, the two
definitions do not exactly coincide, and for this reason events fulfilling the event
selection criteria at the reconstruction level can originate from either inside or outside the
 fiducial phase space.
To minimize the effect of events migrating, the selection criteria at the reconstruction level and the definition of this phase space are aligned as closely as possible.

The fiducial phase space for the analysis is defined by requiring that the generator-level ratio between the \pt of the \pt-leading (\pt-subleading) photon and \mgg, $\pt^{\gamma_1}/\mgg$ ($\pt^{\gamma_2}/\mgg$), be greater than 1/3 (1/4), and that the
absolute pseudorapidity of both photons be less than 2.5.

In addition,
the sum of the generator-level transverse energy of stable particles in a cone of radius $\DeltaR=0.3$ around each photon candidate, $\text{Iso}_{\text{gen}}^{\gamma}$, is required to be less than 10\GeV. This requirement mimics at generator level the requirement on the output of the photon identification classifier applied on reconstructed quantities, as described in Section~\ref{sec:selection}.
Further requirements, that depend on the observable under study, can be applied on top of this ``baseline'' phase space definition.
 For observables involving only one jet, events with at least one jet  with $\abs{\eta^j}<2.5$, selected as described in Section~\ref{sec:selection}, are retained, corresponding to $\sim$35\% of the signal events in the baseline phase space. Observables involving two jets are studied by requiring at least two jets with $\abs{\eta^j}<4.7$ and defined as in Section~\ref{sec:selection}, further restricting the region of the phase  space to $\sim$16\% of the baseline selection.
A VBF-enriched region of the fiducial phase space, where a subset of the two-jet observables is measured, is defined by requiring the presence of two reconstructed and selected jets within $\abs{\eta^j}<4.7$, with a combined invariant mass $\mjonejtwo$ greater than $200$\GeV and a pseudorapidity gap between the jets $\absDetajonejtwo$ greater than $3.5$, exploiting the main kinematic features of the VBF production mode. This set of criteria selects $\sim$3.8\% of the signal events contained in the baseline phase space.
The definition of the four regions of the fiducial phase space is summarized in Table~\ref{tab:observables}, which also gives
a summary of the observables under study and the bins chosen in each phase space.
The symbol $\jone$ ($\jtwo$) indicates the \pt-leading (subleading) hadronic jet in the event, while $y$ is used to denote the rapidity of a particle or a system of particles. The transverse momentum and the rapidity of the diphoton system, indicated with $\ptgg$ and $\absrapgg$, respectively, are sensitive probes of the Higgs boson production mechanism, the modeling of the QCD radiation, and the PDFs of the proton.
The cosine of the polar angle in the Collins--Soper reference frame of the diphoton system~\cite{Collins:1977iv}, $\abscosthetast$, probes the spin and CP properties of the diphoton resonance.
Observables involving jets are sensitive to the QCD parameters relevant to Higgs boson production. The separation in the azimuthal angle between the diphoton and the two-jet systems, $\absDphiggjonejtwo$, and the Zeppenfeld variable, $\zeppenf$~\cite{Rainwater:1996ud},  probe specifically the properties of the VBF production mechanism.
The number of jets within $\abs{\eta}<2.5$, {\cPqb} jets, and leptons are indicated with \njet, \nbjet, and \nlep, respectively.

\begin{table}[htbp]
    \centering
  \topcaption{The differential observables studied with the corresponding bins chosen, grouped by the region of the fiducial phase space where the measurements are performed. \label{tab:observables}
    }
\cmsTable{
        \begin{tabular}{lllllllllll}
          \hline
          Phase space region & Observable & \multicolumn{8}{c}{Bin boundaries} \\
          \hline
          \multirow{10}{*}{\parbox{4.0cm}{Baseline \newline $\pt^{\gamma_1}/\mgg > 1/3$ \newline $\pt^{\gamma_2}/\mgg > 1/4$ \newline$ \abs{\eta^\gamma} < 2.5$ \newline $\textrm{Iso}_{\textrm{gen}}^{\gamma}<10$\GeV}} & $\ptgg$ (\GeVns) & 0  & 15 & 30 & 45 & 80 & 120 & 200 & 350 & $\infty$ \\
           & $\njet$ & 0 & 1 & 2 & 3 & 4 & $\infty$ & & & \\
           & $\absrapgg$ & 0 & 0.15 & 0.3 & 0.6 & 0.9 & 2.5 & & & \\
           & $\abscosthetast$ & 0 & 0.1 & 0.25 & 0.35 & 0.55 & 1 & & & \\
           & $\ptgg$ (\GeVns), $\njet=0$ & 0  & 20 & 60 & $\infty$ & & & & & \\
           & $\ptgg$ (\GeVns), $\njet=1$ & 0  & 60 & 120 & $\infty$ & & & & & \\
           & $\ptgg$ (\GeVns), $\njet>1$ & 0  & 150 & 300 & $\infty$ & & & & & \\
           & $\nbjet$ & 0  & 1 & 2 & $\infty$ & & & & & \\
           & $\nlep$ & 0  & 1 & 2 & $\infty$ & & & & & \\
           & $\ptmiss$ (\GeVns) & 0  & 100 & 200  & $\infty$ & & & & & \\[\cmsTabSkip]
          \multirow{4}{*}{\parbox{4.0cm}{1-jet \newline Baseline + $\geq$1 jet \newline $\pt^{j}>30\GeV$, $\abs{\eta^{j}}<2.5$}} & $\ptjone$ (\GeVns) & 0 & 45 & 70 & 110 & 200 & $\infty$ & & & \\
          & $\absrapjone$ & 0 & 0.5 & 1.2 & 2 & 2.5  &  & & & \\
          & $\absDphiggjone$ & 0 & 2.6 & 2.9 & 3.03 & $\pi$ & & & & \\
          & $\absDrapggjone$ & 0 & 0.6 & 1.2 & 1.9 & $\infty$ & & & & \\[\cmsTabSkip]
          \multirow{7}{*}{\parbox{4.0cm}{2-jets \newline Baseline + $\geq$2 jets \newline $\pt^{j}>30\GeV$, $\abs{\eta^{j}}<4.7$}} & $\ptjtwo$ (\GeVns) & 0 & 45 & 90 & $\infty$ & & & & & \\
           & $\absrapjtwo$ & 0 & 1.2 & 2.5 & 4.7 & & & & & \\
           & $\absDphijonejtwo$ & 0 & 0.9 & 1.8 & $\pi$ & & & & & \\
           & $\absDphiggjonejtwo$ & 0 & 2.9 & 3.05 & $\pi$ & & & & & \\
           & $\zeppenf$ & 0 & 0.5 & 1.2 & $\infty$ & & & & & \\
           & $\mjonejtwo$ (\GeVns) & 0 & 100 & 150 & 450 & 1000 & $\infty$ & & & \\
           & $\absDetajonejtwo$ & 0 & 1.6 & 4.3 & $\infty$ & & & & & \\[\cmsTabSkip]
          \multirow{3}{*}{\parbox{4.0cm}{VBF-enriched \newline 2-jets + $\absDetajonejtwo>3.5$, $\mjonejtwo>200$\GeV }} & $\ptjtwo$ (\GeVns) & 0 & 45 & 90 & $\infty$ & & & & & \\
           & $\absDphijonejtwo$ & 0 & 0.9 & 1.8 & $\pi$ & & & & & \\
           & $\absDphiggjonejtwo$ & 0 & 2.9 & 3.05 & $\pi$ & & & & & \\
          \hline
        \end{tabular}
}
\end{table}

The inclusive fiducial cross section is also measured in restricted regions of the fiducial phase space, defined using additional criteria as follows:
\begin{itemize}
\item at least one lepton, at least one \cPqb-tagged jet, referred to as the \emph{$\geq$1-lepton, $\geq$1-\emph{\cPqb}-jet} fiducial cross section ($\sim$1.7$\times 10^{-3}$ of the baseline phase space);
\item exactly one lepton, \ptmiss $\geq$100\GeV, referred to as the \emph{1-lepton, high-\emph{\ptmiss}} fiducial cross section ($\sim$1.5$\times 10^{-3}$ of the baseline phase space);
\item exactly one lepton, \ptmiss $<$100\GeV, referred to as the \emph{1-lepton, low-\emph{\ptmiss}} fiducial cross section ($\sim$7.4$\times 10^{-3}$ of the baseline phase space).
\end{itemize}

The first and second of these definitions loosely reproduce the event selections described in Ref.~\cite{CMS-PAS-HIG-16-040}, which respectively target \ttH and $\PW\PH$ production mechanisms, with the {\PW} boson decaying leptonically.
The third definition selects a region complementary to the second, populated mostly by events where the Higgs boson is produced in association with either a {\PW} or a {\cPZ} boson.

For all the regions of the baseline phase space, the events contained in the baseline phase space that fail the additional requirements of a given region are collected in an additional bin (referred to as the ``underflow'') and used to provide an additional constraint on the measurements, in particular to correctly account for migrations occurring between the baseline phase space and the region and to allow the profiling of the value of the Higgs boson mass in the signal-extraction fit, described in Section~\ref{sec:results}.

\section{Statistical analysis}

\label{sec:analysis}

The events fulfilling the selection criteria are grouped into three categories, according to their \sigmaMoMdecorr, as described in Section~\ref{sec:categorization}. For each category, the final categorization employed for the signal extraction is obtained by further splitting the events  into the bins defined for each observable, as described in Section~\ref{sec:observables}.
The signal production cross section is extracted through a simultaneous extended maximum likelihood fit to the diphoton invariant mass spectrum in all the analysis categories. The likelihood in a given \sigmaMoMdecorr category $i$ and in given kinematic bin $j$ of an observable is reported in Eq.~(\ref{eq:likelihood1}):

\begin{equation}
  \label{eq:likelihood1}
  \begin{split}
&   \mathcal{L}_{ij}(\text{data} | \Delta \vec{\sigma}^{\fid}, \vec{{n}}_{\text{bkg}} , \vec{\theta_{\mathrm{S}}},
      \vec{\theta_{\mathrm{B}}}) =  \\
&   \prod_{l=1}^{{n}_{\mgg}}
          \left ( \frac{\sum_{k=1}^{{n}_{b}} \Delta \sigma_{k}^{\fid} K_{k}^{ij}(\vec{\theta_{\mathrm{S}}})
              S_{k}^{ij}(\mgg^l|\vec{\theta_{\mathrm{S}}})  L + n_{\mathrm{OOA}}^{ij}
              S_{\mathrm{OOA}}^{ij}(\mgg^l|\vec{\theta_{\mathrm{S}}})
              + n_{\text{bkg}}^{ij}  B^{ij}
              (\mgg^l|\vec{\theta_{\mathrm{B}}})}{ n_{\text{sig}}^{ij} + n_{\text{bkg}}^{ij} } \right )^{{n}^{lij}_{\text{ev}}},
  \end{split}
\end{equation}

where:
\begin{itemize}
\item  ${n}_{\mgg}$ is the number of bins of the \mgg distribution and ${n}_{\mathrm{b}}$ is the number of kinematic bins for the given observable;
\item $\Delta \vec{\sigma}^{\fid} = (\Delta \sigma_1^{\fid}, \dots ,\Delta \sigma_{n_{\mathrm{b}}}^{\fid})$ is the
  vector of fiducial cross sections being measured,  multiplied by the branching fraction of the diphoton decay channel;
\item $K_{\mathrm{k}}^{ij}$ are the response matrices, which represent the efficiency that
  an event in the $k$-th kinematic bin at generator level is reconstructed
  in the $ij$-th reconstruction-level category (with the index $i$ running over the
  \sigmaMoMdecorr categories and the index $j$ running on the kinematic bins);
\item the functions $S_{k}^{ij}$ and $B^{ij}$ are the signal and background
  probability distribution functions in \mgg for the bin $ijk$, which are described in the Sections~\ref{sec:sigmodel} and~\ref{sec:bgmodel}, respectively;
\item $L$ is the total integrated luminosity analyzed;
\item ${n}^{ij}_{\text{ev}}$, ${n}_{\text{sig}}^{ij}$, ${n}_{\text{bkg}}^{ij}$ are the numbers of observed, signal and background events in the $ij$th reconstruction-level category, respectively;
\item the terms ${n}_{\mathrm{OOA}}^{ij} S_{\mathrm{OOA}}^{ij}$ represent
  the contributions to the Higgs boson signal originating outside of the fiducial phase space.
  The contribution of the out-of-acceptance (OOA) Higgs boson signal is estimated from simulation to be approximately 1\% of the
  total expected SM signal;
\item the parameters $\vec{\theta_{\mathrm{S}}}$ and $\vec{\theta_{\mathrm{B}}}$ are the nuisance
  parameters associated with the signal and background models, respectively.
\end{itemize}
The complete likelihood is given in Eq.~(\ref{eq:likelihood2}):

\begin{equation}
  \label{eq:likelihood2}
       \mathcal{L}(\text{data} | \Delta \vec{\sigma}^{\fid}, \vec{{n}}_{\text{bkg}} , \vec{\theta_{\mathrm{S}}},
      \vec{\theta_{\mathrm{B}}}) =
         \prod_{i=1}^{\mathrm{n_\text{cat}}}
          \prod_{j=1}^{{n}_{\mathrm{b}}}
          \mathcal{L}_{ij}
       \mathrm{Pois}( n^{ij}_{\text{ev}} | n_{\text{sig}}^{ij} + n_{\text{bkg}}^{ij})  \mathrm{Pdf}(\vec{\theta_{\mathrm{S}}})  \mathrm{Pdf}(\vec{\theta_{\mathrm{B}}}) ,
\end{equation}
where:
\begin{itemize}
\item ${n}_{\text{cat}}$ is the number of categories in \sigmaMoMdecorr;
\item $\mathrm{Pois}$ and $\mathrm{Pdf}$ indicate the Poisson distribution and the nuisance parameters probability density function, respectively.
\end{itemize}

The unfolding to the particle-level cross sections is achieved by extracting the vector $\Delta \vec{\sigma}^{\fid}$ directly from the likelihood fit, providing unfolded unregularized cross sections. No regularization of the results is applied, since the bins chosen are sufficiently larger than the resolution for a given observable.  The uncertainties and the correlation matrices are obtained from the test statistic $\mathrm{q}(\Delta\vec{\sigma}^{\fid})$ defined  below and asymptotically distributed as a $\chi^2$ with $n_{\mathrm{b}}$
degrees of freedom~\cite{Cowan:2010st}:
\begin{equation}
  \mathrm{q}(\Delta \vec{\sigma}^\fid) = - 2 \log \left (
  \frac{ \mathcal{L}( \Delta \vec{\sigma}^\fid       | \hat{\vec{\theta}}_{\Delta \vec{\sigma}^\fid} ) }
       { \mathcal{L}( \Delta \hat{\vec{\sigma}}^\fid | \hat{\vec{\theta}} )  } \right ),
\end{equation}
where $\vec{\theta} = ( n_{\text{bkg}}, \vec{\theta}_{\mathrm{S}},\vec{\theta}_{\mathrm{B}} )$.
The notations $\hat{\vec{\theta}}$ and $\Delta \hat{\vec{\sigma}}^\fid$ represent the best fit estimate of $\vec{\theta}$ and $\Delta \vec{\sigma}^\fid$, respectively, and $\hat{\vec{\theta}}_{\Delta \vec{\sigma}^\fid}$ indicates the best fit estimate of  $\vec{\theta}$, conditional on the value of $\Delta \vec{\sigma}^\fid$.
 The nuisance parameters, including the Higgs boson mass, are profiled in the fit across all the bins.

\subsection{Signal model}
\label{sec:sigmodel}
For each observable, a parametric signal model is constructed separately for each fiducial-level bin (including an extra bin collecting the OOA events), reconstruction-level bin, and category in \sigmaMoMdecorr.
Since the shape of the \mgg distribution is significantly different for events where the vertex has been correctly identified  compared to other events, these two components are modeled separately.
The model is built as a fit to a sum of up to five Gaussian distributions of the simulated invariant mass shape, modified by the trigger, reconstruction, and identification efficiency corrections estimated from data control samples, for each of the three values of \mH\ $\in\{120, 125, 130\}\GeV$. Signal models for other nominal values of \mH\ between $120$ and $130\GeV$ are produced by interpolating the fitted parameters. The final signal model for a given category and a reconstruction-level bin is obtained by summing the functions, normalized to the expected signal yields, for each fiducial-level bin and vertex identification scenario.

\subsection{Background model}
\label{sec:bgmodel}

A background model is produced for every bin of the observable and for each of the three categories in \sigmaMoMdecorr. A discrete profiling method~\cite{Dauncey:2014xga}, originally developed for the \Hgg decay observation analysis \cite{Khachatryan:2014ira}, is used. The background is evaluated by fitting to the $\mgg$ distribution in data over the range $100<\mgg<180\GeV$.

The choice of the function used to fit the background in a particular event class is included as a discrete nuisance parameter in the formulation of the likelihood. Exponentials, power-law functions, polynomials in the Bernstein basis, and Laurent polynomials are used to represent $B(\mgg|\vec{\theta_{B}})$ in Eq.~(\ref{eq:likelihood1}).
A signal-plus-background hypothesis is fit  to data by minimizing the value of twice the negative logarithm of the likelihood. All functions  are tried, with a ``penalty term'' added to account for the number of free parameters in the fit. The penalized likelihood function $\Lt_B$ for a single fixed background fitting function $B$ is defined as:
\begin{equation}
  -2\,\ln\Lt_B=-2\,\ln\Lk_B+N_{B},
\end{equation}
where $\Lk_B$ is the ``unpenalized'' likelihood function and $N_{B}$ is the number of free parameters in $B$. When fitting the complete likelihood, the number of degrees of freedom (number of exponentials, number of terms in the series, degree of the polynomial, etc.) is increased until no significant improvement
occurs in the likelihood between $N+1$ and $N$ degrees of freedom for the fit to the data distribution. The improvement is quantified by extracting the $p$-value from the F-distribution between the fits  using $N+1$ and $N$ degrees of freedom and requiring it to be smaller than $0.05$.

\section{Systematic uncertainties}
\label{sec:systematics}

Systematic uncertainties listed in this section are included in the likelihood as nuisance parameters and are profiled during the minimization. Unless specified otherwise, the sources of uncertainty refer to the individual quantity studied, and not to the final yield.
 The total uncertainty in the inclusive and differential measurements is dominated by the statistical uncertainties.

The systematic uncertainties affecting the shape of the \mgg distribution are treated as Gaussian variations. Those considered in this analysis are as follows:
\begin{itemize}

\item \textit{Vertex finding efficiency:} the largest contribution to the
  uncertainty comes from the modeling of the underlying event, plus the
  uncertainty in the measurement of the ratio of data and
  simulation efficiencies obtained using \Zmm events. It is handled
  as an additional nuisance parameter built into the signal model that allows
  the fraction of events in the right vertex/wrong vertex scenarios to
  change. The size of the uncertainty in the vertex selection efficiency is
  1.5\%;
\item \textit{Energy scale and resolution:} these corrections are studied with
  electrons from \Zee and then applied to photons. The main source of
  systematic uncertainty is the different interactions of electrons and photons
  with the material upstream from the ECAL.  Uncertainties are assessed by changing the
  $\RNINE$ distribution, the energy regression training (using electrons instead of
  photons), and the electron selection used to derive the corrections. The uncertainties in the different $\abs{\eta}$ and $\RNINE$ bins are propagated to the Higgs boson
  signal phase space in order to estimate the
  uncertainty in the additional energy smearing.
    In both cases, dedicated nuisance parameters are included
  as additional systematic terms in the signal model and amount to less than
  about 0.5\%, depending on the photon category.
\end{itemize}

The sources of systematic uncertainty having an impact mainly on the category yield, while leaving the shape of the \mgg distribution largely unaffected, are treated as log-normal uncertainties. In this analysis, the following are considered:
\begin{itemize}
\item \textit{Integrated luminosity:} the systematic uncertainty is estimated from data to be 2.5\%~\cite{CMS:2017sdi};
\item \textit{Trigger efficiency:} the trigger efficiency is measured from \Zee
  events using the tag-and-probe technique~\cite{CMS:2011aa}; the size of the uncertainty is about 1\%;
\item \textit{Photon selection:} the systematic uncertainty is taken as the
  uncertainty in the ratio between the efficiency measured in data and in
  simulation; it ranges from 0.3 to 3.2\%  and
  results in an event yield variation from 0.7 to 4.0\% depending on the photon
  category;
\item \textit{Photon identification BDT score:} the uncertainties in the signal
  yields in the different categories of the analysis are estimated conservatively
  by propagating the uncertainty in the BDT inputs, which are estimated from the observed discrepancies between data and simulation,
  to the final photon identification BDT shape. This uncertainty has an effect of 3--5\% on the signal yield, depending on the category;
\item \textit{Per-photon energy resolution estimate:} this is parametrized as a
  rescaling of the resolution estimate by $\pm5$\% about the nominal value;
\item \textit{Jet energy scale and resolution corrections:}
  the uncertainties in these quantities are propagated to the final signal yields and induce
  event migrations between jet bins. The size of such migrations is in the 10--20\%
  range, depending on the jet bin;
\item \textit{Pileup identification for jets:}
this uncertainty is estimated in events with a {\cPZ} boson and one balanced jet. The full discrepancy between data and simulation in the identification score of jets is taken as the estimated uncertainty. It results in migrations from one jet bin to another, whose size is $<$1\%;
\item \textit{Background modeling:} the choice of the background parametrization is
  handled using the discrete profiling method. This is automatically included as a statistical
  uncertainty in the shape of the background function and  no additional systematic uncertainty needs to be added;
\item \textit{{\cPqb} tagging efficiency:} this is evaluated by varying the ratio between
  the measured {\cPqb} tagging efficiency in data and simulation within its
  uncertainty~\cite{Chatrchyan:2012jua}. The resulting uncertainty in the signal
  yield is $<$1\%;
\item \textit{Lepton identification:} for both electrons and muons, the uncertainty
  is computed by varying the ratio of the efficiency measured in data and
  simulation by its uncertainty. The resulting differences in the selection
  efficiency, for observables involving leptons, is less
  than 1\%;
 \item \textit{Missing transverse momentum:} the size of this uncertainty is computed by shifting the momentum
   scale and resolution of the \pt of every particle-flow candidate entering the
   computation of \ptmiss, by an amount that depends on the type of the reconstructed object,
   as described in Ref. \cite{CMS-PAS-JME-16-004}. This has an effect on the yield per category below 1\%;
This results in events migrating from one bin to another and from one category to another for observables involving \ptmiss;
 \item \textit{PDF uncertainties:}
    the effect of the uncertainty from the
    choice of PDF is assessed by estimating the relative
    yield variation in each bin of the observable variables and category,
    after re-weighting the events of the simulated
    signal sample. The re-weighting is done using the
    PDF4LHC15 combined PDF set and NNPDF3.0~\cite{Ball:2014uwa,Demartin:2010er}
    using the \textsc{mc2hessian} procedure~\cite{Carrazza:2015aoa}.
    The category migrations are found to be less than 0.3\%;
  \item \textit{Renormalization and factorization scale uncertainty:} the size of this uncertainty is estimated by varying the renormalization and
    factorization scales. The effect on category migrations is found to be negligible.
\end{itemize}

\section{Results}
\label{sec:results}
\begin{figure}[!htb]
 \begin{center}
  \includegraphics[width=0.49\textwidth]{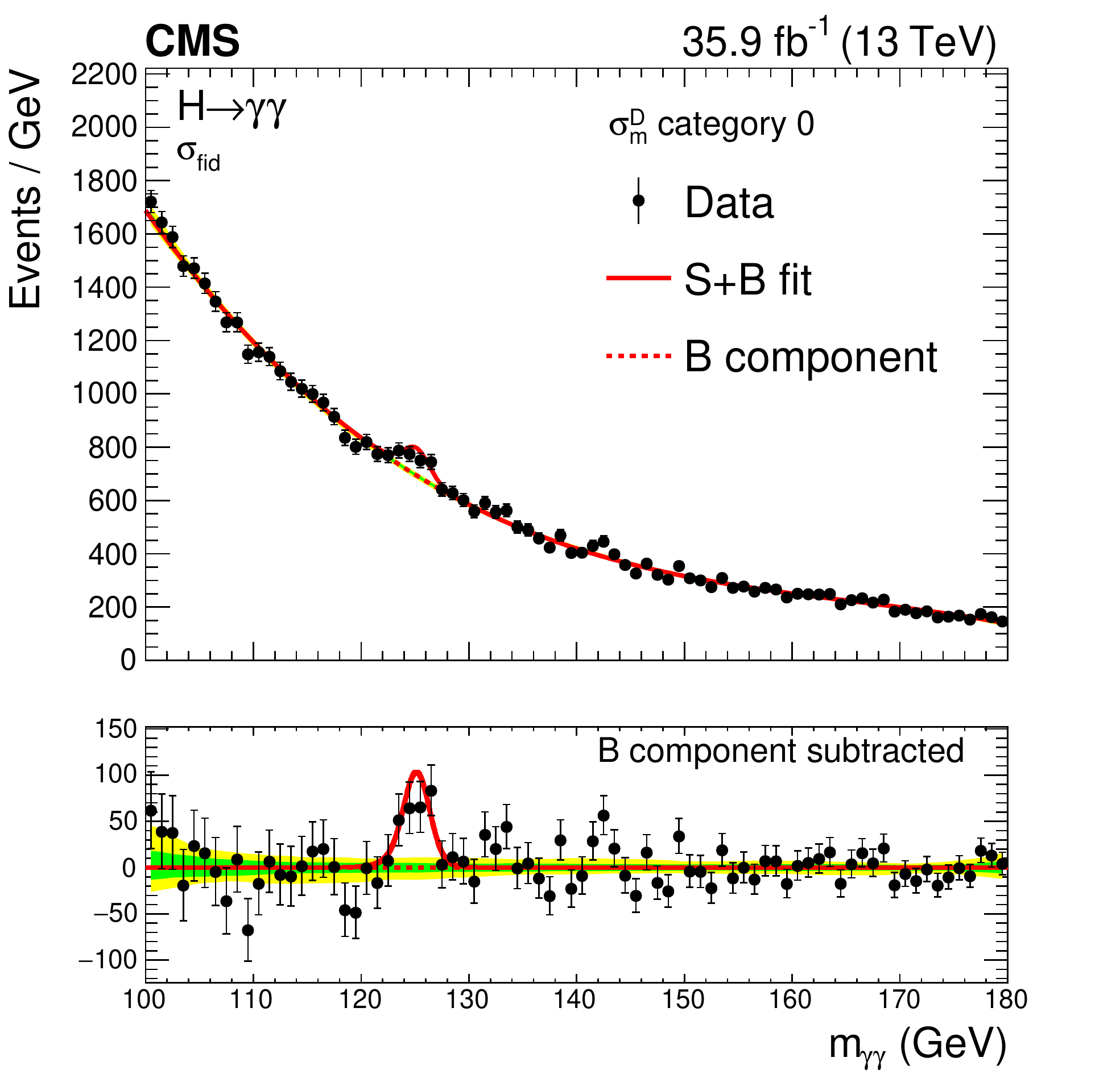}
  \includegraphics[width=0.49\textwidth]{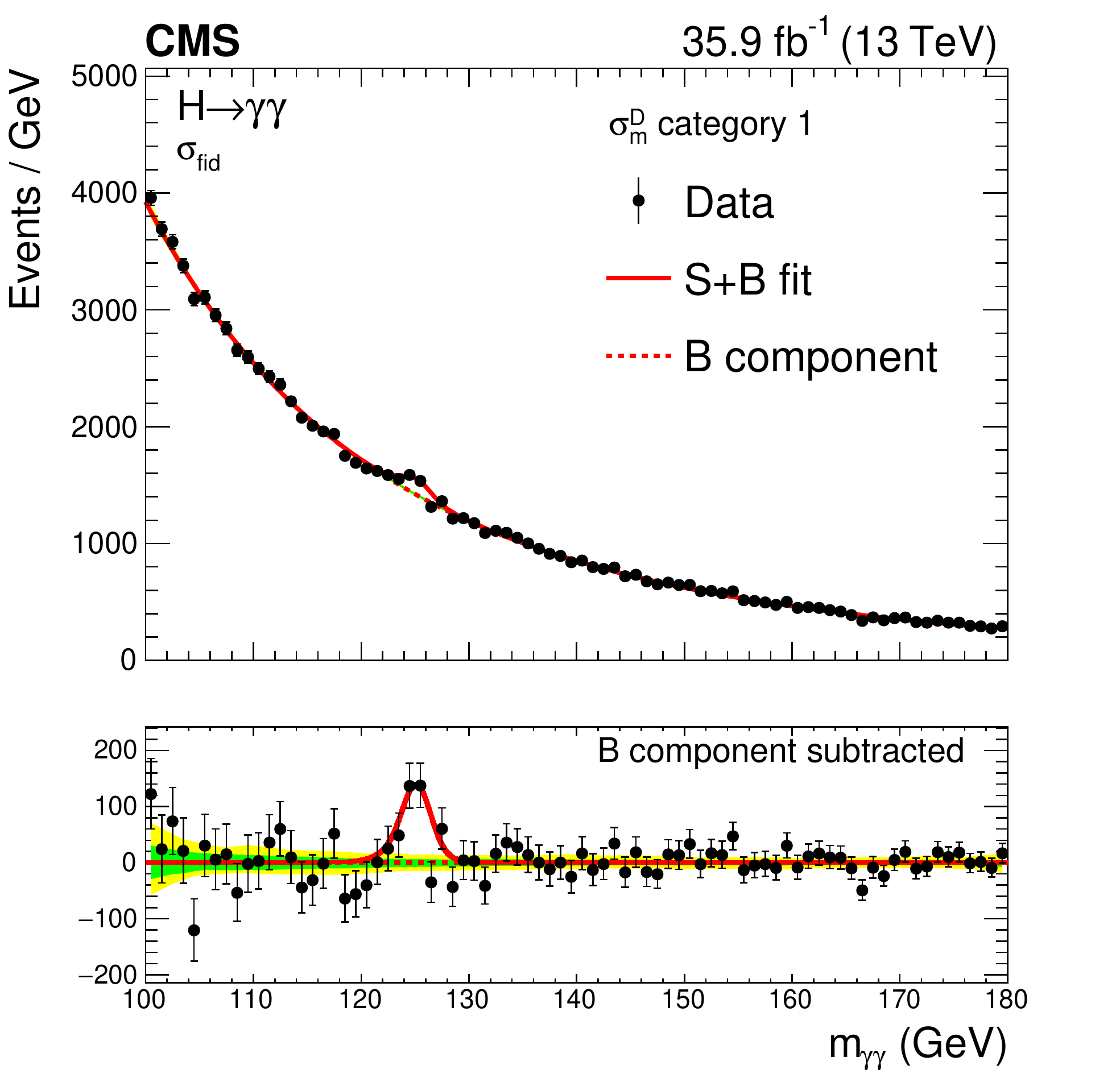}\\
  \includegraphics[width=0.49\textwidth]{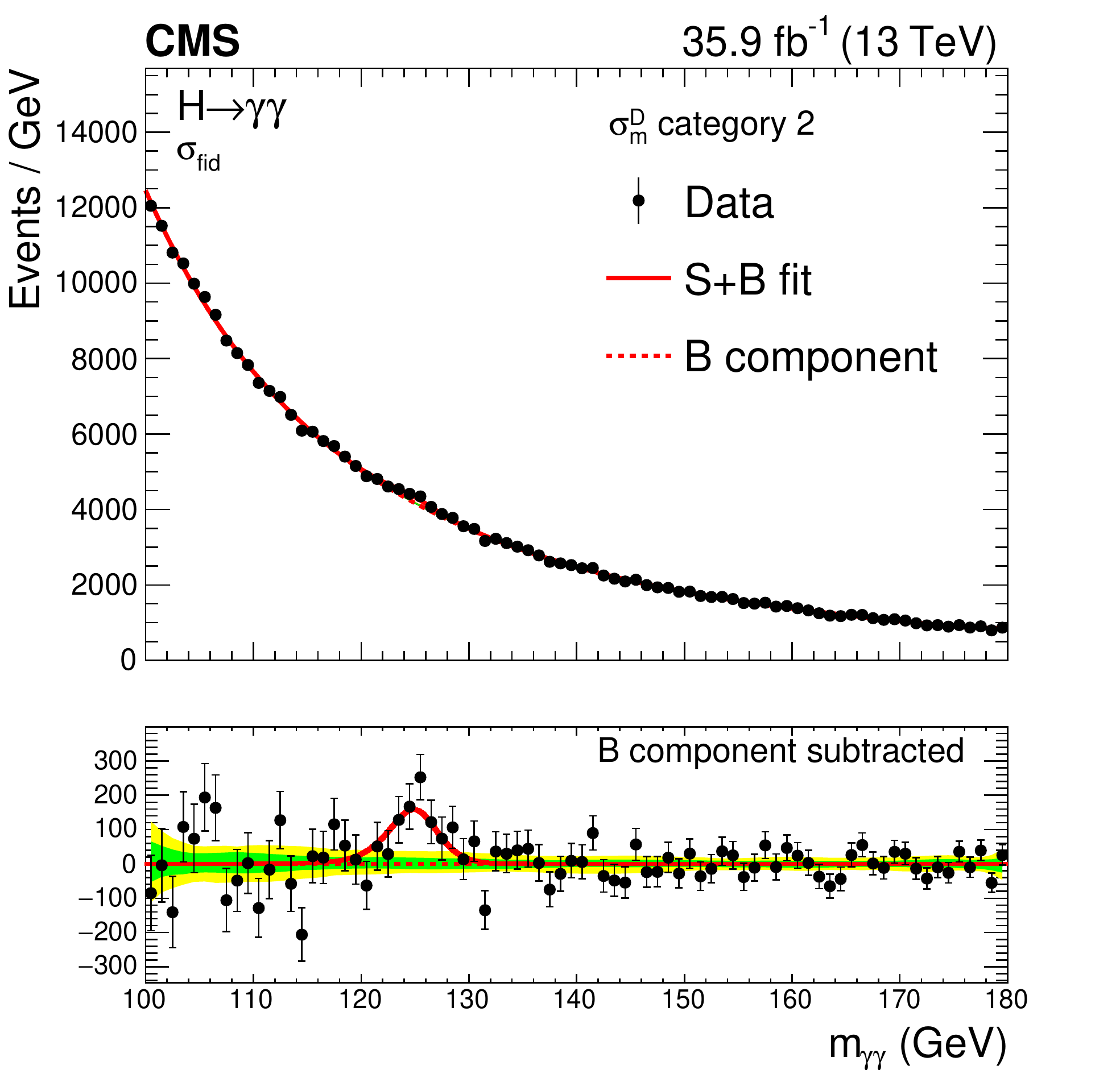}

   \caption{The diphoton mass spectrum in data (black points), together with the best signal-plus-background fit (red lines), for each \sigmaMoMdecorr category employed for the measurement of the inclusive fiducial cross section, as defined in Section~\ref{sec:categorization}. The two bands indicate the one and two standard deviation uncertainty in the background component.}
  \label{fig:dataFitsFiducial}
 \end{center}
\end{figure}

\begin{figure}[!htb]
 \begin{center}
   \includegraphics[width=0.495\textwidth]{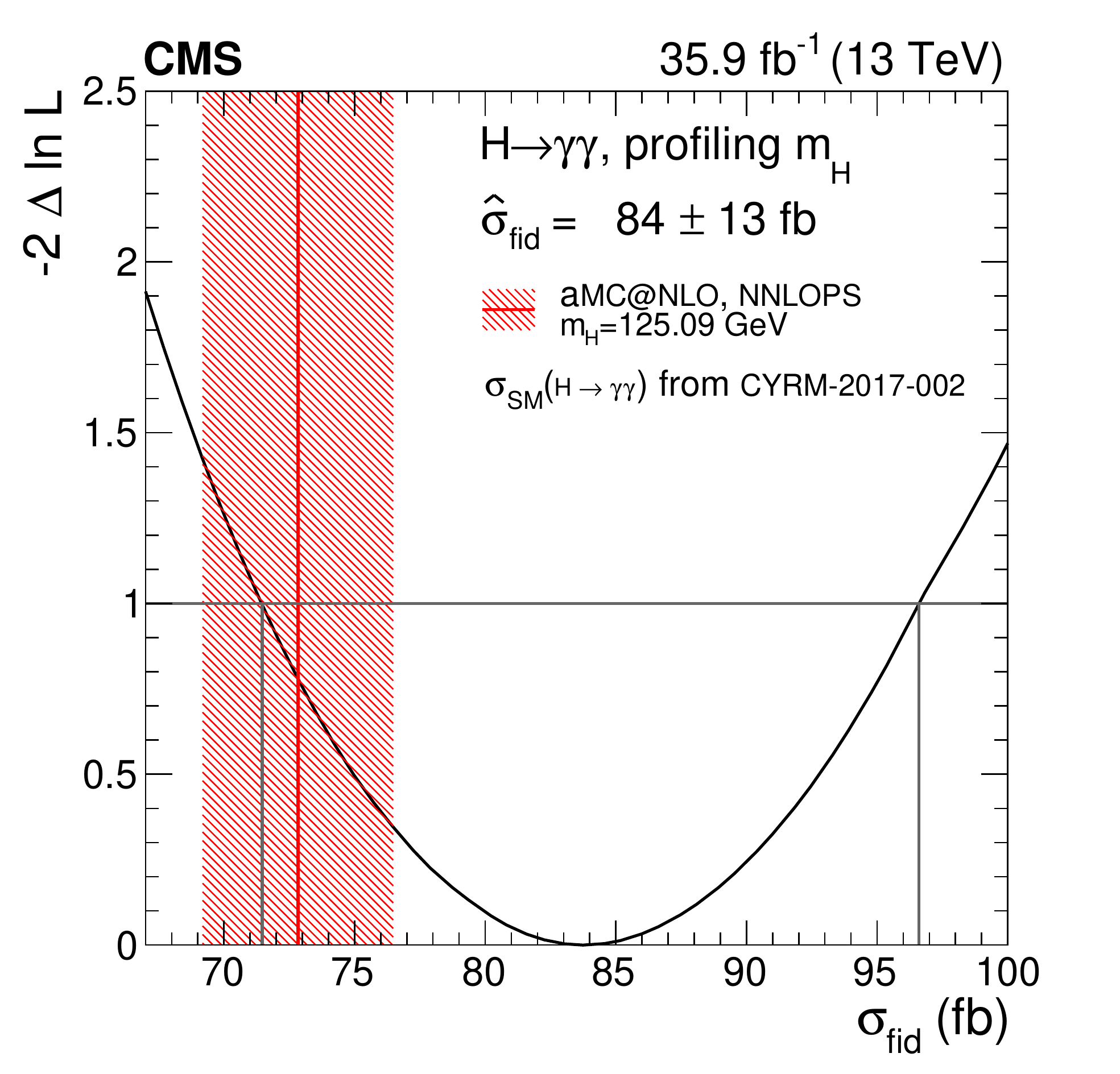}
   \caption{Likelihood scan (black curve) for the fiducial cross section measurement, where the value of the SM Higgs boson mass is profiled in the fit. The measurement is compared to the theoretical prediction (vertical red line), shown with its uncertainty (red hatched area), and it is found in agreement within the uncertainties.}
  \label{fig:fidLikelihoodScan}
 \end{center}
\end{figure}

The reconstructed diphoton invariant mass distributions are shown in Fig.~\ref{fig:dataFitsFiducial} for the three \sigmaMoMdecorr categories. The signal-plus-background fit is performed simultaneously in all three categories to extract the inclusive fiducial cross section.
The best fit value of the inclusive fiducial cross section is:
\begin{equation}
  \hat{\sigma}_{\mathrm{fiducial}}=84\pm11\stat\pm7\syst\unit{fb}=84\pm13 \text{ (stat+syst)}\unit{fb}
\end{equation}
The total uncertainty ($13\unit{fb}$) is dominated by its statistical component ($11\unit{fb}$). The primary contributions to the systematic component ($7\unit{fb}$) arise from the uncertainties in the photon identification BDT score and in the per-photon energy resolution estimate, described in Section~\ref{sec:systematics}.
The corresponding likelihood scan is shown in Fig.~\ref{fig:fidLikelihoodScan}, together with
the theoretical prediction for the cross section.
In the measurement of both inclusive and differential fiducial cross sections,
the Higgs boson mass is treated as a
nuisance parameter and profiled in the likelihood maximization. The value of the profiled
mass is compatible with the world average~\cite{PhysRevLett.114.191803}.

The theoretical prediction for the inclusive cross section is $\sigma^{\text{theory}}_{\text{fiducial}}
= 73 \pm 4\unit{fb}$. The measured value is in agreement with the prediction within 1 standard deviation.
The prediction is computed using simulated events generated with
\MGvATNLO, where each of the Higgs boson production mechanisms is normalized to the
predictions from Ref.~\cite{LHCHXSWG:YR4}.
The simulated events are used to compute the fiducial phase space acceptance
 for the SM Higgs boson with a mass of 125.09\GeV, corresponding to the measured world average value~\cite{PhysRevLett.114.191803},  and this value is then
multiplied by the corresponding total cross section and branching fraction quoted in Ref.~\cite{LHCHXSWG:YR4}.
The uncertainties in the cross section and  branching ratio predictions are also taken from Ref.~\cite{LHCHXSWG:YR4} and propagated to the final prediction.
The fiducial phase space acceptance is estimated to be 0.60 for the SM Higgs boson. This value
amounts to 0.60, 0.60, 0.52, and 0.52 for $\Pg\Pg\PH$, VBF, V\PH, and \ttH production, respectively.
The associated QCD scale uncertainty is estimated by independently varying the renormalization and
factorization scales used in the calculation by a factor of 2 upwards and downwards, excluding the combinations
(1/2, 2) and (2, 1/2), and it amounts to approximately 1\% of the acceptance value.
The acceptance for the $\Pg\Pg\PH$ production mode is estimated using events generated with \POWHEG, both with and without weighting the events to match the prediction from the  {\textsc{nnlops}} program, leading in both cases to a change of about 1\%.

\begin{figure}[!htb]
 \begin{center}
   \includegraphics[width=0.49\textwidth]{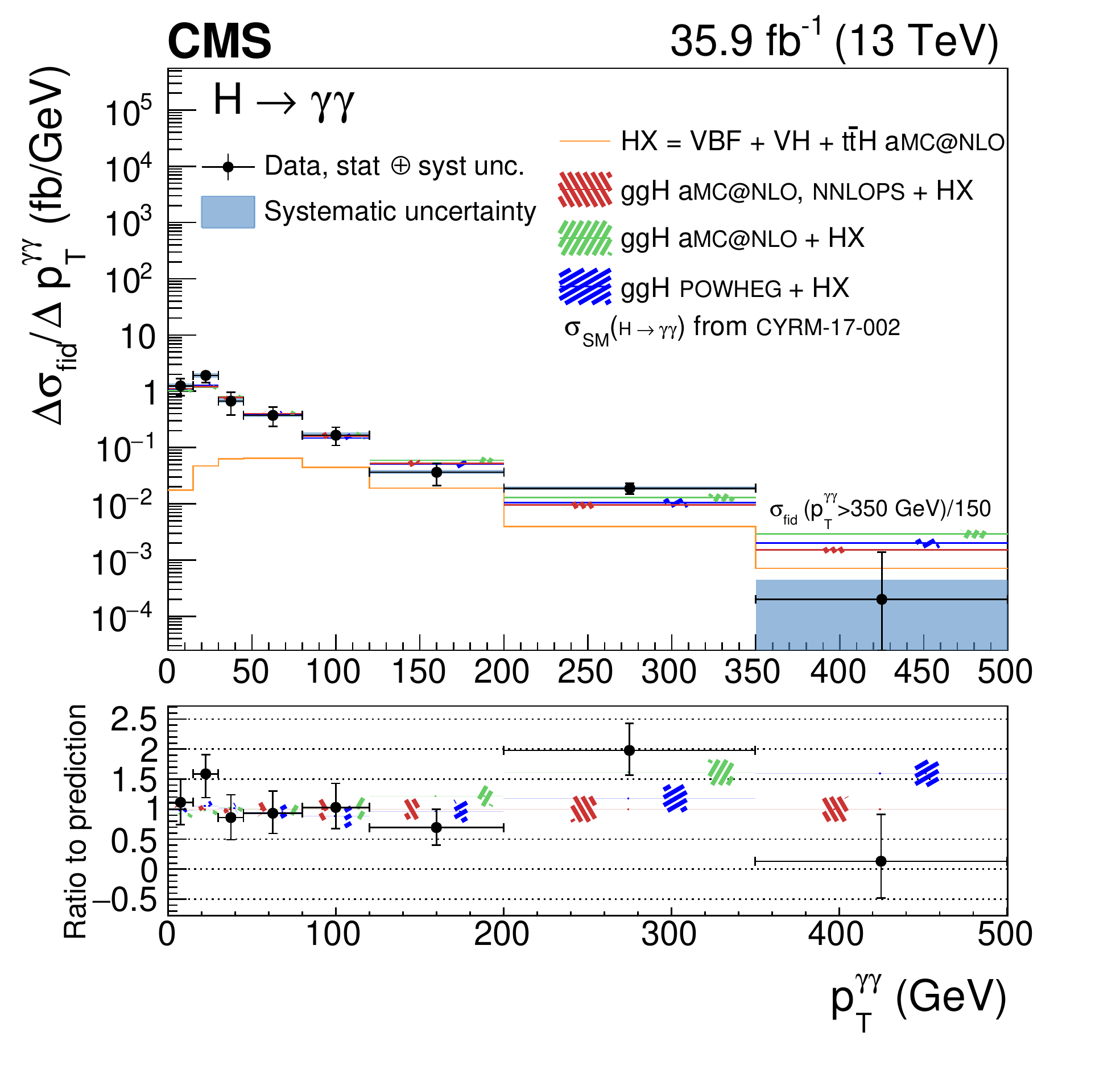}
   \includegraphics[width=0.49\textwidth]{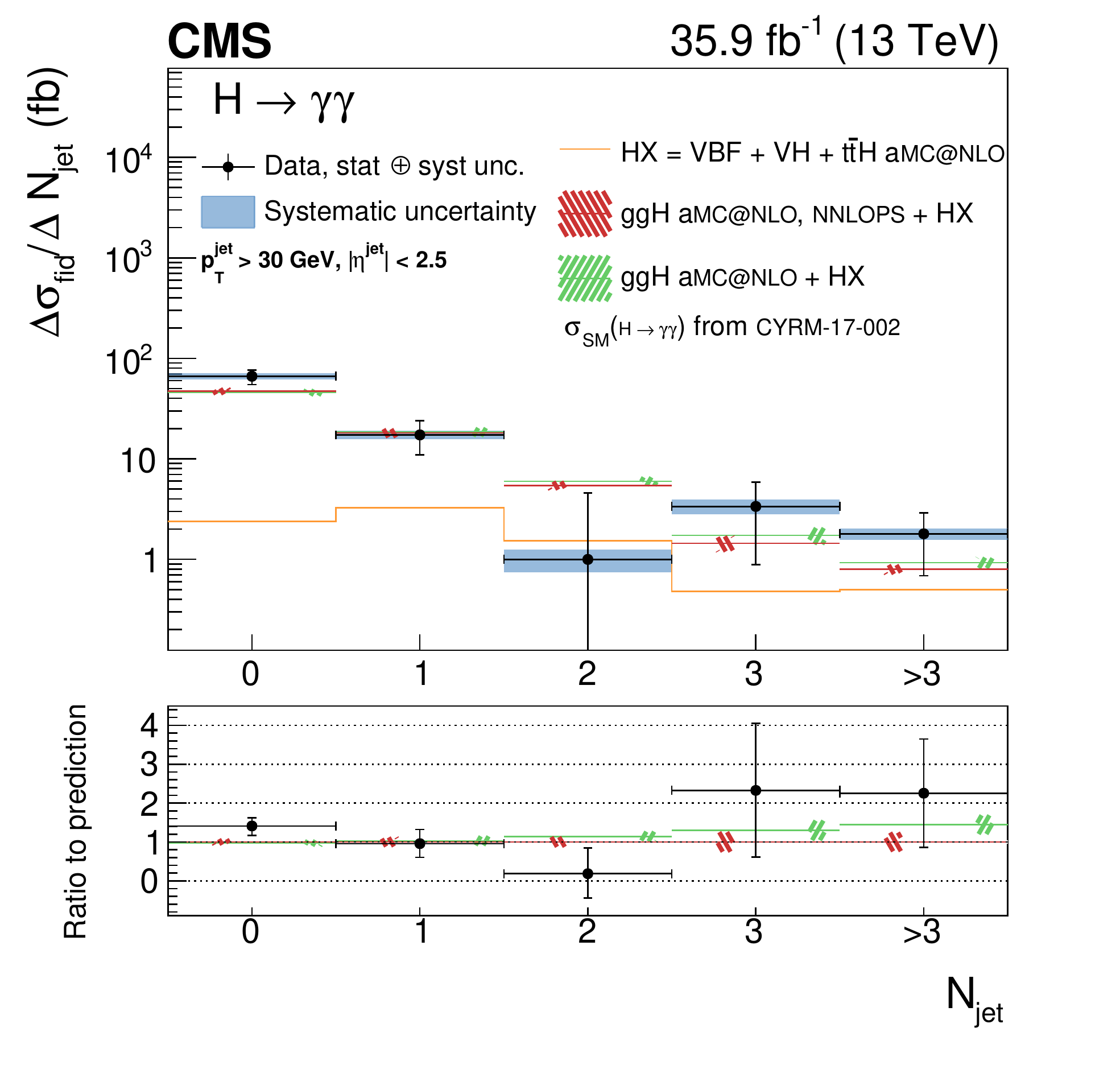}\\
   \includegraphics[width=0.49\textwidth]{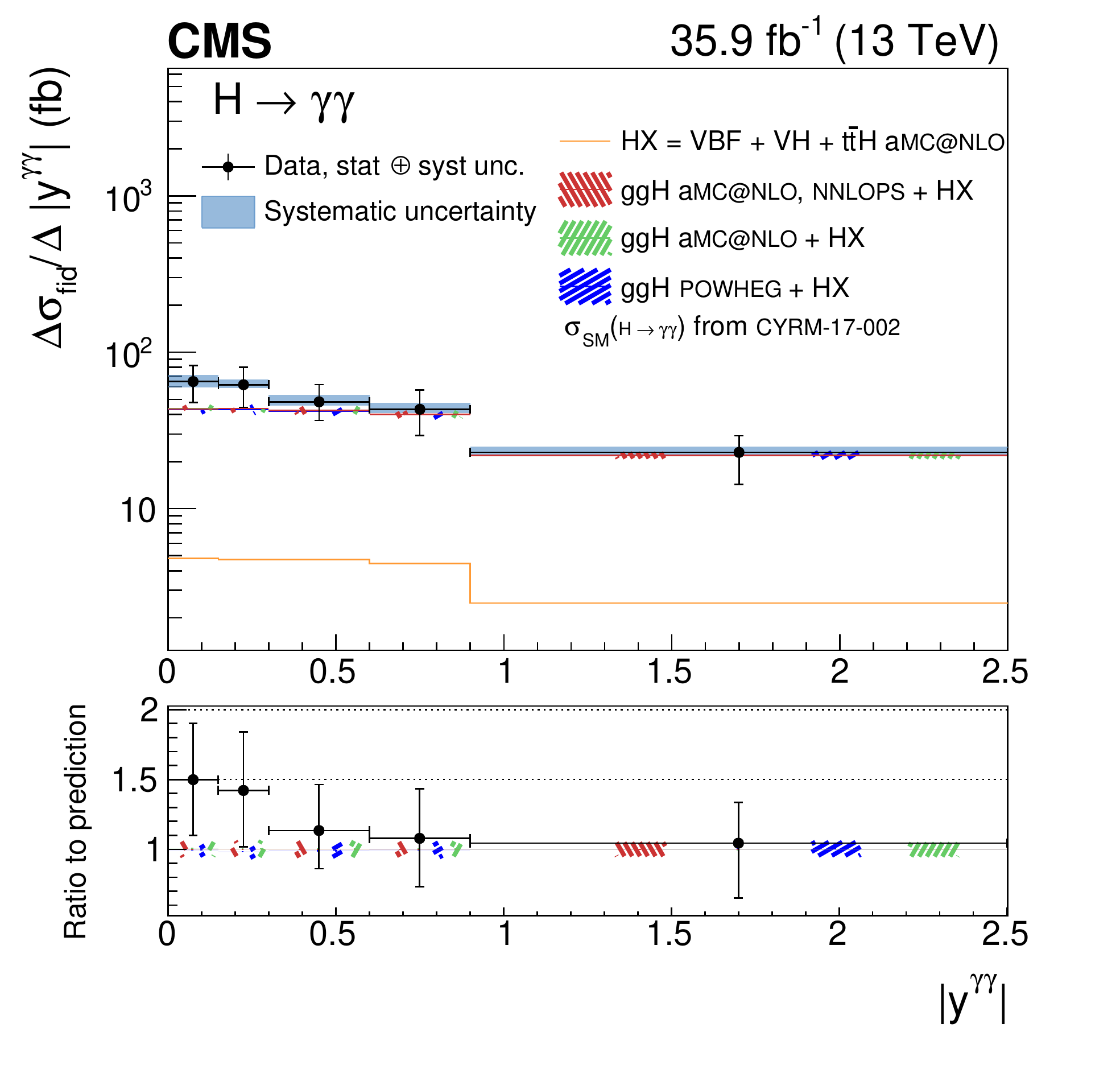}
   \includegraphics[width=0.49\textwidth]{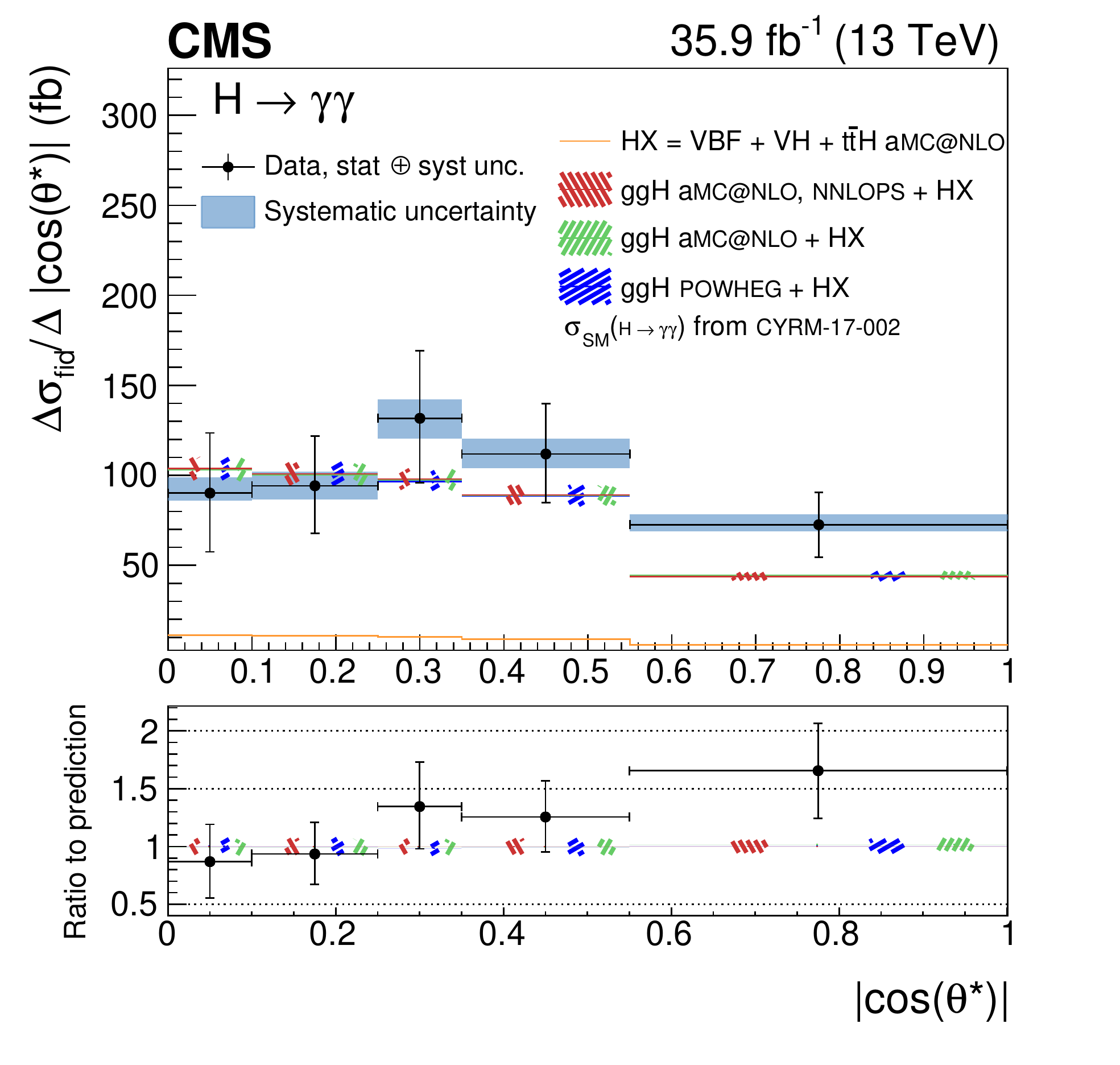}
   \caption{Measurement of the differential cross section (black points) as functions of $\ptgg$, $\njet$, $\absrapgg$, and $\abscosthetast$. The error bars indicate $1$ standard deviation uncertainty. The systematic component of the uncertainty is shown by the blue band. The measurements are compared to different simulation programs (histograms) with their uncertainties (hatched areas), all normalized to the same theoretical predictions from Ref.~\cite{LHCHXSWG:YR4}.
 When the last bin of the distribution is an overflow bin, the normalization of the cross section in that bin is indicated in the figure. }
  \label{fig:expPrecision1}
 \end{center}
\end{figure}

\begin{figure}[!htb]
 \begin{center}
   \includegraphics[width=0.49\textwidth]{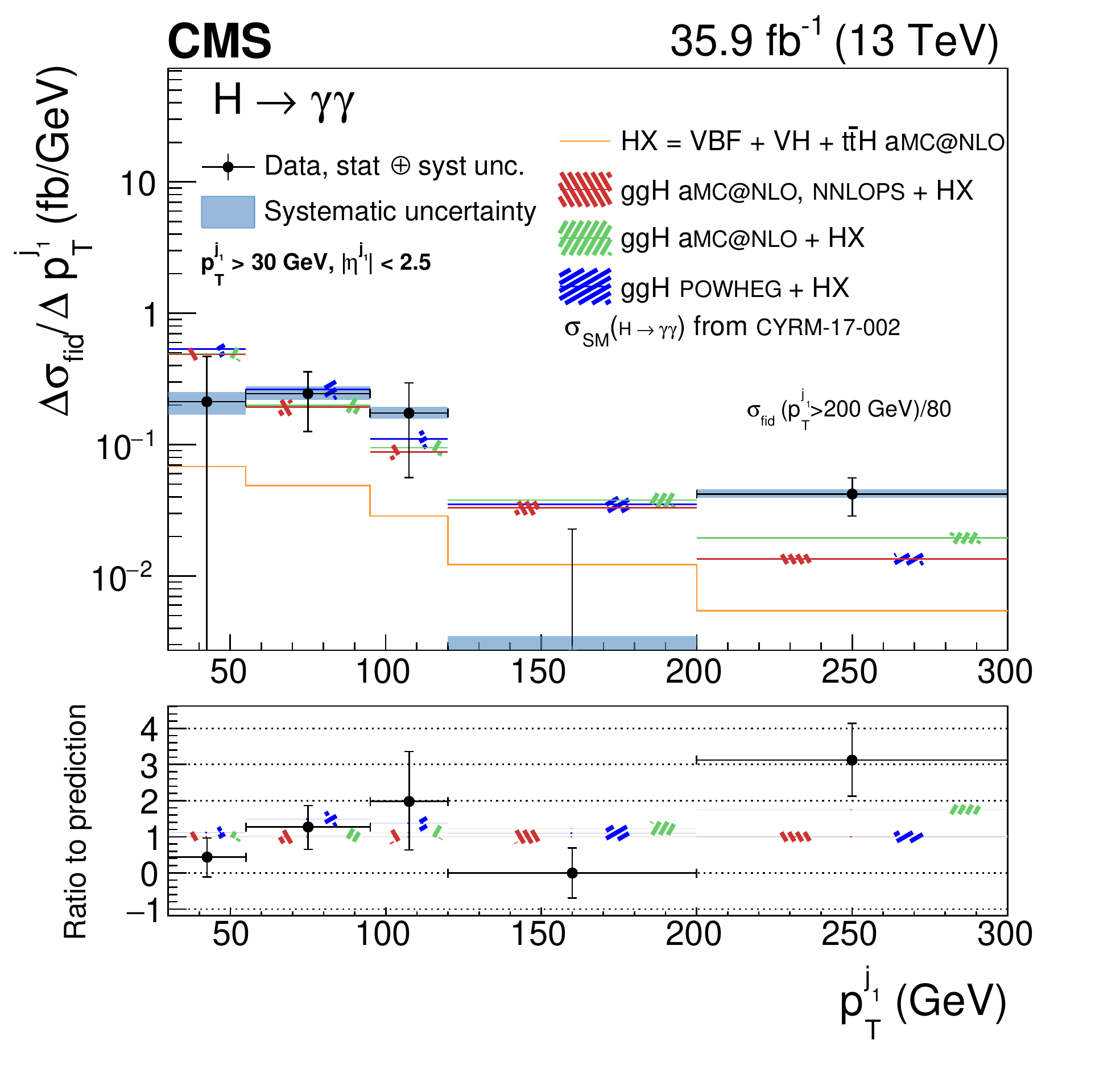}
   \includegraphics[width=0.49\textwidth]{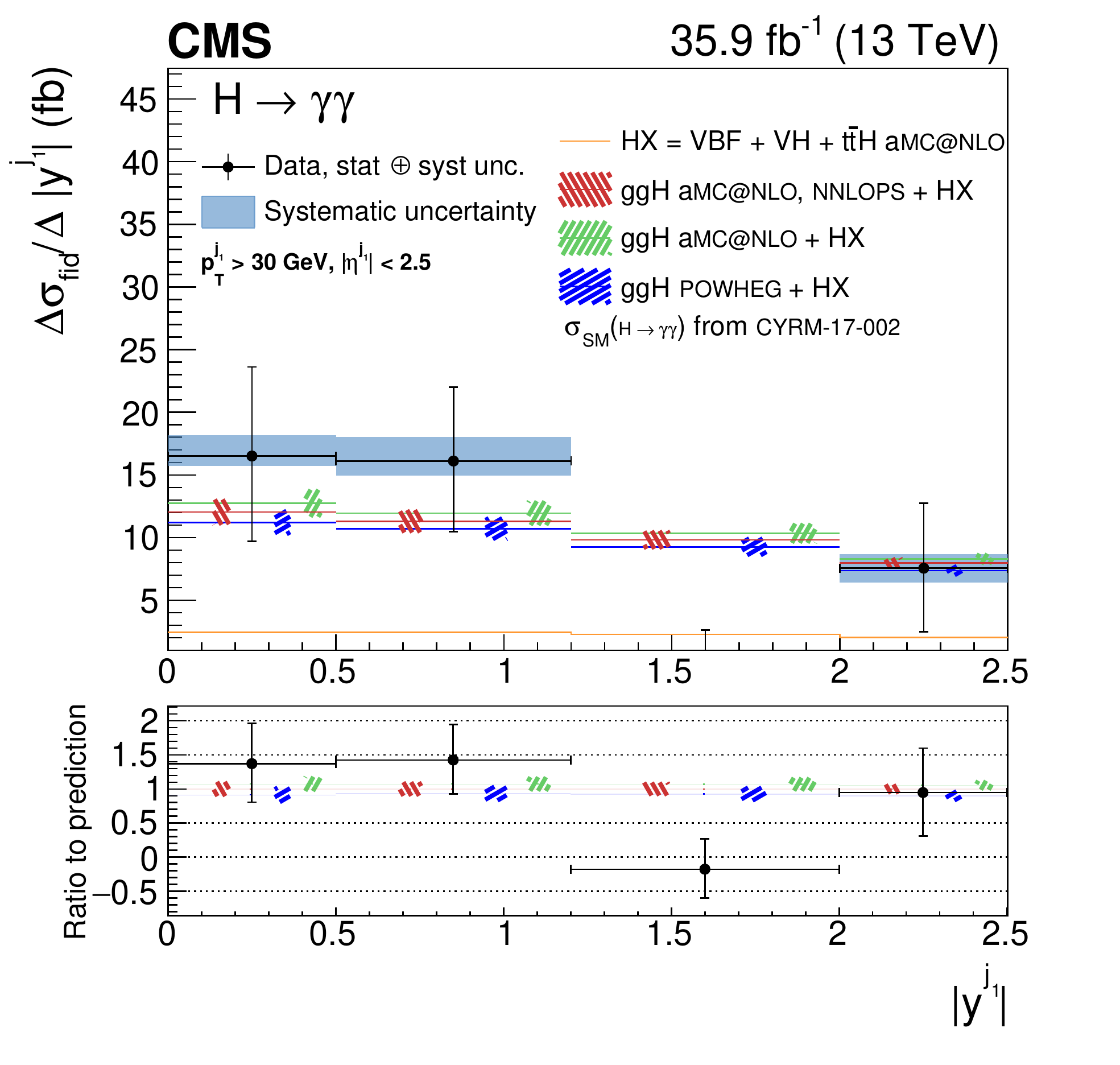}\\
   \includegraphics[width=0.49\textwidth]{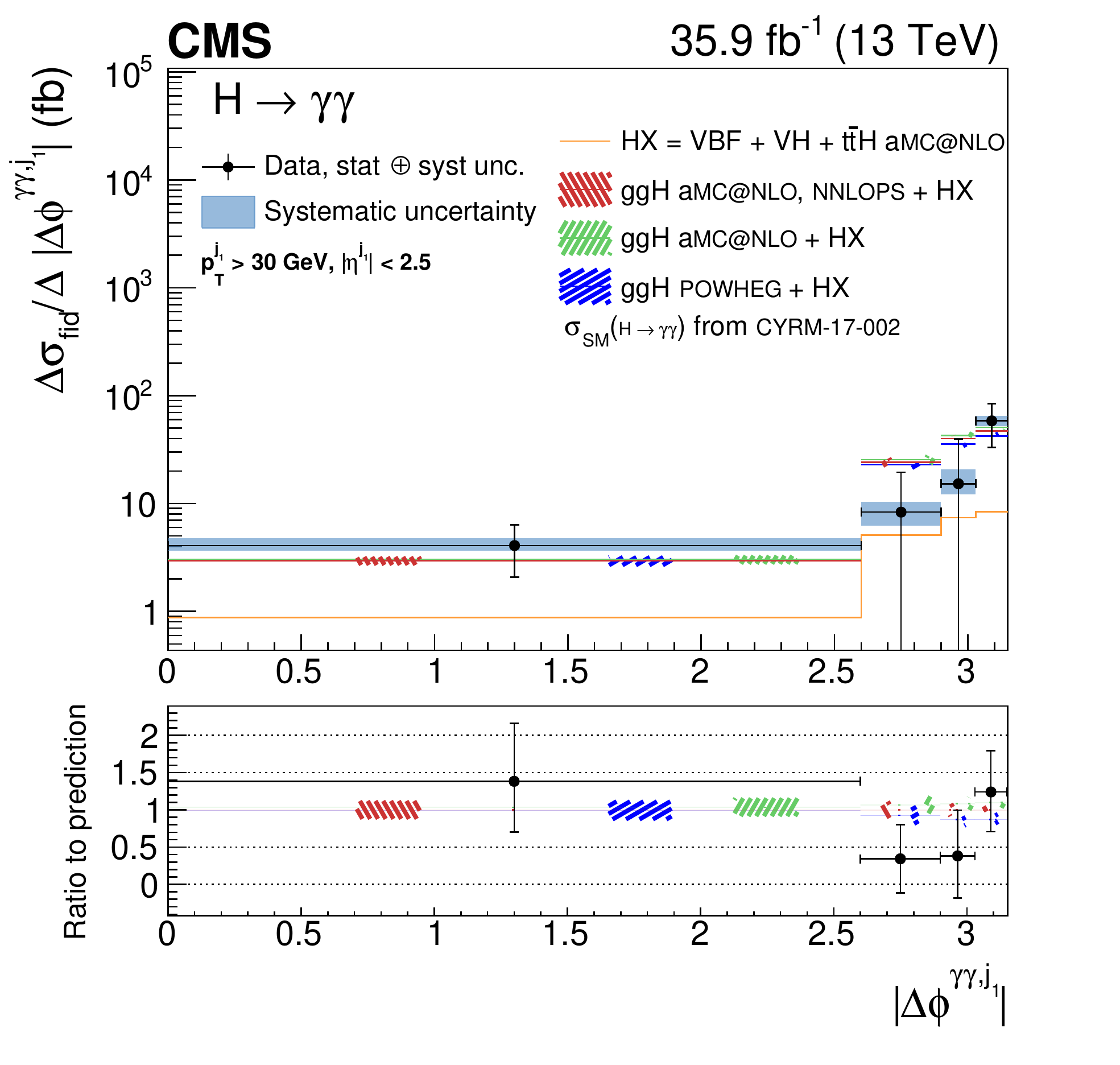}
   \includegraphics[width=0.49\textwidth]{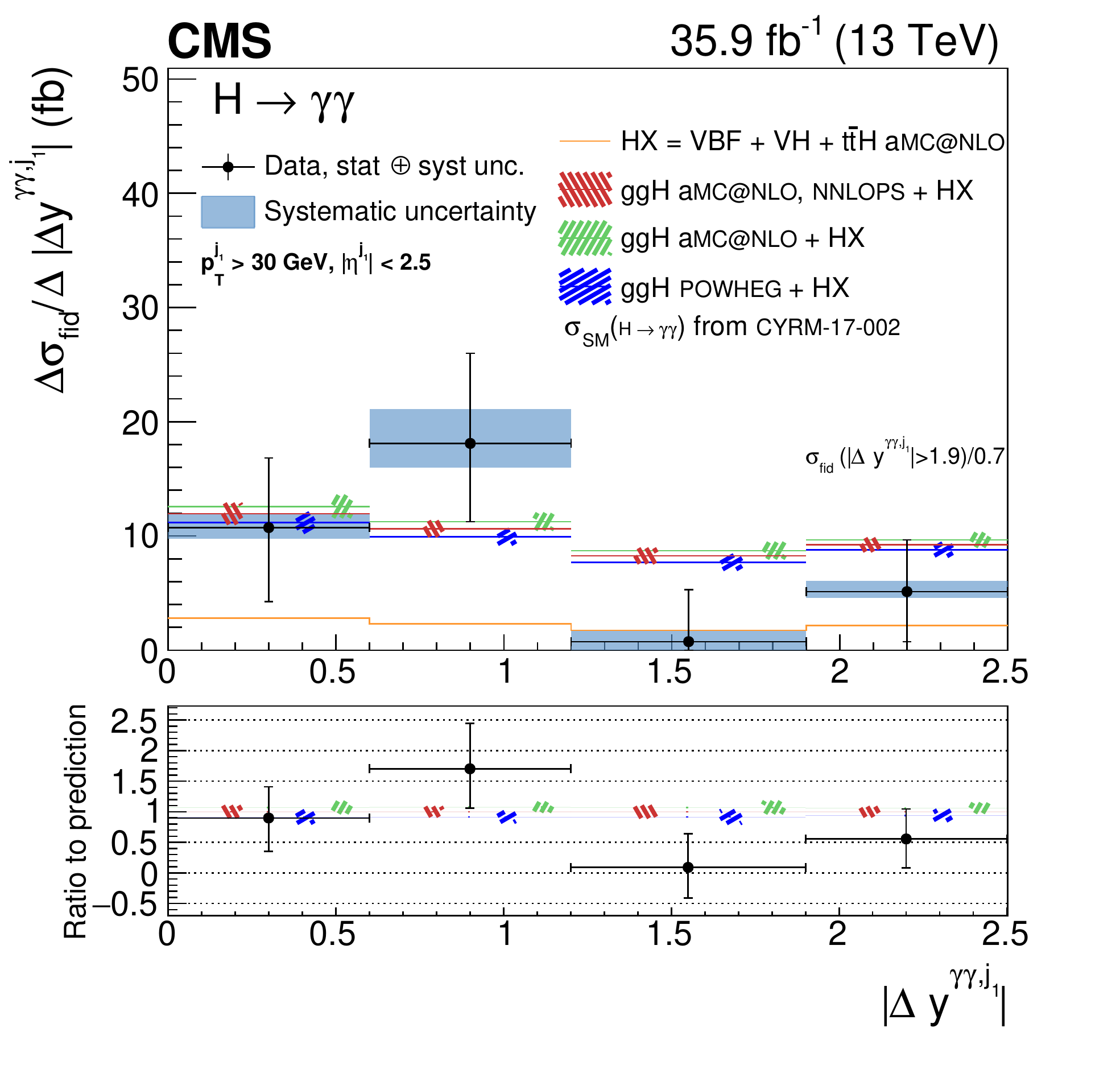}
   \caption{Measurement of the differential cross section (black points) as functions $\ptjone$, $\absrapjone$, $\absDphiggjone$, and $\absDrapggjone$. The error bars indicate $1$ standard deviation uncertainty. The systematic component of the uncertainty is shown by the blue band.
The measurements are compared to different simulation programs (histograms) with their uncertainties (hatched areas), all normalized to the same theoretical predictions from Ref.~\cite{LHCHXSWG:YR4}.
 When the last bin of the distribution is an overflow bin, the normalization of the cross section in that bin is indicated in the figure.}
  \label{fig:expPrecision2}
 \end{center}
\end{figure}

\begin{figure}[!htb]
 \begin{center}
   \includegraphics[width=0.49\textwidth]{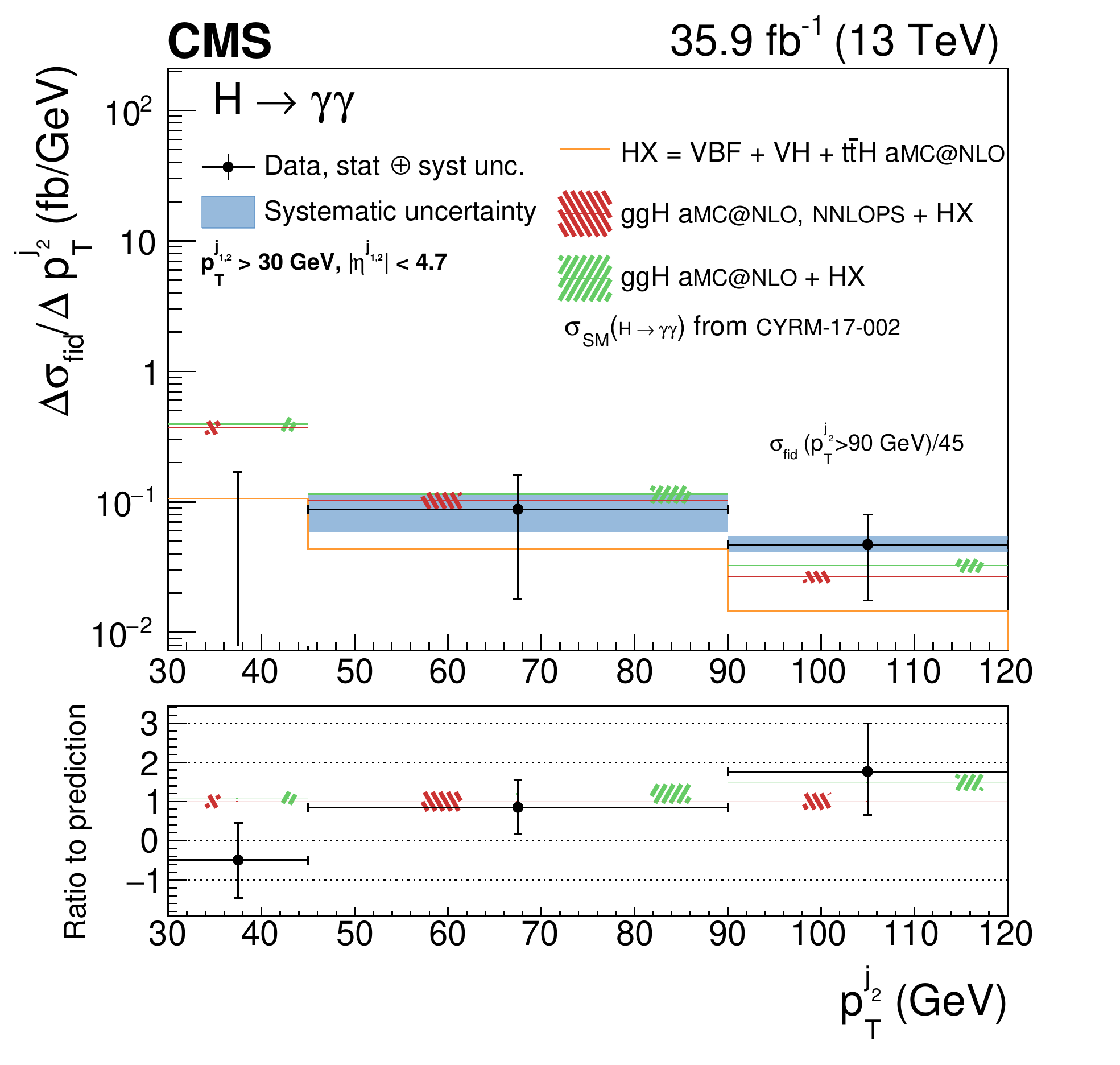}
   \includegraphics[width=0.49\textwidth]{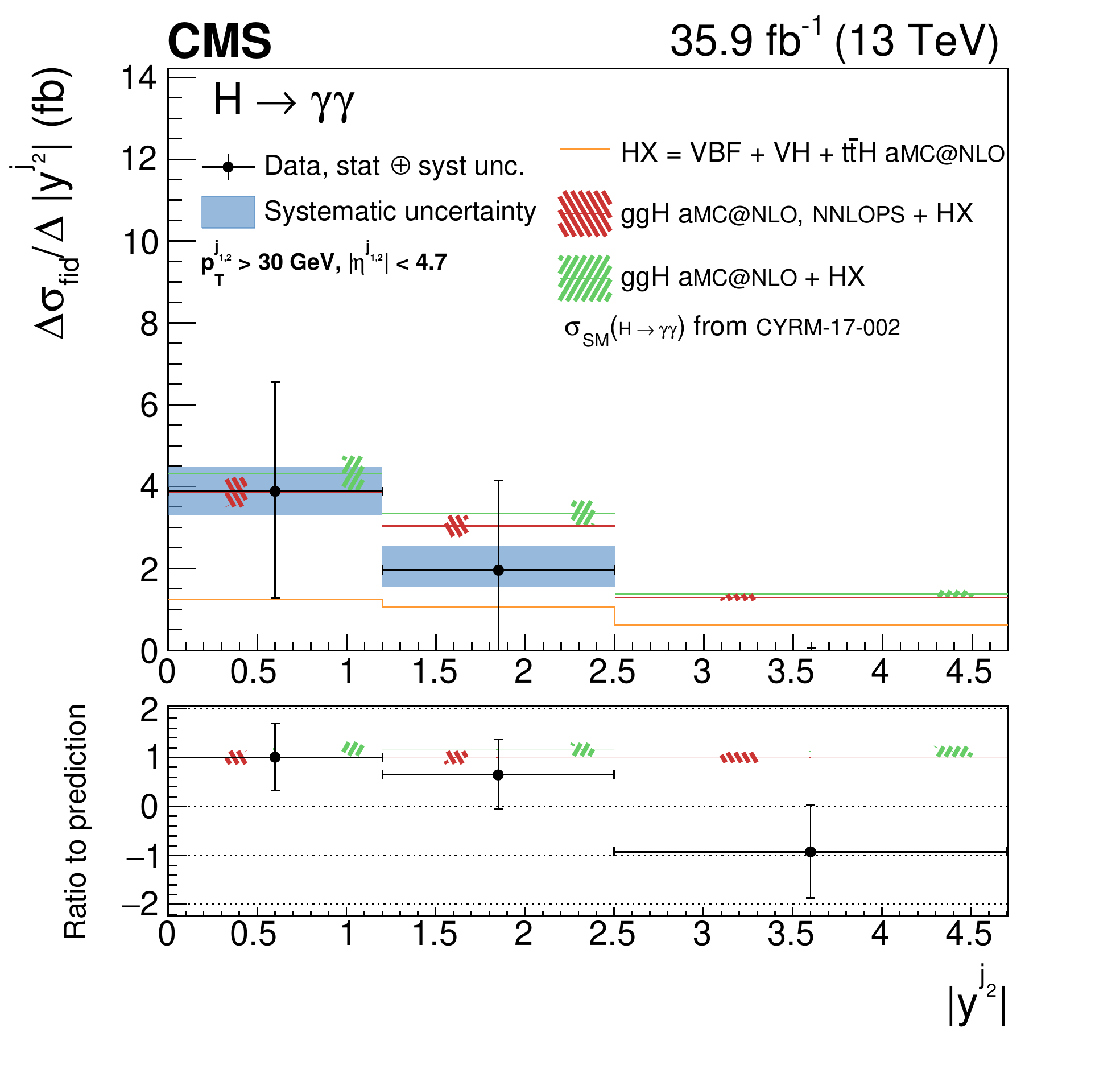}\\
   \includegraphics[width=0.49\textwidth]{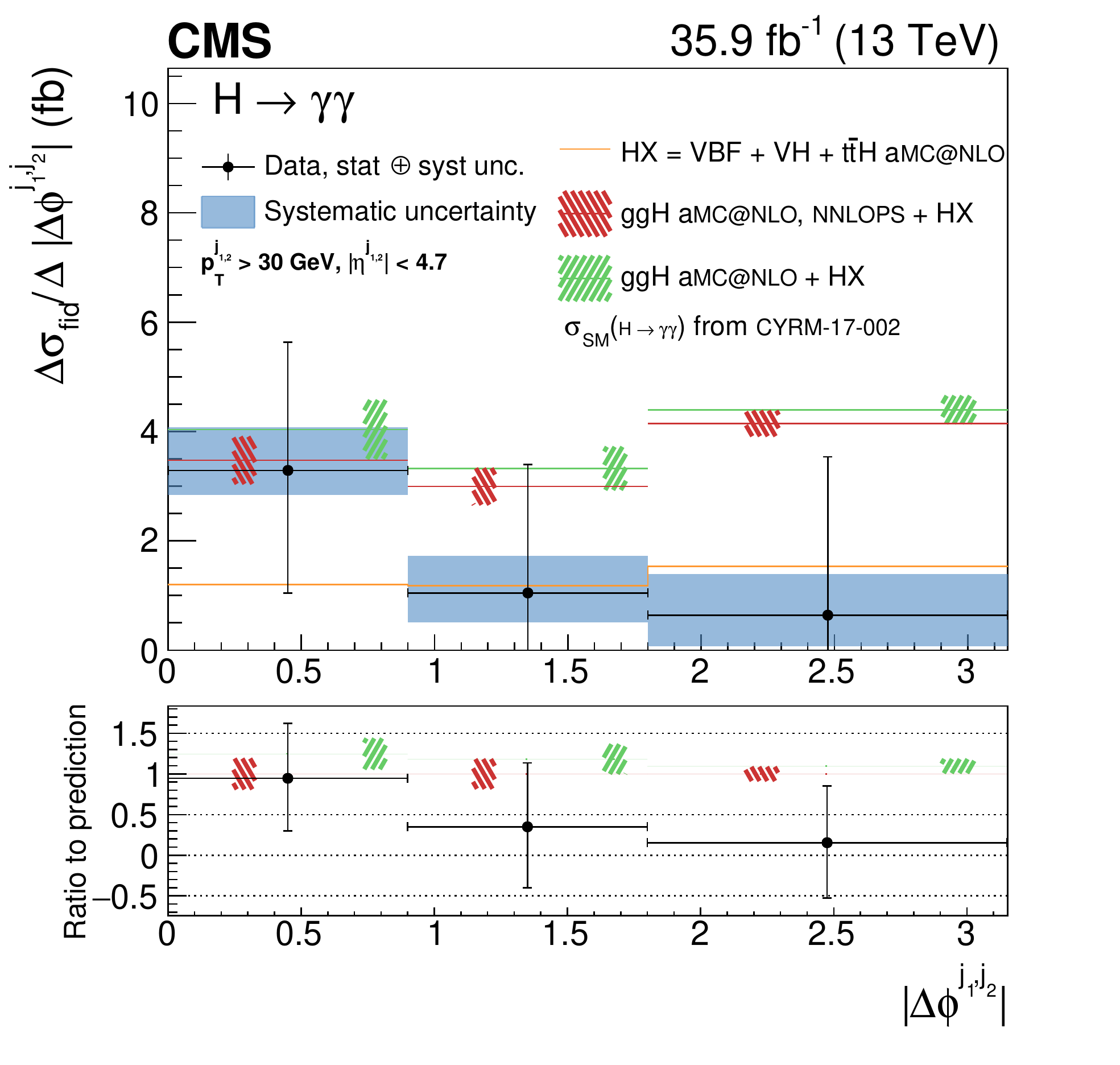}
   \includegraphics[width=0.49\textwidth]{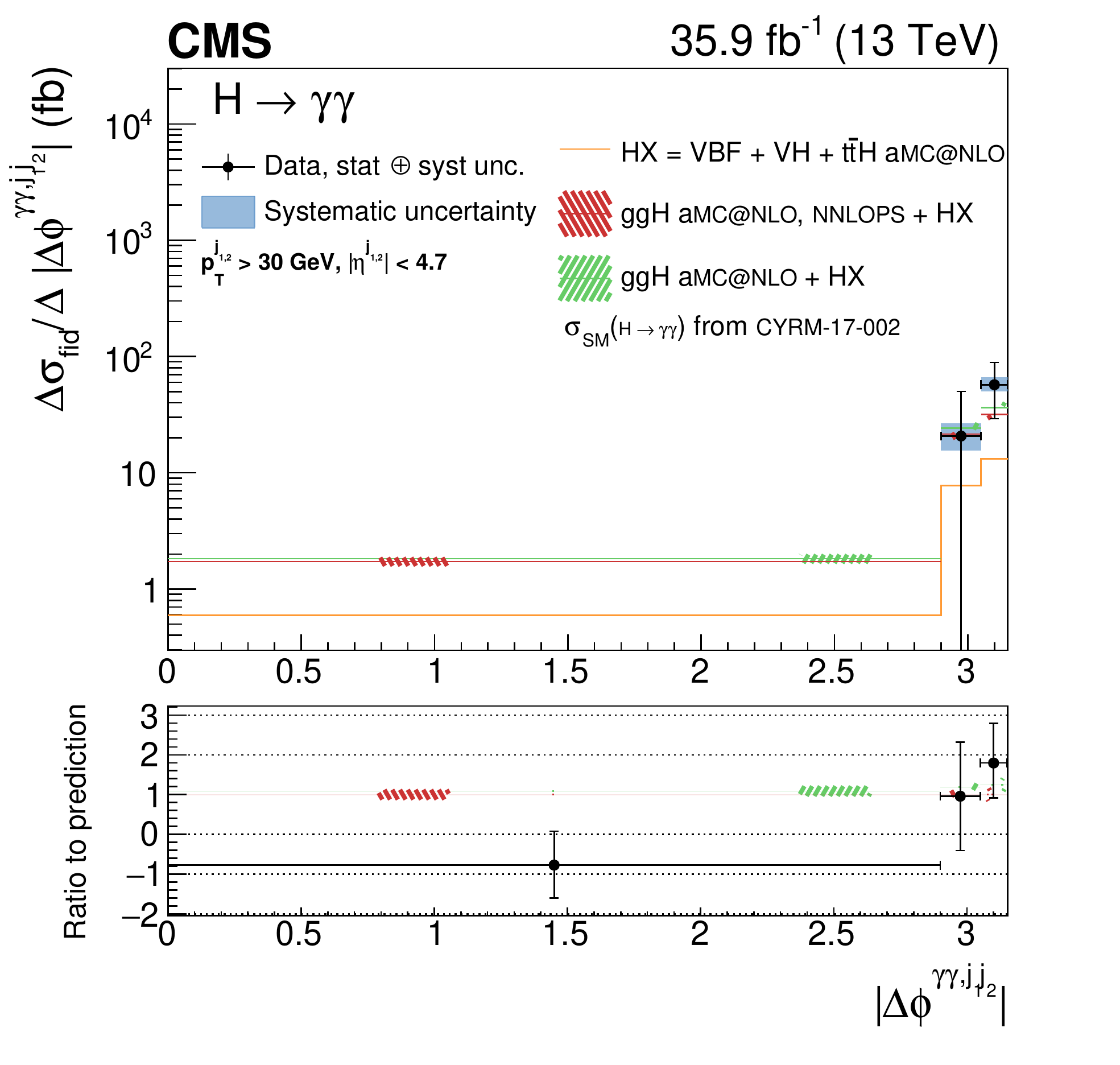}
   \caption{Measurement of the differential cross section (black points) as functions of $\ptjtwo$, $\absrapjtwo$, $\absDphijonejtwo$, and $\absDphiggjonejtwo$. The error bars indicate $1$ standard deviation uncertainty. The systematic component of the uncertainty is shown by the blue band.
The measurements are compared to two different simulation programs (histograms) with their uncertainties (hatched areas), both normalized to the same theoretical predictions from Ref.~\cite{LHCHXSWG:YR4}.
When the last bin of the distribution is an overflow bin, the normalization of the cross section in that bin is indicated in the figure. }
  \label{fig:expPrecision3}
 \end{center}
\end{figure}

\begin{figure}[!htb]
 \begin{center}
   \includegraphics[width=0.49\textwidth]{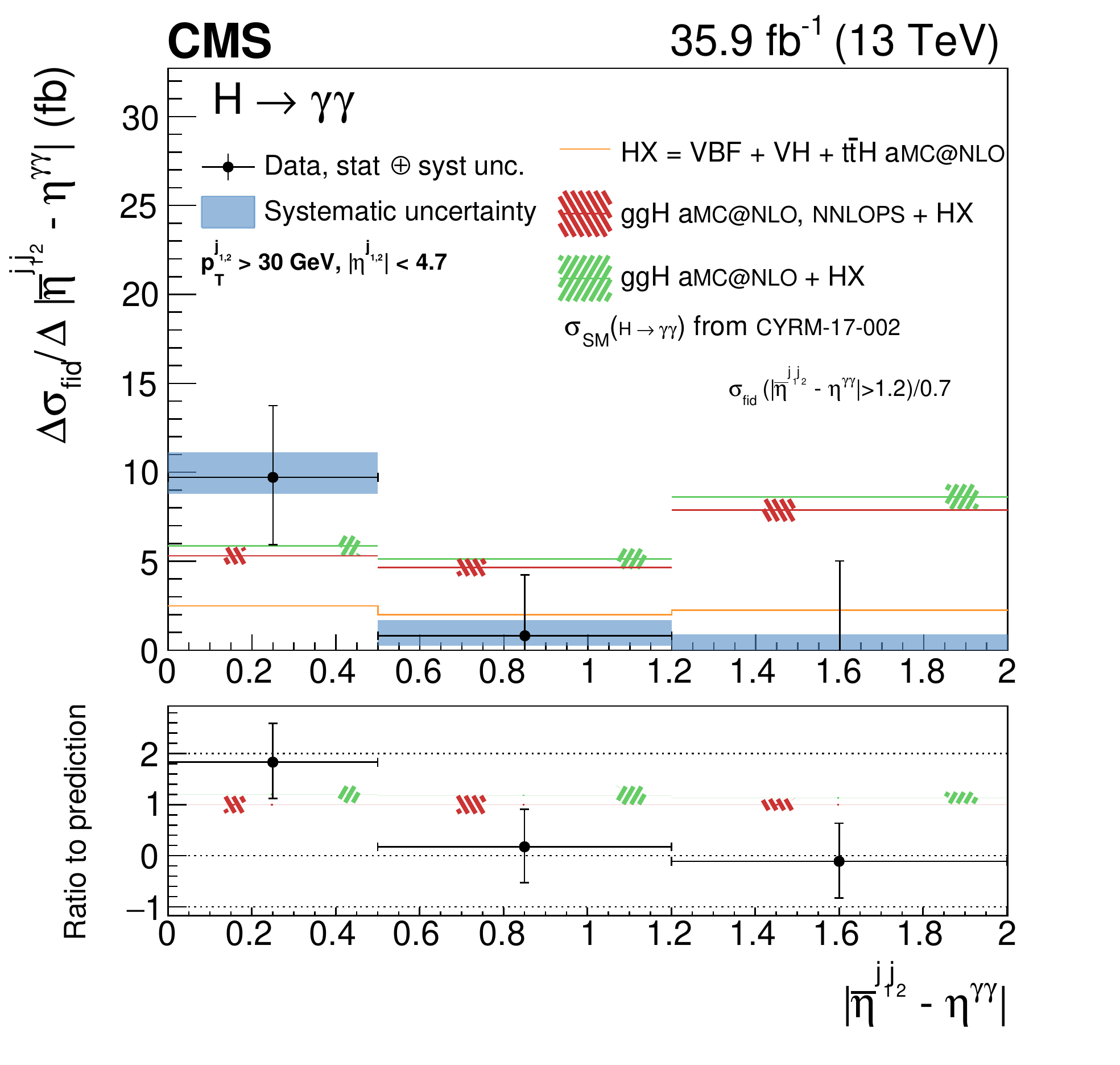}
   \includegraphics[width=0.49\textwidth]{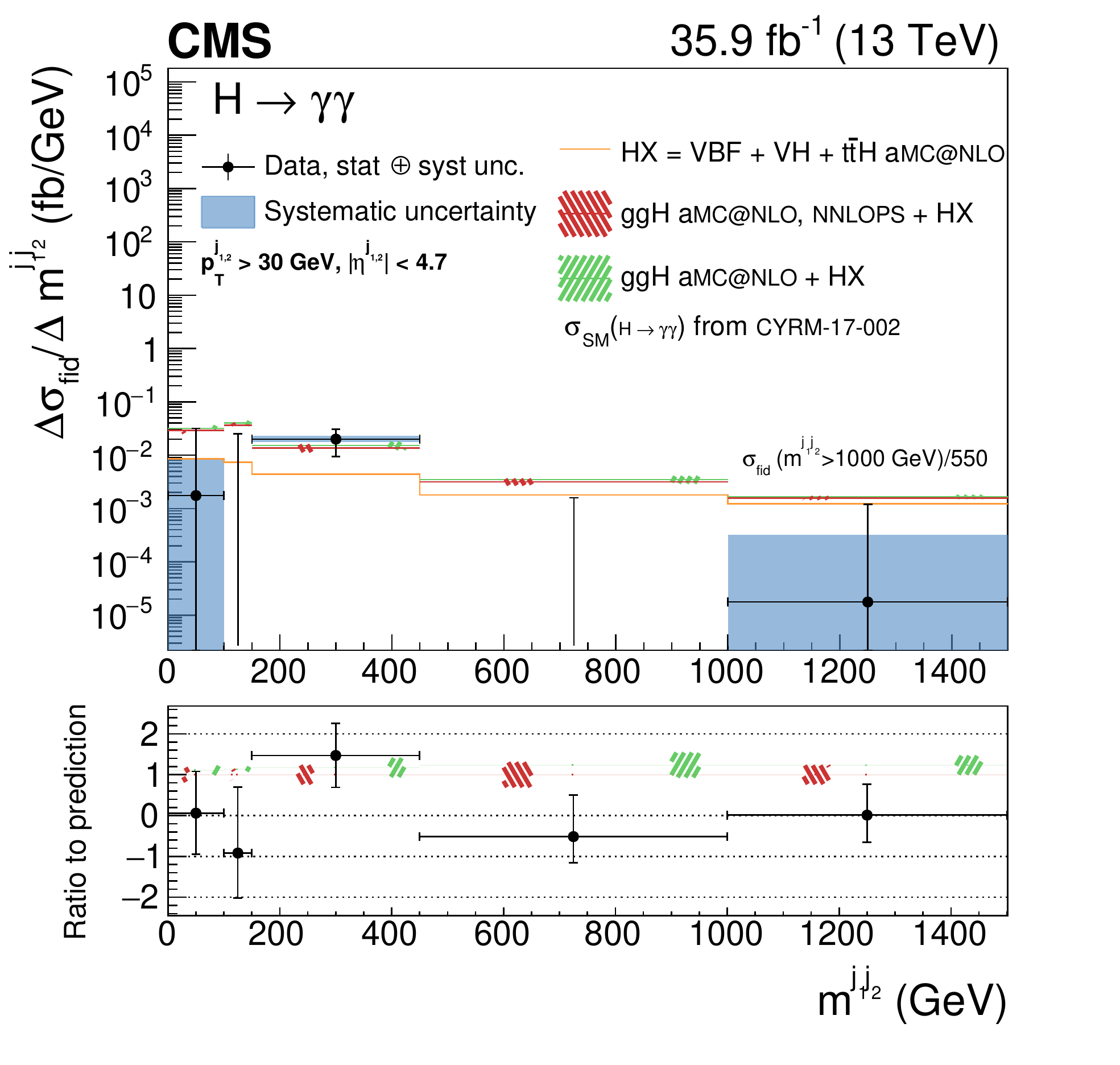}\\
   \includegraphics[width=0.49\textwidth]{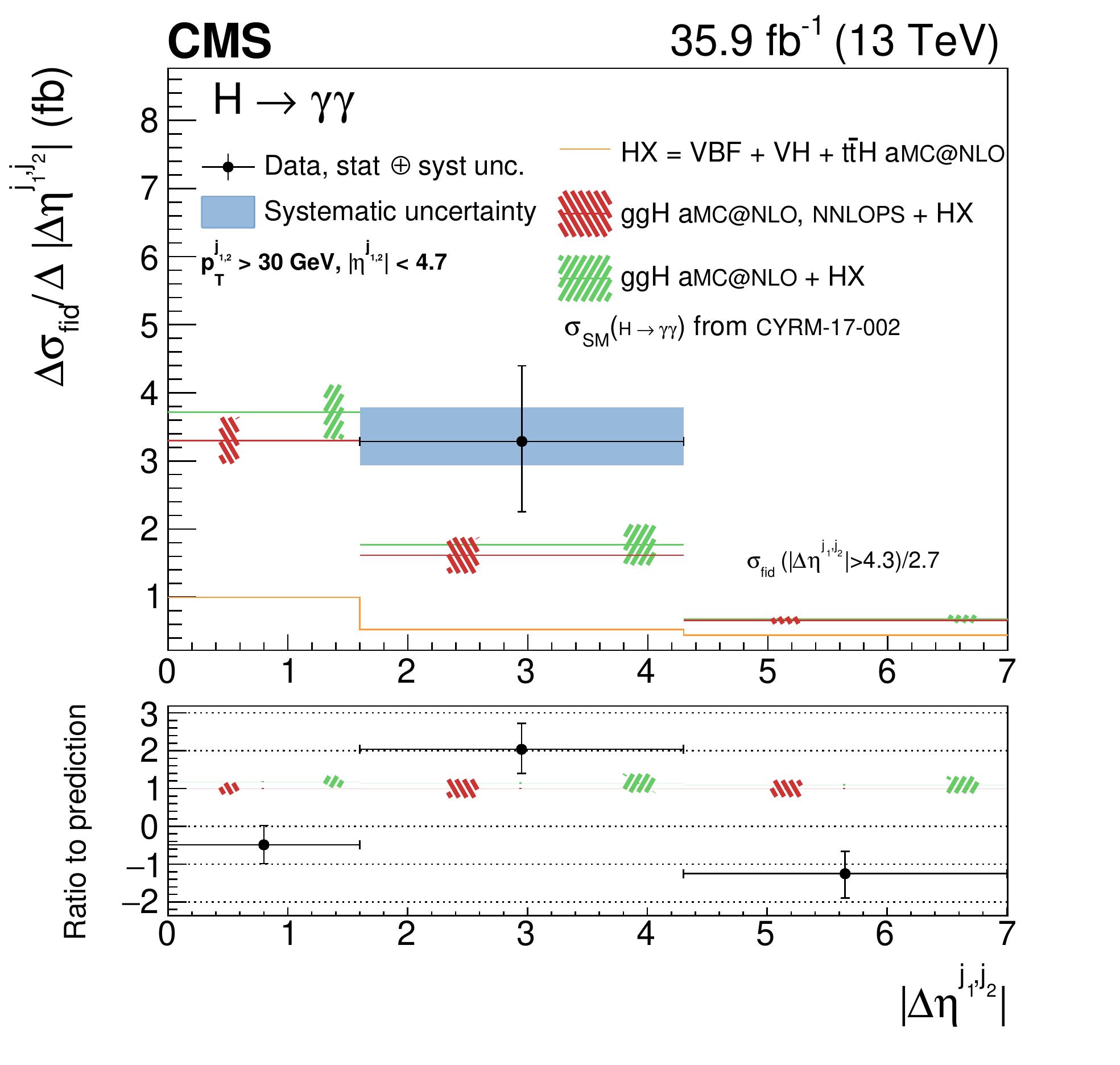}

   \caption{Measurement of the differential cross section (black points) as functions of $\zeppenf$, $\mjonejtwo$, and $\absDetajonejtwo$. The error bars indicate $1$ standard deviation uncertainty. The systematic component of the uncertainty is shown by the blue band.
The measurements are compared to two different simulation programs (histograms) with their uncertainties (hatched areas), both normalized to the same theoretical predictions from Ref.~\cite{LHCHXSWG:YR4}.
 When the last bin of the distribution is an overflow bin, the normalization of the cross section in that bin is indicated in the figure.}
  \label{fig:expPrecision4}
 \end{center}
\end{figure}

\begin{figure}[!htb]
 \begin{center}
   \includegraphics[width=0.49\textwidth]{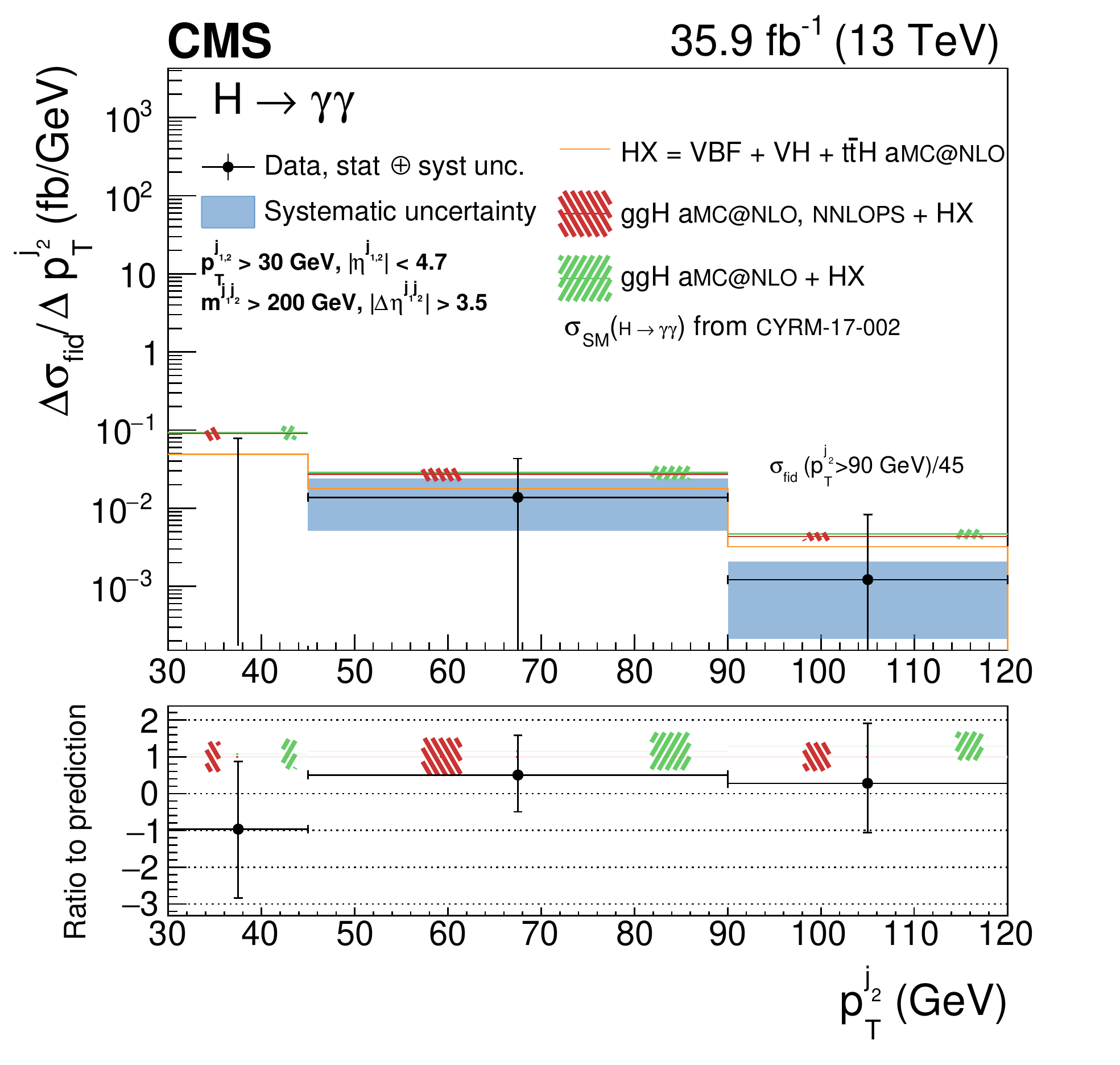}
   \includegraphics[width=0.49\textwidth]{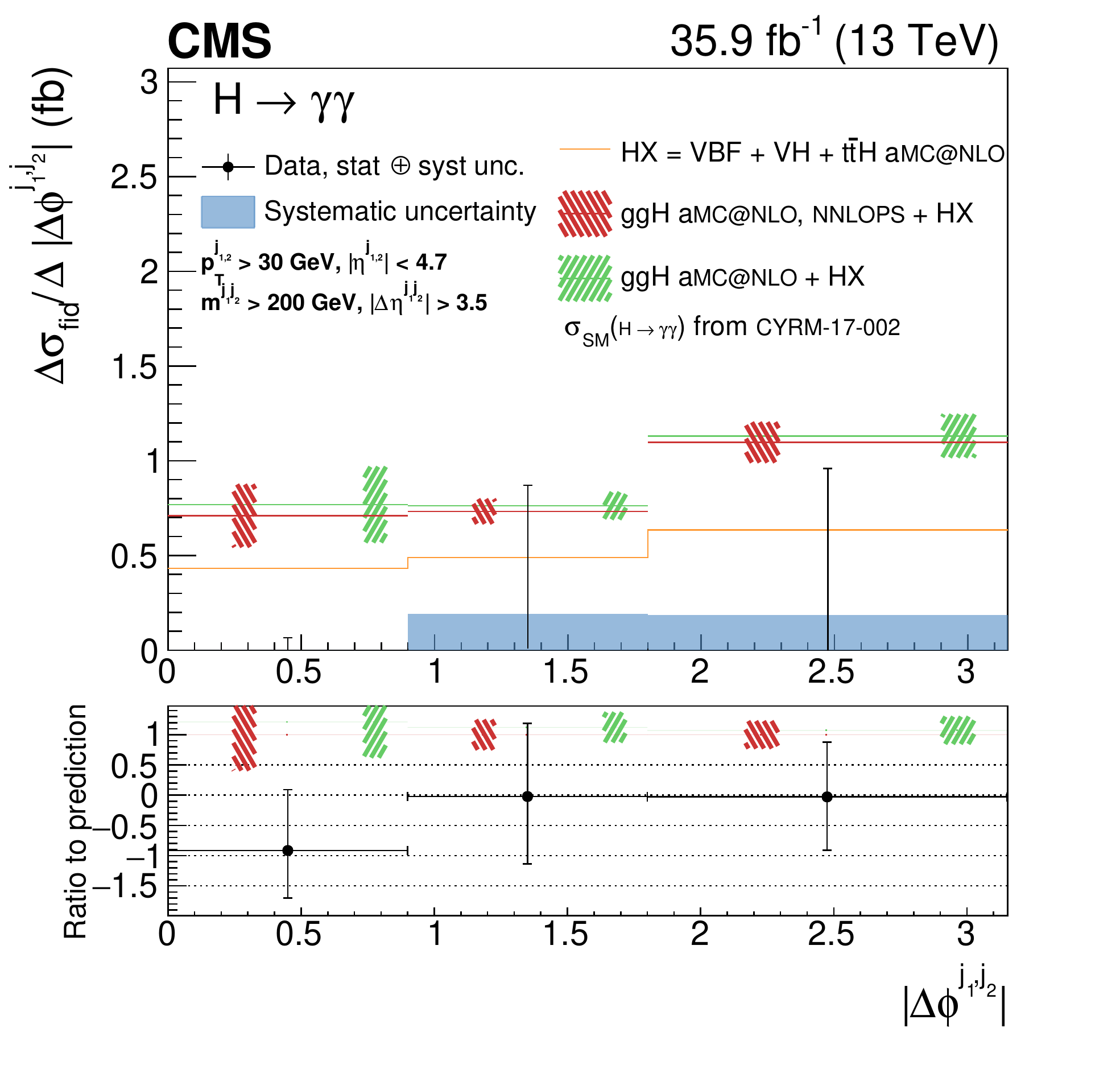}\\
   \includegraphics[width=0.49\textwidth]{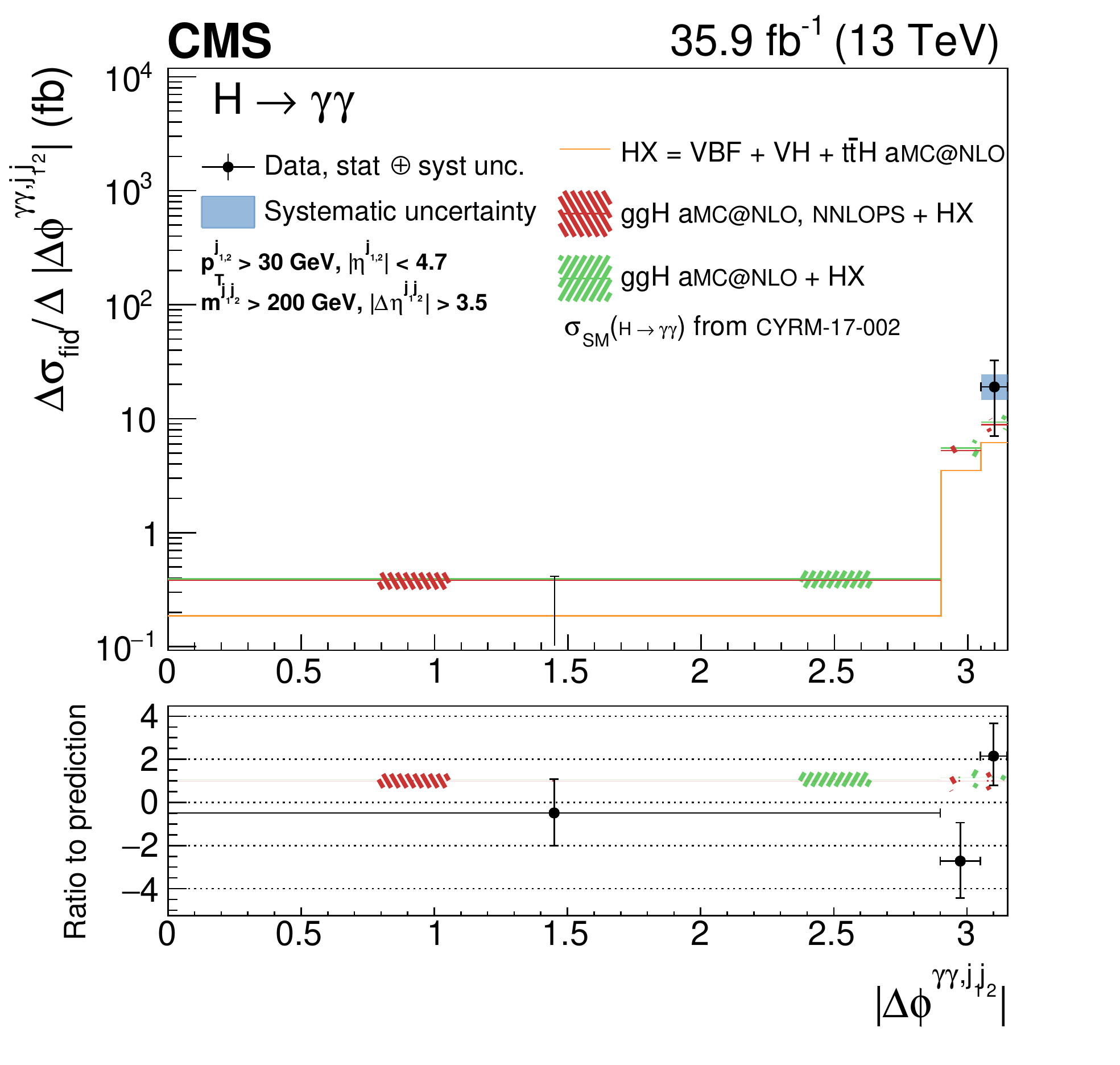}
   \caption{Measurement in a VBF-enriched region of the fiducial phase space of the differential cross section (black points) as functions of  $\ptjtwo$, $\absDphijonejtwo$, and $\absDphiggjonejtwo$. The error bars indicate $1$ standard deviation uncertainty. The systematic component of the uncertainty is shown by the blue band.
The measurements are compared to two different simulation programs (histograms) with their uncertainties (hatched areas), both normalized to the same theoretical predictions from Ref.~\cite{LHCHXSWG:YR4}.
  When the last bin of the distribution is an overflow bin, the normalization of the cross section in that bin is indicated in the figure.}
  \label{fig:expPrecision5}
 \end{center}
\end{figure}

\begin{figure}[!htb]
 \begin{center}
   \includegraphics[width=0.49\textwidth]{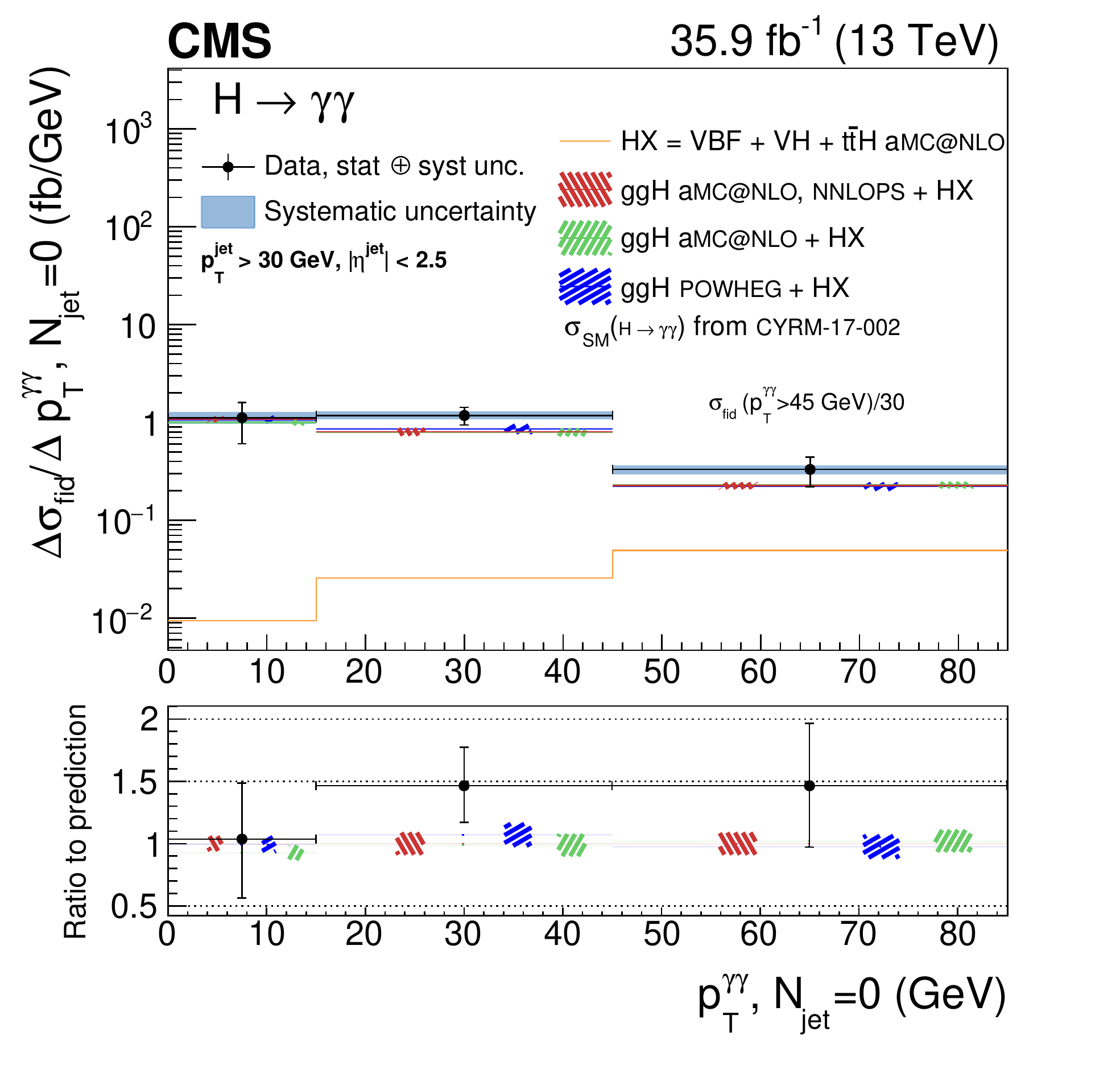}
   \includegraphics[width=0.49\textwidth]{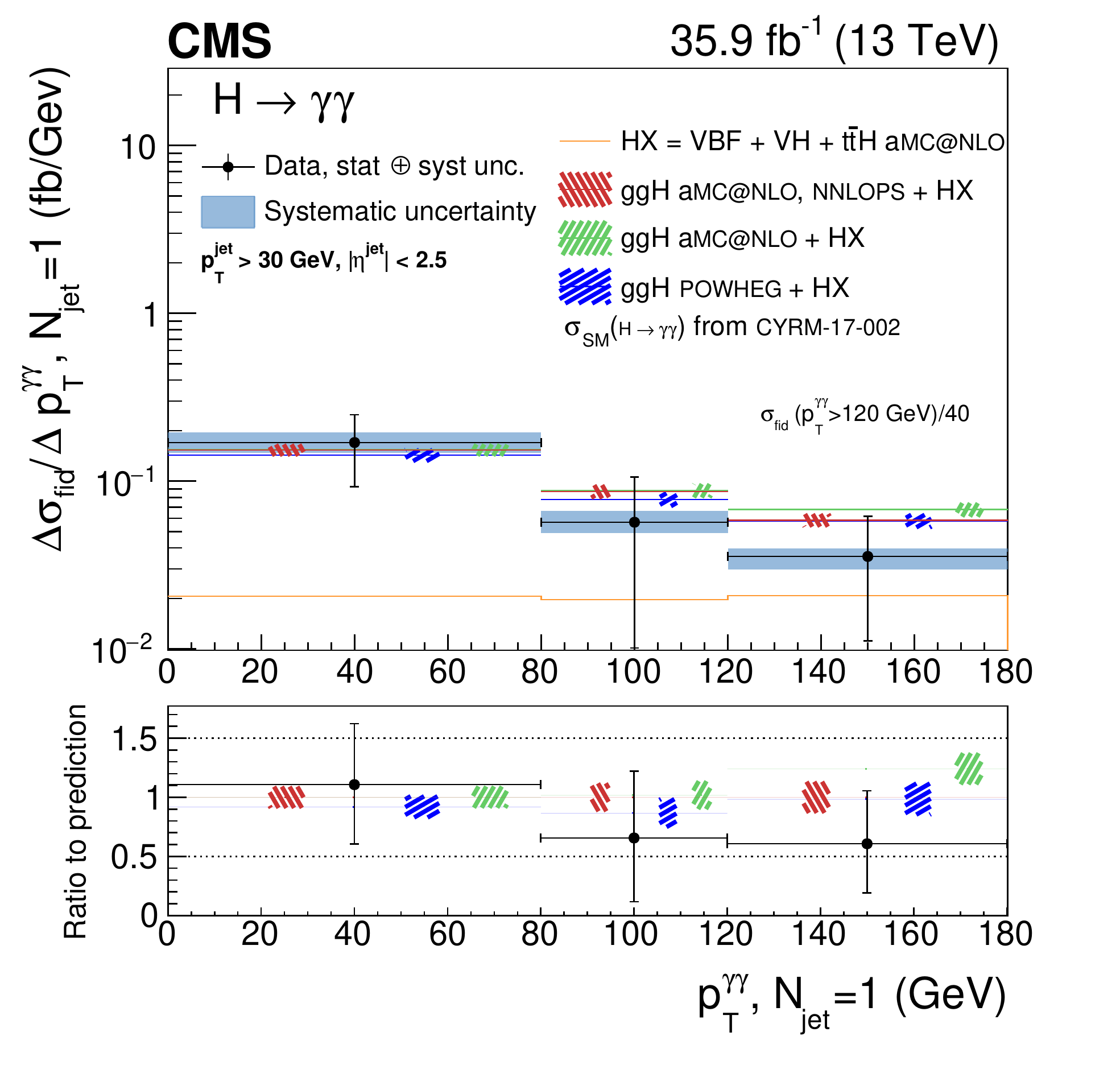}\\
   \includegraphics[width=0.49\textwidth]{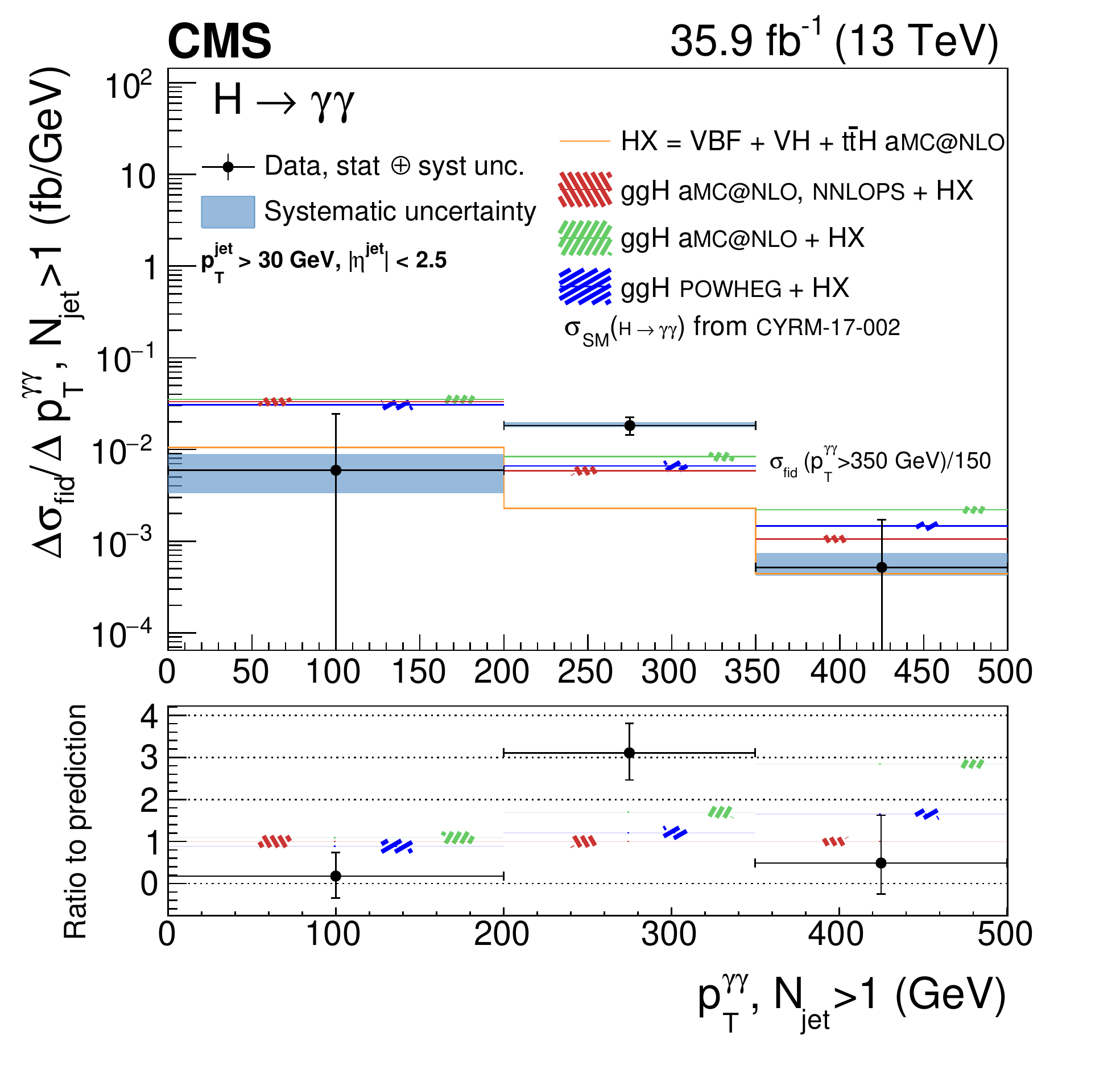}

   \caption{Measurement of the differential cross section (black points) as function of $\ptgg$ and $\njet$ simultaneously. The error bars indicate $1$ standard deviation uncertainty. The systematic component of the uncertainty is shown by the blue band.
The measurements are compared to different simulation programs (histograms) with their uncertainties (hatched areas), all normalized to the same theoretical predictions from Ref.~\cite{LHCHXSWG:YR4}.
   The normalization of the cross section in last, overflow bin is indicated in the figure.}
  \label{fig:expPrecision8}
 \end{center}
\end{figure}

\begin{figure}[!htb]
 \begin{center}
   \includegraphics[width=0.49\textwidth]{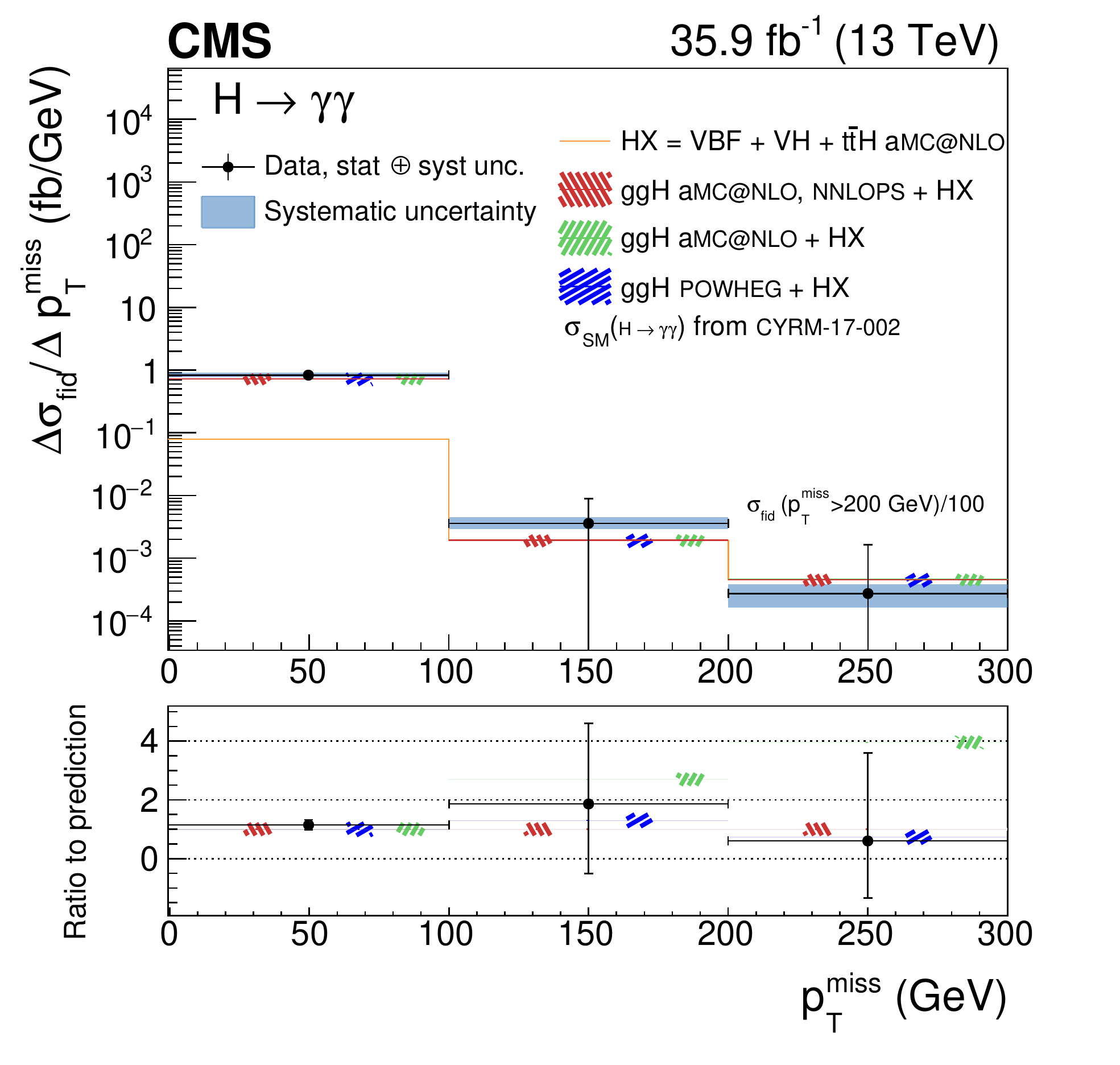}
   \includegraphics[width=0.49\textwidth]{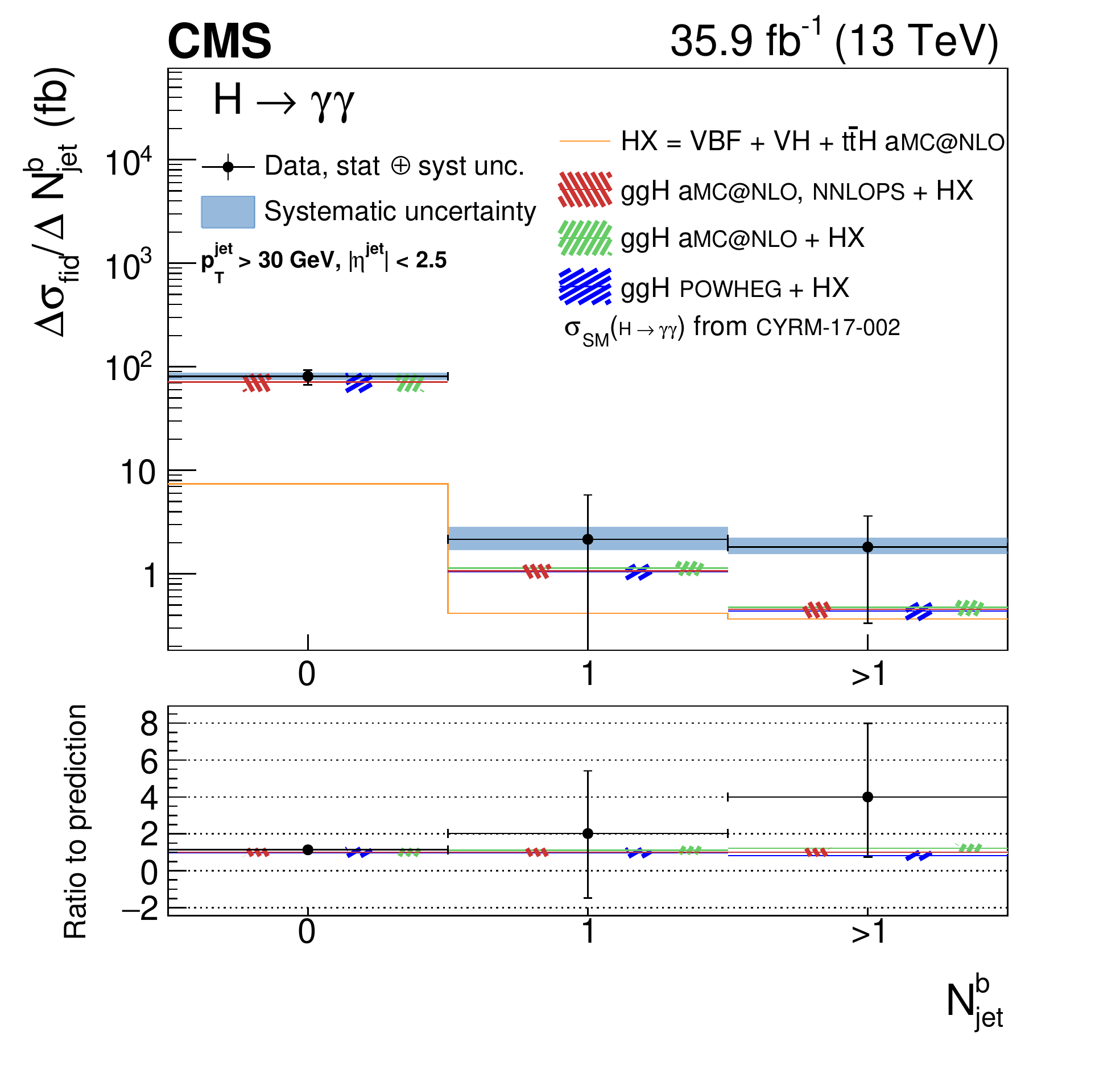}\\
   \includegraphics[width=0.49\textwidth]{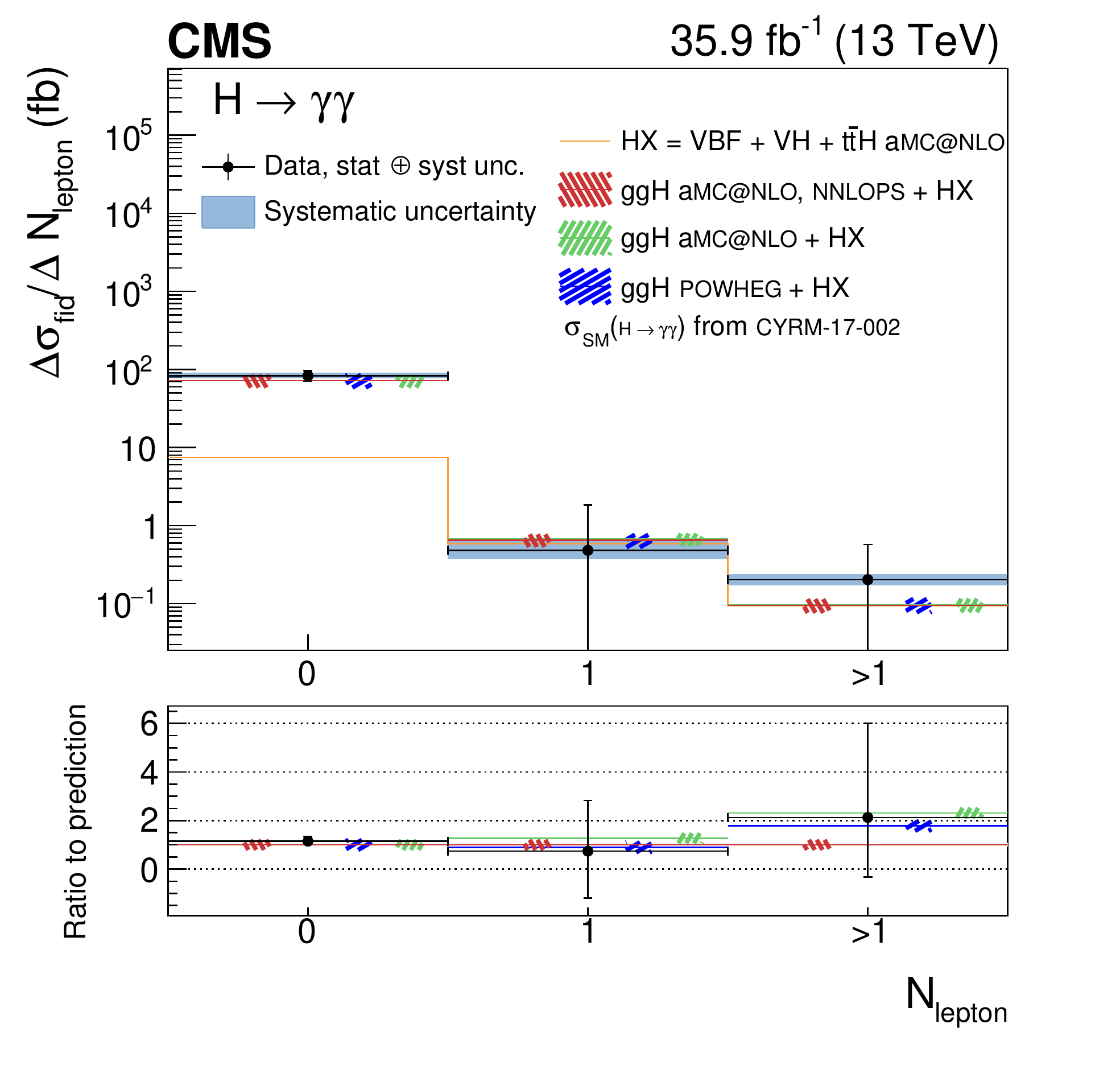}

   \caption{Measurement of the differential cross section (black points) as functions of $\ptmiss$, $\nbjet$, and $\nlep$. The error bars indicate $1$ standard deviation uncertainty. The systematic component of the uncertainty is shown by the blue band.
The measurements are compared to different simulation programs (histograms) with their uncertainties (hatched areas), all normalized to the same theoretical predictions from Ref.~\cite{LHCHXSWG:YR4}.
  When the last bin of the distribution is an overflow bin, the normalization of the cross section in that bin is indicated in the figure.}
  \label{fig:expPrecision6}
 \end{center}
\end{figure}

The measurements of the differential cross sections as functions of the observables under study are reported in Figs.~\ref{fig:expPrecision1}--\ref{fig:expPrecision6}.
The figures show the best fit value, the  $1$ standard deviation uncertainty resulting from the likelihood
scans for each bin of each observable, and  the systematic contribution to the total uncertainty. The measurements are compared to theoretical predictions obtained using different generators for the calculation of the spectrum of the observables, with the cross section and branching fraction values taken from Ref.~\cite{LHCHXSWG:YR4}. The contributions from the VBF, V\PH, and \ttH production mechanisms are simulated with the \MGvATNLO program. For the $\Pg\Pg\PH$ contribution, three different predictions are calculated and each of these in turn is added to the VBF, V\PH, and \ttH contributions.
The $\Pg\Pg\PH$ contribution is simulated with the \MGvATNLO program and its events are weighted to match the {\textsc{nnlops}} prediction, as explained in Section~\ref{sec:samples}. For the observables inclusive in the number of jets or describing the kinematic observables of the first jet, the prediction for the $\Pg\Pg\PH$ contribution is also simulated using the \POWHEG program.
The theoretical prediction for the $\absDphiggjonejtwo$ spectrum is known to be not infrared-safe for values close to $\pi$ \cite{LHCHXSWG:YR3}, with large uncertainties related to soft jet production in $\Pg\Pg\PH$ events. In this regime the theoretical uncertainties obtained with scale variations tend to be underestimated. This effect is particularly relevant  in the last bin of the spectrum corresponding to the values $3.05$--$\pi$.

The precision in the measurement of the differential fiducial cross sections varies widely depending on the observable under study. The observable that allows the most precise measurement and the largest number of bins is $\ptgg$, where 8 bins are defined and the measurements have uncertainties around 40\% on average, as shown in Fig.~\ref{fig:expPrecision1} (top left). The observables $\absrapgg$ and $\abscosthetast$ yield measurements with uncertainties at the level of $\sim$35\% in 5  bins, reported in Fig. ~\ref{fig:expPrecision1} (bottom left and right, respectively). The uncertainties in the measurement as a function of the jet multiplicity, $\njet$, presented in Fig.~\ref{fig:expPrecision1} (top right), range from $\sim$25\% for the $0$-jet bin up to $\gtrsim$100\% for the high jet multiplicity bins. For the observables describing the properties of the first additional jet $\jone$, shown in Fig.~\ref{fig:expPrecision2}, the average uncertainty is $\sim$50\% with four bins, with the exception of $\ptjone$, where $5$ bins are used and the uncertainties are around 70\%. The spectrum of the observables involving two jets, displayed in Figs.~\ref{fig:expPrecision3} and ~\ref{fig:expPrecision4}, is measured with uncertainties ranging between $\sim$70 and $\sim$90\% and employing three bins, except for $\mjonejtwo$ for which 5 bins are defined.
As the measurements as functions of $\ptjtwo$, $\absDphijonejtwo$, and  $\absDphiggjonejtwo$ are restricted to the VBF-enriched region of the phase space, the uncertainties are between 110 and 150\%, as shown in Fig.~\ref{fig:expPrecision5}.
The double differential measurement as a function of $\ptgg$ and $\njet$, reported in Fig.~\ref{fig:expPrecision8}, allows the extraction of the cross section in 9 bins with uncertainties ranging from $\sim$35 to $\sim$60\%.
The measurements as a function of $\nbjet$, $\nlep$, and $\ptmiss$, presented in Fig.~\ref{fig:expPrecision6}, have uncertainties, in all bins except the first, of 200--250\%. In the first bin, which contains the vast majority of the selected events,  the uncertainties are comparable to the uncertainty in the inclusive cross section measurement.
The results are found to be in agreement with the SM predictions within the uncertainties.

\begin{figure}
 \begin{center}
   \includegraphics[width=0.75\textwidth]{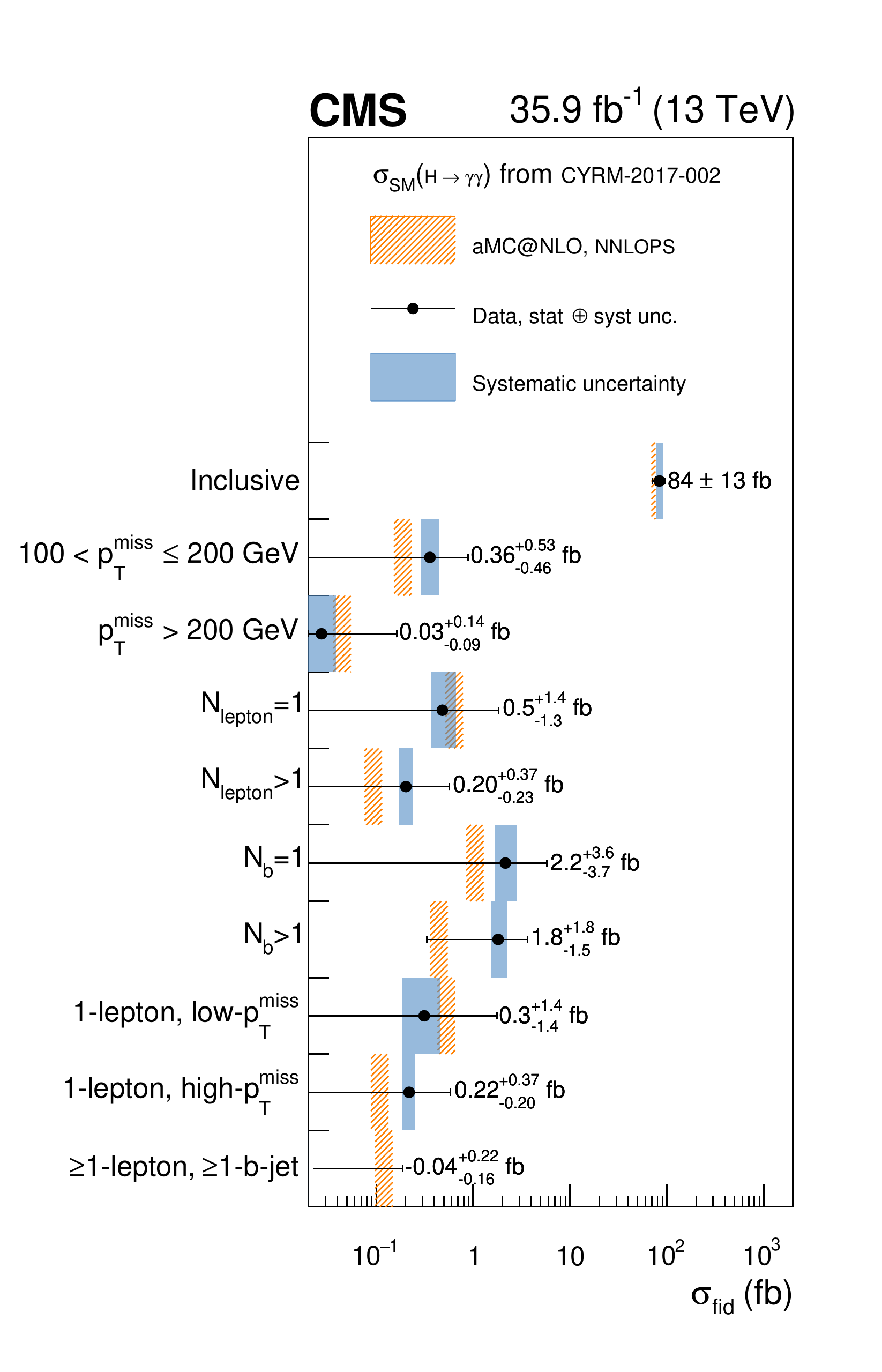}
   \caption{The measurement of the differential cross section (black points) for different regions of the phase space, listed on the vertical axis. The black error bars indicate the $1$ standard deviation uncertainty and its systematic component is shown by the blue band. The measurements are found in agreement with the theoretical predictions (orange hatched area), normalized to the predictions from Ref.~\cite{LHCHXSWG:YR4}. The measured value of some of the cross sections is found to be compatible with the background-only hypothesis.}
  \label{fig:summaryInclusive}
 \end{center}
\end{figure}

The measurement of the inclusive fiducial cross section is also performed in regions of the fiducial phase space. These regions, as described in Section~\ref{sec:observables}, represent a very limited fraction ($\sim$10$^{-3}$) of the baseline phase space and target individual production mechanisms of the Higgs boson. The results of these measurements are summarized in Fig.~\ref{fig:summaryInclusive}, where selected bins of the differential measurements are also reported, in order to provide a more comprehensive summary. The measurements are compared to the corresponding theoretical predictions, obtained using  \MGvATNLO
 simulated signal events, with the $\Pg\Pg\PH$ simulated events weighted to match the {\textsc{nnlops}} program prediction.  The values of  the cross section and  the branching fraction are  taken from Ref.~\cite{LHCHXSWG:YR4}. The uncertainties in the measurements are around $250$\% for the \emph{1-lepton, high \ptmiss} and \emph{1-lepton, low \ptmiss} cross sections, and $\sim$350\% for the \emph{$\geq$1-lepton, $\geq$1-b-jet} cross section. The measurements are found to be compatible with the SM prediction.

\section{Summary}
\label{sec:conclusions}

Measurements of the inclusive and differential fiducial  cross sections for production of the Higgs boson in the diphoton decay channel have been performed using an integrated luminosity of \mylumi of proton-proton collision data collected by the CMS experiment at a center-of-mass energy of $13\TeV$. The measurements of the differential cross sections are reported as functions of a set of observables characterizing the diphoton system and particles produced in association with the Higgs boson.
The measurements are performed for isolated photons in the fiducial phase space defined by requiring that both photons are isolated and within the pseudorapidity $\abs{\eta^{\gamma}}<2.5$ and $\pt/m_{\gamma\gamma}>1/3 (1/4)$ for the leading (subleading) photon.
In this fiducial phase space, the cross section is measured to be  $84\pm13\unit{fb}$, compared with a theoretical prediction of $73\pm4\unit{fb}$.
The double-differential measurement is performed as a function of the transverse momentum of the diphoton system and the jet multiplicity in the event.

A subset of the differential observables describing the kinematics of the system of two additional jets is studied in a vector-boson-fusion enriched fiducial phase space.
The inclusive cross section is also measured in three regions of the fiducial phase space, additionally requiring the presence of one selected lepton and  missing transverse momentum $\ptmiss<100$\GeV, or one selected lepton and $\ptmiss\geq100$\GeV, or at least one selected lepton and at least one \cPqb-tagged jet, respectively.
The measurements  are in agreement within the uncertainties with the predictions for the production of a standard model Higgs boson.

\clearpage

\begin{acknowledgments}
We congratulate our colleagues in the CERN accelerator departments for the excellent performance of the LHC and thank the technical and administrative staffs at CERN and at other CMS institutes for their contributions to the success of the CMS effort. In addition, we gratefully acknowledge the computing centers and personnel of the Worldwide LHC Computing Grid for delivering so effectively the computing infrastructure essential to our analyses. Finally, we acknowledge the enduring support for the construction and operation of the LHC and the CMS detector provided by the following funding agencies: BMWFW and FWF (Austria); FNRS and FWO (Belgium); CNPq, CAPES, FAPERJ, FAPERGS, and FAPESP (Brazil); MES (Bulgaria); CERN; CAS, MoST, and NSFC (China); COLCIENCIAS (Colombia); MSES and CSF (Croatia); RPF (Cyprus); SENESCYT (Ecuador); MoER, ERC IUT, and ERDF (Estonia); Academy of Finland, MEC, and HIP (Finland); CEA and CNRS/IN2P3 (France); BMBF, DFG, and HGF (Germany); GSRT (Greece); NKFIA (Hungary); DAE and DST (India); IPM (Iran); SFI (Ireland); INFN (Italy); MSIP and NRF (Republic of Korea); MES (Latvia); LAS (Lithuania); MOE and UM (Malaysia); BUAP, CINVESTAV, CONACYT, LNS, SEP, and UASLP-FAI (Mexico); MOS (Montenegro); MBIE (New Zealand); PAEC (Pakistan); MSHE and NSC (Poland); FCT (Portugal); JINR (Dubna); MON, RosAtom, RAS, RFBR, and NRC KI (Russia); MESTD (Serbia); SEIDI, CPAN, PCTI, and FEDER (Spain); MOSTR (Sri Lanka); Swiss Funding Agencies (Switzerland); MST (Taipei); ThEPCenter, IPST, STAR, and NSTDA (Thailand); TUBITAK and TAEK (Turkey); NASU and SFFR (Ukraine); STFC (United Kingdom); DOE and NSF (USA).

\hyphenation{Rachada-pisek} Individuals have received support from the Marie-Curie program and the European Research Council and Horizon 2020 Grant, contract No. 675440 (European Union); the Leventis Foundation; the A. P. Sloan Foundation; the Alexander von Humboldt Foundation; the Belgian Federal Science Policy Office; the Fonds pour la Formation \`a la Recherche dans l'Industrie et dans l'Agriculture (FRIA-Belgium); the Agentschap voor Innovatie door Wetenschap en Technologie (IWT-Belgium); the F.R.S.-FNRS and FWO (Belgium) under the ``Excellence of Science - EOS" - be.h project n. 30820817; the Ministry of Education, Youth and Sports (MEYS) of the Czech Republic; the Lend\"ulet (``Momentum") Program and the J\'anos Bolyai Research Scholarship of the Hungarian Academy of Sciences, the New National Excellence Program \'UNKP, the NKFIA research grants 123842, 123959, 124845, 124850 and 125105 (Hungary); the Council of Science and Industrial Research, India; the HOMING PLUS program of the Foundation for Polish Science, cofinanced from European Union, Regional Development Fund, the Mobility Plus program of the Ministry of Science and Higher Education, the National Science Center (Poland), contracts Harmonia 2014/14/M/ST2/00428, Opus 2014/13/B/ST2/02543, 2014/15/B/ST2/03998, and 2015/19/B/ST2/02861, Sonata-bis 2012/07/E/ST2/01406; the National Priorities Research Program by Qatar National Research Fund; the Programa Estatal de Fomento de la Investigaci{\'o}n Cient{\'i}fica y T{\'e}cnica de Excelencia Mar\'{\i}a de Maeztu, grant MDM-2015-0509 and the Programa Severo Ochoa del Principado de Asturias; the Thalis and Aristeia programs cofinanced by EU-ESF and the Greek NSRF; the Rachadapisek Sompot Fund for Postdoctoral Fellowship, Chulalongkorn University and the Chulalongkorn Academic into Its 2nd Century Project Advancement Project (Thailand); the Welch Foundation, contract C-1845; and the Weston Havens Foundation (USA). \end{acknowledgments}

\bibliography{auto_generated}

\cleardoublepage \appendix\section{The CMS Collaboration \label{app:collab}}\begin{sloppypar}\hyphenpenalty=5000\widowpenalty=500\clubpenalty=5000\input{HIG-17-025-authorlist.tex}\end{sloppypar}
\end{document}

%% file: HIG-17-025-authorlist.tex
\vskip\cmsinstskip
\textbf{Yerevan Physics Institute, Yerevan, Armenia}\\*[0pt]
A.M.~Sirunyan, A.~Tumasyan
\vskip\cmsinstskip
\textbf{Institut f\"{u}r Hochenergiephysik, Wien, Austria}\\*[0pt]
W.~Adam, F.~Ambrogi, E.~Asilar, T.~Bergauer, J.~Brandstetter, M.~Dragicevic, J.~Er\"{o}, A.~Escalante~Del~Valle, M.~Flechl, R.~Fr\"{u}hwirth\cmsAuthorMark{1}, V.M.~Ghete, J.~Hrubec, M.~Jeitler\cmsAuthorMark{1}, N.~Krammer, I.~Kr\"{a}tschmer, D.~Liko, T.~Madlener, I.~Mikulec, N.~Rad, H.~Rohringer, J.~Schieck\cmsAuthorMark{1}, R.~Sch\"{o}fbeck, M.~Spanring, D.~Spitzbart, A.~Taurok, W.~Waltenberger, J.~Wittmann, C.-E.~Wulz\cmsAuthorMark{1}, M.~Zarucki
% ...
\vskip\cmsinstskip
% ...
1:  Also at Vienna University of Technology, Vienna, Austria\\
% ...


%% file: HIG-17-025_temp.bbl
\providecommand{\href}[2]{#2}\begingroup\raggedright\begin{thebibliography}{10}%
\makeatletter
\providecommand{\hrefCMSnoop }[0]{\@secondoftwo}%
\makeatother
\providecommand{\doi}{\texttt{doi:}\begingroup \urlstyle{tt}\Url}

\bibitem{Aad:2012tfa}
\hrefCMSnoop {}{{ATLAS Collaboration}, ``{Observation of a new particle in the
  search for the standard model Higgs boson with the ATLAS detector at the
  LHC}'',} \textit{ Phys. Lett. B} \textbf{ 716} (2012) 1,
  \href{http://dx.doi.org/10.1016/j.physletb.2012.08.020}{\doi{10.1016/j.physletb.2012.08.020}},
\href{http://www.arXiv.org/abs/1207.7214}{\texttt{arXiv:1207.7214}}.

\bibitem{Chatrchyan:2012ufa}
\hrefCMSnoop {}{{CMS Collaboration}, ``{Observation of a new boson at a mass of
  125 GeV with the CMS experiment at the LHC}'',} \textit{ Phys. Lett. B}
  \textbf{ 716} (2012) 30,
\href{http://dx.doi.org/10.1016/j.physletb.2012.08.021}{\doi{10.1016/j.physletb.2012.08.021}}.

\bibitem{Chatrchyan:2013lba}
\hrefCMSnoop {}{{CMS Collaboration}, ``{Observation of a new boson with mass
  near 125 GeV in pp collisions at $\sqrt{s}$ = 7 and 8 TeV}'',} \textit{ JHEP}
  \textbf{ 06} (2013) 081,
  \href{http://dx.doi.org/10.1007/JHEP06(2013)081}{\doi{10.1007/JHEP06(2013)081}},
\href{http://www.arXiv.org/abs/1303.4571}{\texttt{arXiv:1303.4571}}.

\bibitem{Aad2016}
\hrefCMSnoop {}{{ATLAS and CMS Collaborations}, ``{Measurements of the Higgs
  boson production and decay rates and constraints on its couplings from a
  combined ATLAS and CMS analysis of the LHC pp collision data at $\sqrt{s}=7$
  and 8 TeV}'',} \textit{ JHEP} \textbf{ 08} (2016) 045,
  \href{http://dx.doi.org/10.1007/JHEP08(2016)045}{\doi{10.1007/JHEP08(2016)045}},
  \href{http://www.arXiv.org/abs/1606.02266}{\texttt{arXiv:1606.02266}}.

\bibitem{Aad2014}
\hrefCMSnoop {}{{{ATLAS}} Collaboration, ``{Measurements of fiducial and
  differential cross sections for Higgs boson production in the diphoton decay
  channel at $\sqrt{s}= 8$ TeV with ATLAS}'',} \textit{ JHEP} \textbf{ 09}
  (2014) 112,
  \href{http://dx.doi.org/10.1007/JHEP09(2014)112}{\doi{10.1007/JHEP09(2014)112}},
  \href{http://www.arXiv.org/abs/1407.4222}{\texttt{arXiv:1407.4222}}.

\bibitem{Khachatryan2016}
\hrefCMSnoop {}{{CMS Collaboration}, ``{Measurement of differential cross
  sections for Higgs boson production in the diphoton decay channel in pp
  collisions at $\sqrt{s} = 8$ \TeV}'',} \textit{ Eur. Phys. J. C} \textbf{ 76}
  (2016) 13,
  \href{http://dx.doi.org/10.1140/epjc/s10052-015-3853-3}{\doi{10.1140/epjc/s10052-015-3853-3}},
  \href{http://www.arXiv.org/abs/1508.07819}{\texttt{arXiv:1508.07819}}.

\bibitem{2014234}
\hrefCMSnoop {}{{{ATLAS}} Collaboration, ``{Fiducial and differential cross
  sections of Higgs boson production measured in the four-lepton decay channel
  in pp collisions at $\sqrt{s} =$ 8 TeV with the ATLAS detector}'',} \textit{
  Phys. Lett. B} \textbf{ 738} (2014) 234,
  \href{http://dx.doi.org/10.1016/j.physletb.2014.09.054}{\doi{10.1016/j.physletb.2014.09.054}},
  \href{http://www.arXiv.org/abs/1408.3226}{\texttt{arXiv:1408.3226}}.

\bibitem{Khachatryan:2118088}
\hrefCMSnoop {}{{CMS Collaboration}, ``{Measurement of differential and
  integrated fiducial cross sections for Higgs boson production in the
  four-lepton decay channel in pp collisions at $\sqrt{s} =$ 7 and 8 TeV}'',}
  \textit{ JHEP} \textbf{ 04} (2015) 005,
  \href{http://dx.doi.org/10.1007/JHEP04(2016)005}{\doi{10.1007/JHEP04(2016)005}},
  \href{http://www.arXiv.org/abs/1512.08377}{\texttt{arXiv:1512.08377}}.

\bibitem{Aad:2145362}
\hrefCMSnoop {}{{ATLAS Collaboration}, ``{Measurement of fiducial differential
  cross sections of gluon-fusion production of Higgs bosons decaying to
  $WW^{\ast}{\rightarrow\,}e\nu\mu\nu$ with the ATLAS detector at $\sqrt{s}=8$
  TeV.}'',} \textit{ JHEP} \textbf{ 08} (2016) 104,
  \href{http://dx.doi.org/10.1007/JHEP08(2016)104}{\doi{10.1007/JHEP08(2016)104}},
  \href{http://www.arXiv.org/abs/1604.02997}{\texttt{arXiv:1604.02997}}.

\bibitem{Khachatryan:2158105}
\hrefCMSnoop {}{{CMS Collaboration}, ``{Measurement of the transverse momentum
  spectrum of the Higgs boson produced in pp collisions at $ \sqrt{s} = $ 8 TeV
  using $\mathrm{ H }\to\mathrm{ W }\mathrm{ W }$ decays}'',} \textit{ JHEP}
  \textbf{ 03} (2017) 032,
  \href{http://dx.doi.org/10.1007/JHEP03(2017)032}{\doi{10.1007/JHEP03(2017)032}},
  \href{http://www.arXiv.org/abs/1606.01522}{\texttt{arXiv:1606.01522}}.

\bibitem{Aaboud:2277731}
\hrefCMSnoop {}{{ATLAS Collaboration}, ``{Measurement of inclusive and
  differential cross sections in the $H \rightarrow ZZ^* \rightarrow 4\ell$
  decay channel in $pp$ collisions at $\sqrt{s}$ = 13 TeV with the ATLAS
  detector.}'',} \textit{ JHEP} \textbf{ 10} (2017) 132,
  \href{http://dx.doi.org/10.1007/JHEP10(2017)132}{\doi{10.1007/JHEP10(2017)132}},
  \href{http://www.arXiv.org/abs/1708.02810}{\texttt{arXiv:1708.02810}}.

\bibitem{Sirunyan:2272260}
\hrefCMSnoop {}{{CMS Collaboration}, ``{Measurements of properties of the Higgs
  boson decaying into the four-lepton final state in pp collisions at $\sqrt{s}
  = 13$ TeV.}'',} \textit{ JHEP} \textbf{ 11} (2017) 047,
  \href{http://dx.doi.org/10.1007/JHEP11(2017)047}{\doi{10.1007/JHEP11(2017)047}},
  \href{http://www.arXiv.org/abs/1706.09936}{\texttt{arXiv:1706.09936}}.

\bibitem{LHCHXSWG:YR4}
\hrefCMSnoop {}{{LHC Higgs Cross Section Working Group}, ``{Handbook of LHC
  Higgs cross sections: 4. Deciphering the nature of the Higgs sector}'',}
  \textit{ CERN} (2016)
  \href{http://dx.doi.org/10.23731/CYRM-2017-002}{\doi{10.23731/CYRM-2017-002}},
\href{http://www.arXiv.org/abs/1610.07922}{\texttt{arXiv:1610.07922}}.

\bibitem{CMS-PAS-HIG-16-040}
\hrefCMSnoop {}{{CMS Collaboration}, ``{Measurements of Higgs boson properties
  in the diphoton decay channel in proton-proton collisions at $\sqrt{s} =$ 13
  TeV}'',} \textit{ JHEP} \textbf{ 11} (2018) 185,
  \href{http://dx.doi.org/10.1007/JHEP11(2018)185}{\doi{10.1007/JHEP11(2018)185}},
\href{http://www.arXiv.org/abs/1804.02716}{\texttt{arXiv:1804.02716}}.

\bibitem{Khachatryan:2016bia}
\hrefCMSnoop {}{{CMS Collaboration}, ``{The CMS trigger system}'',} \textit{
  JINST} \textbf{ 12} (2017) P01020,
  \href{http://dx.doi.org/10.1088/1748-0221/12/01/P01020}{\doi{10.1088/1748-0221/12/01/P01020}},
\href{http://www.arXiv.org/abs/1609.02366}{\texttt{arXiv:1609.02366}}.

\bibitem{Chatrchyan:2008zzk}
\hrefCMSnoop {}{{CMS Collaboration}, ``The {CMS} experiment at the {CERN}
  {LHC}'',} \textit{ JINST} \textbf{ 3} (2008) S08004,
\href{http://dx.doi.org/10.1088/1748-0221/3/08/S08004}{\doi{10.1088/1748-0221/3/08/S08004}}.

\bibitem{Agostinelli:2002hh}
\hrefCMSnoop {}{{GEANT4} Collaboration, ``{GEANT4} --- a simulation toolkit'',}
  \textit{ Nucl. Instrum. Meth. A} \textbf{ 506} (2003) 250,
\href{http://dx.doi.org/10.1016/S0168-9002(03)01368-8}{\doi{10.1016/S0168-9002(03)01368-8}}.

\bibitem{Alwall:2014hca}
J.~Alwall\hrefCMSnoop {}{ {et~al.}, ``{The automated computation of tree-level
  and next-to-leading order differential cross sections, and their matching to
  parton shower simulations}'',} \textit{ JHEP} \textbf{ 07} (2014) 079,
  \href{http://dx.doi.org/10.1007/JHEP07(2014)079}{\doi{10.1007/JHEP07(2014)079}},
\href{http://www.arXiv.org/abs/1405.0301}{\texttt{arXiv:1405.0301}}.

\bibitem{Frederix:2012ps}
\hrefCMSnoop {}{R.~Frederix and S.~Frixione, ``Merging meets matching in
  {MC@NLO}'',} \textit{ JHEP} \textbf{ 12} (2012) 061,
  \href{http://dx.doi.org/10.1007/JHEP12(2012)061}{\doi{10.1007/JHEP12(2012)061}},
\href{http://www.arXiv.org/abs/1209.6215}{\texttt{arXiv:1209.6215}}.

\bibitem{Sjostrand:2014zea}
T.~Sj{\"o}strand\hrefCMSnoop {}{ {et~al.}, ``{An Introduction to PYTHIA
  8.2}'',} \textit{ Comput. Phys. Commun.} \textbf{ 191} (2015) 159,
  \href{http://dx.doi.org/10.1016/j.cpc.2015.01.024}{\doi{10.1016/j.cpc.2015.01.024}},
\href{http://www.arXiv.org/abs/1410.3012}{\texttt{arXiv:1410.3012}}.

\bibitem{CUETP8M1}
\hrefCMSnoop {}{P.~Skands, S.~Carrazza, and J.~Rojo, ``{Tuning PYTHIA 8.1: the
  Monash 2013 Tune}'',} \textit{ Eur. Phys. J. C} \textbf{ 74} (2014) 3024,
  \href{http://dx.doi.org/10.1140/epjc/s10052-014-3024-y}{\doi{10.1140/epjc/s10052-014-3024-y}},
\href{http://www.arXiv.org/abs/1404.5630}{\texttt{arXiv:1404.5630}}.

\bibitem{Hamilton:2013fea}
\hrefCMSnoop {}{K.~Hamilton, P.~Nason, E.~Re, and G.~Zanderighi, ``{NNLOPS
  simulation of Higgs boson production}'',} \textit{ JHEP} \textbf{ 10} (2013)
  222,
  \href{http://dx.doi.org/10.1007/JHEP10(2013)222}{\doi{10.1007/JHEP10(2013)222}},
\href{http://www.arXiv.org/abs/1309.0017}{\texttt{arXiv:1309.0017}}.

\bibitem{PowhegMinlo}
\hrefCMSnoop {}{K.~Hamilton, P.~Nason, and G.~Zanderighi, ``{MINLO: multi-scale
  improved NLO}'',} \textit{ JHEP} \textbf{ 10} (2012) 155,
  \href{http://dx.doi.org/10.1007/JHEP10(2012)155}{\doi{10.1007/JHEP10(2012)155}},
\href{http://www.arXiv.org/abs/1206.3572}{\texttt{arXiv:1206.3572}}.

\bibitem{Kardos:2014dua}
\hrefCMSnoop {}{A.~Kardos, P.~Nason, and C.~Oleari, ``{Three-jet production in
  POWHEG}'',} \textit{ JHEP} \textbf{ 04} (2014) 043,
  \href{http://dx.doi.org/10.1007/JHEP04(2014)043}{\doi{10.1007/JHEP04(2014)043}},
\href{http://www.arXiv.org/abs/1402.4001}{\texttt{arXiv:1402.4001}}.

\bibitem{powheg1}
\hrefCMSnoop {}{P.~Nason, ``{A new method for combining NLO QCD with shower
  Monte Carlo algorithms}'',} \textit{ JHEP} \textbf{ 11} (2004) 040,
  \href{http://dx.doi.org/10.1088/1126-6708/2004/11/040}{\doi{10.1088/1126-6708/2004/11/040}},
\href{http://www.arXiv.org/abs/hep-ph/0409146}{\texttt{arXiv:hep-ph/0409146}}.

\bibitem{powheg2}
\hrefCMSnoop {}{S.~Frixione, P.~Nason, and C.~Oleari, ``{Matching NLO QCD
  computations with parton shower simulations: the POWHEG method}'',} \textit{
  JHEP} \textbf{ 11} (2007) 070,
  \href{http://dx.doi.org/10.1088/1126-6708/2007/11/070}{\doi{10.1088/1126-6708/2007/11/070}},
\href{http://www.arXiv.org/abs/0709.2092}{\texttt{arXiv:0709.2092}}.

\bibitem{powheg3}
\hrefCMSnoop {}{S.~Alioli, P.~Nason, C.~Oleari, and E.~Re, ``{A general
  framework for implementing NLO calculations in shower Monte Carlo programs:
  the POWHEG BOX}'',} \textit{ JHEP} \textbf{ 06} (2010) 043,
  \href{http://dx.doi.org/10.1007/JHEP06(2010)043}{\doi{10.1007/JHEP06(2010)043}},
\href{http://www.arXiv.org/abs/1002.2581}{\texttt{arXiv:1002.2581}}.

\bibitem{powheg-ggH}
\hrefCMSnoop {}{S.~Alioli, P.~Nason, C.~Oleari, and E.~Re, ``{NLO Higgs boson
  production via gluon fusion matched with shower in POWHEG}'',} \textit{ JHEP}
  \textbf{ 04} (2009) 002,
  \href{http://dx.doi.org/10.1088/1126-6708/2009/04/002}{\doi{10.1088/1126-6708/2009/04/002}},
\href{http://www.arXiv.org/abs/0812.0578}{\texttt{arXiv:0812.0578}}.

\bibitem{Bagnaschi2012}
\hrefCMSnoop {}{E.~Bagnaschi, G.~Degrassi, P.~Slavich, and A.~Vicini, ``{Higgs
  production via gluon fusion in the POWHEG approach in the SM and in the
  MSSM}'',} \textit{ JHEP} \textbf{ 02} (2012) 088,
  \href{http://dx.doi.org/10.1007/JHEP02(2012)088}{\doi{10.1007/JHEP02(2012)088}},
  \href{http://www.arXiv.org/abs/1111.2854}{\texttt{arXiv:1111.2854}}.

\bibitem{Ball:2014uwa}
\hrefCMSnoop {}{{NNPDF} Collaboration, ``{Parton distributions for the LHC Run
  II}'',} \textit{ JHEP} \textbf{ 04} (2015) 040,
  \href{http://dx.doi.org/10.1007/JHEP04(2015)040}{\doi{10.1007/JHEP04(2015)040}},
\href{http://www.arXiv.org/abs/1410.8849}{\texttt{arXiv:1410.8849}}.

\bibitem{Gleisberg:2008ta}
T.~Gleisberg\hrefCMSnoop {}{ {et~al.}, ``{Event generation with SHERPA 1.1}'',}
  \textit{ JHEP} \textbf{ 02} (2009) 007,
  \href{http://dx.doi.org/10.1088/1126-6708/2009/02/007}{\doi{10.1088/1126-6708/2009/02/007}},
\href{http://www.arXiv.org/abs/0811.4622}{\texttt{arXiv:0811.4622}}.

\bibitem{CMS:EGM-14-001}
\hrefCMSnoop {}{{CMS Collaboration}, ``{Performance of photon reconstruction
  and identification with the CMS detector in proton-proton collisions at
  $\sqrt{s} = 8$\TeV}'',} \textit{ JINST} \textbf{ 10} (2015) P08010,
  \href{http://dx.doi.org/10.1088/1748-0221/10/08/P08010}{\doi{10.1088/1748-0221/10/08/P08010}},
\href{http://www.arXiv.org/abs/1502.02702}{\texttt{arXiv:1502.02702}}.

\bibitem{1748-0221-10-06-P06005}
\hrefCMSnoop {}{{CMS Collaboration}, ``{Performance of electron reconstruction
  and selection with the CMS detector in proton-proton collisions at $\sqrt{s}
  = 8$\TeV}'',} \textit{ JINST} \textbf{ 10} (2015) P06005,
  \href{http://dx.doi.org/10.1088/1748-0221/10/06/P06005}{\doi{10.1088/1748-0221/10/06/P06005}},
\href{http://www.arXiv.org/abs/1502.02701}{\texttt{arXiv:1502.02701}}.

\bibitem{Chatrchyan:2013dga}
\hrefCMSnoop {}{{CMS Collaboration}, ``{Energy calibration and resolution of
  the CMS electromagnetic calorimeter in pp collisions at $\sqrt{s}$ = 7
  TeV}'',} \textit{ JINST} \textbf{ 8} (2013) P09009,
  \href{http://dx.doi.org/10.1088/1748-0221/8/09/P09009}{\doi{10.1088/1748-0221/8/09/P09009}},
  \href{http://www.arXiv.org/abs/1306.2016}{\texttt{arXiv:1306.2016}}.

\bibitem{Gaiser:1982yw}
\href
  {http://www-public.slac.stanford.edu/sciDoc/docMeta.aspx?slacPubNumber=slac-r-236.html}{M.~Oreglia,
  ``{A Study of the Reactions $\psi^\prime \to \gamma \gamma \psi$}''}.
\newblock PhD thesis, {Stanford University, Stanford U.S.A.}, 1980.
\newblock
{SLAC} report {SLAC-R-0236} and online at
  http://www.slac.stanford.edu/pubs/slacreports/slac-r-236.html.

\bibitem{CMS-PRF-14-001}
\hrefCMSnoop {}{{CMS Collaboration}, ``{Particle-flow reconstruction and global
  event description with the CMS detector}'',} \textit{ JINST} \textbf{ 12}
  (2017) P10003,
  \href{http://dx.doi.org/10.1088/1748-0221/12/10/P10003}{\doi{10.1088/1748-0221/12/10/P10003}},
\href{http://www.arXiv.org/abs/1706.04965}{\texttt{arXiv:1706.04965}}.

\bibitem{Cacciari:2008gp}
\hrefCMSnoop {}{M.~Cacciari, G.~P. Salam, and G.~Soyez, ``The anti-$\kt$ jet
  clustering algorithm'',} \textit{ JHEP} \textbf{ 04} (2008) 063,
  \href{http://dx.doi.org/10.1088/1126-6708/2008/04/063}{\doi{10.1088/1126-6708/2008/04/063}},
  \href{http://www.arXiv.org/abs/0802.1189}{\texttt{arXiv:0802.1189}}.

\bibitem{Cacciari:2011ma}
\hrefCMSnoop {}{M.~Cacciari, G.~P. Salam, and G.~Soyez, ``{FastJet user
  manual}'',} \textit{ Eur. Phys. J. C} \textbf{ 72} (2012) 1896,
  \href{http://dx.doi.org/10.1140/epjc/s10052-012-1896-2}{\doi{10.1140/epjc/s10052-012-1896-2}},
\href{http://www.arXiv.org/abs/1111.6097}{\texttt{arXiv:1111.6097}}.

\bibitem{CMS-PAS-JME-16-003}
\href {https://cds.cern.ch/record/2256875}{{CMS Collaboration}, ``{Jet
  algorithms performance in 13 TeV data}'',} CMS Physics Analysis Summary
  CMS-PAS-JME-16-003, 2017.

\bibitem{Khachatryan:2016kdb}
\hrefCMSnoop {}{{CMS Collaboration}, ``Jet energy scale and resolution in the
  {CMS} experiment in pp collisions at 8 {TeV}'',} \textit{ JINST} \textbf{ 12}
  (2017) P02014,
  \href{http://dx.doi.org/10.1088/1748-0221/12/02/P02014}{\doi{10.1088/1748-0221/12/02/P02014}},
\href{http://www.arXiv.org/abs/1607.03663}{\texttt{arXiv:1607.03663}}.

\bibitem{Sirunyan:2017ezt}
\hrefCMSnoop {}{{CMS Collaboration}, ``{Identification of heavy-flavour jets
  with the CMS detector in pp collisions at 13 TeV}'',} \textit{ JINST}
  \textbf{ 13} (2018), no.~05, P05011,
  \href{http://dx.doi.org/10.1088/1748-0221/13/05/P05011}{\doi{10.1088/1748-0221/13/05/P05011}},
\href{http://www.arXiv.org/abs/1712.07158}{\texttt{arXiv:1712.07158}}.

\bibitem{Khachatryan:2014ira}
\hrefCMSnoop {}{{CMS Collaboration}, ``{Observation of the diphoton decay of
  the Higgs boson and measurement of its properties}'',} \textit{ Eur. Phys. J.
  C} \textbf{ 74} (2014) 3076,
  \href{http://dx.doi.org/10.1140/epjc/s10052-014-3076-z}{\doi{10.1140/epjc/s10052-014-3076-z}},
\href{http://www.arXiv.org/abs/1407.0558}{\texttt{arXiv:1407.0558}}.

\bibitem{CMS-PAS-JME-13-005}
\href {http://cdsweb.cern.ch/record/1581583}{{CMS Collaboration}, ``Pileup jet
  identification'',} CMS Physics Analysis Summary CMS-PAS-JME-13-005, 2013.

\bibitem{CMS-DP-2015-067}
\href {https://cds.cern.ch/record/2118397}{{CMS Collaboration}, ``{Electron and
  photon performance using data collected by CMS at $\sqrt{s}$ = 13 TeV and 25
  ns}'',} CMS Detector Performance Summary CMS-DP-2015-067, 2015.

\bibitem{CMS-DP-2017-004}
\href {https://cds.cern.ch/record/2255497}{{CMS Collaboration}, ``{Electron and
  photon performance in CMS with the full 2016 data sample}'',} CMS Detector
  Performance Summary CMS-DP-2017-004, 2017.

\bibitem{Collins:1977iv}
\hrefCMSnoop {}{J.~C. Collins and D.~E. Soper, ``{Angular distribution of
  dileptons in high-energy hadron collisions}'',} \textit{ Phys. Rev. D}
  \textbf{ 16} (1977) 2219,
\href{http://dx.doi.org/10.1103/PhysRevD.16.2219}{\doi{10.1103/PhysRevD.16.2219}}.

\bibitem{Rainwater:1996ud}
\hrefCMSnoop {}{D.~L. Rainwater, R.~Szalapski, and D.~Zeppenfeld, ``{Probing
  color singlet exchange in $Z$ + two jet events at the CERN LHC}'',} \textit{
  Phys. Rev. D} \textbf{ 54} (1996) 6680,
  \href{http://dx.doi.org/10.1103/PhysRevD.54.6680}{\doi{10.1103/PhysRevD.54.6680}},
\href{http://www.arXiv.org/abs/hep-ph/9605444}{\texttt{arXiv:hep-ph/9605444}}.

\bibitem{Cowan:2010st}
\hrefCMSnoop {}{G.~Cowan, K.~Cranmer, E.~Gross, and O.~Vitells, ``Asymptotic
  formulae for likelihood-based tests of new physics'',} \textit{ Eur. Phys. J.
  C} \textbf{ 71} (2011) 1554,
  \href{http://dx.doi.org/10.1140/epjc/s10052-011-1554-0}{\doi{10.1140/epjc/s10052-011-1554-0}},
  \href{http://www.arXiv.org/abs/1007.1727}{\texttt{arXiv:1007.1727}}.
  [Erratum: \DOI{10.1140/epjc/s10052-013-2501-z}].

\bibitem{Dauncey:2014xga}
\hrefCMSnoop {}{P.~D. Dauncey, M.~Kenzie, N.~Wardle, and G.~J. Davies,
  ``{Handling uncertainties in background shapes: the discrete profiling
  method}'',} \textit{ JINST} \textbf{ 10} (2015) P04015,
  \href{http://dx.doi.org/10.1088/1748-0221/10/04/P04015}{\doi{10.1088/1748-0221/10/04/P04015}},
\href{http://www.arXiv.org/abs/1408.6865}{\texttt{arXiv:1408.6865}}.

\bibitem{CMS:2017sdi}
\href {https://cds.cern.ch/record/2257069}{{CMS Collaboration}, ``{CMS
  luminosity measurements for the 2016 data taking period}'',} CMS Physics
  Analysis Summary CMS-PAS-LUM-17-001, 2017.

\bibitem{CMS:2011aa}
\hrefCMSnoop {}{{CMS Collaboration}, ``{Measurement of the inclusive $W$ and
  $Z$ production cross sections in pp collisions at $\sqrt{s}=7$ TeV}'',}
  \textit{ JHEP} \textbf{ 10} (2011) 132,
  \href{http://dx.doi.org/10.1007/JHEP10(2011)132}{\doi{10.1007/JHEP10(2011)132}},
\href{http://www.arXiv.org/abs/1107.4789}{\texttt{arXiv:1107.4789}}.

\bibitem{Chatrchyan:2012jua}
\hrefCMSnoop {}{{CMS Collaboration}, ``{Identification of b-quark jets with the
  CMS experiment}'',} \textit{ JINST} \textbf{ 8} (2013) P04013,
  \href{http://dx.doi.org/10.1088/1748-0221/8/04/P04013}{\doi{10.1088/1748-0221/8/04/P04013}},
  \href{http://www.arXiv.org/abs/1211.4462}{\texttt{arXiv:1211.4462}}.

\bibitem{CMS-PAS-JME-16-004}
\href {https://cds.cern.ch/record/2205284}{{{CMS}} Collaboration,
  ``{Performance of missing energy reconstruction in 13 TeV pp collision data
  using the CMS detector}'',} CMS Physics Analysis Summary CMS-PAS-JME-16-004,
  2016.

\bibitem{Demartin:2010er}
F.~Demartin\hrefCMSnoop {}{ {et~al.}, ``{The impact of PDF and $\alpha_S$
  uncertainties on Higgs production in gluon fusion at hadron colliders}'',}
  \textit{ Phys. Rev. D} \textbf{ 82} (2010) 014002,
  \href{http://dx.doi.org/10.1103/PhysRevD.82.014002}{\doi{10.1103/PhysRevD.82.014002}},
\href{http://www.arXiv.org/abs/1004.0962}{\texttt{arXiv:1004.0962}}.

\bibitem{Carrazza:2015aoa}
S.~Carrazza\hrefCMSnoop {}{ {et~al.}, ``{An unbiased Hessian representation for
  Monte Carlo PDFs}'',} \textit{ Eur. Phys. J. C} \textbf{ 75} (2015) 369,
  \href{http://dx.doi.org/10.1140/epjc/s10052-015-3590-7}{\doi{10.1140/epjc/s10052-015-3590-7}},
\href{http://www.arXiv.org/abs/1505.06736}{\texttt{arXiv:1505.06736}}.

\bibitem{PhysRevLett.114.191803}
\hrefCMSnoop {}{{ATLAS and CMS Collaborations}, ``{Combined measurement of the
  Higgs boson mass in $pp$ collisions at $\sqrt{s}=7$ and 8 TeV with the ATLAS
  and CMS experiments}'',} \textit{ Phys. Rev. Lett.} \textbf{ 114} (2015)
  191803,
  \href{http://dx.doi.org/10.1103/PhysRevLett.114.191803}{\doi{10.1103/PhysRevLett.114.191803}},
\href{http://www.arXiv.org/abs/1503.07589}{\texttt{arXiv:1503.07589}}.

\bibitem{LHCHXSWG:YR3}
\hrefCMSnoop {}{{LHC Higgs Cross Section Working Group}, ``{Handbook of LHC
  Higgs cross sections: 3. Higgs Properties}'',} \textit{ CERN} (2013)
  \href{http://dx.doi.org/10.5170/CERN-2013-004}{\doi{10.5170/CERN-2013-004}},
\href{http://www.arXiv.org/abs/1307.1347}{\texttt{arXiv:1307.1347}}.

\end{thebibliography}\endgroup
